\documentclass[11pt]{article}
\pdfoutput=1

\usepackage{cite}
\usepackage{booktabs}
\usepackage[english]{babel}
\usepackage{amsmath,amssymb,amsbsy,amstext, amsthm, simplewick}
\usepackage{hyperref} 
 \hypersetup{ 
     colorlinks=true, 
     linkcolor=blue, 
     filecolor=blue, 
     citecolor = black,       
     urlcolor=cyan, 
     }
\usepackage{graphicx}
\usepackage{amsfonts}
\usepackage{amssymb}
\usepackage[small]{caption}
\usepackage{upgreek}
\usepackage[svgnames,dvipsnames,x11names,table]{xcolor}
\usepackage{multirow} 
\usepackage{geometry}
\usepackage[hang,flushmargin]{footmisc}
\usepackage{bm}
\usepackage{braket}
\usepackage{subcaption}
\usepackage{simpler-wick}
\usepackage{mathtools}

\usepackage{colortbl}

\makeatletter
\newlength{\apb@width}
\newcommand{\autoparbox}[2][c]{\settowidth{\apb@width}{#2}\parbox[#1]{\apb@width}{#2}}

\makeatother

\definecolor{lightgray}{gray}{0.9}

\usepackage[framemethod=default]{mdframed}
\newmdenv[skipabove=7pt,
skipbelow=7pt,
rightline=false,
leftline=false,
topline=false,
bottomline=false,
backgroundcolor=gray!10,
linecolor=gray,
innerleftmargin=5pt,
innerrightmargin=5pt,
innertopmargin=5pt,
innerbottommargin=5pt,
leftmargin=0cm,
rightmargin=0cm,
linewidth=4pt]{eBox}

\numberwithin{equation}{section}

\def\beq{\begin{equation}}
\def\eeq{\end{equation}}

\def\bea{\begin{eqnarray}}
\def\eea{\end{eqnarray}}

\def\Neff{N_{\rm eff}}
\def\beq{\begin{equation}}
\def\eeq{\end{equation}}
\def\bea{\begin{eqnarray}}
\def\eea{\end{eqnarray}}

\def\T{{\mathcal T}}

\def\vl{{\vec{\ell}}}
\def\L{{\vec{L}}}

\def\Mpl{M_{\rm pl}}
\def\Mp{M_{\rm pl}}

\def\fnl{f_{\rm NL}}
\def\fnlloc{f^{\rm loc}_{\rm NL}}
\def\fnleq{f^{\rm eq}_{\rm NL}}

\def\H{{\cal H}}

\def\n{{\hat n}}

\def\k{{\vec k}}

\def\p{{\vec p}}

\def\vc{{\vec v}}
\def\x{{\vec x}}

\DeclareRobustCommand{\SkipTocEntry}[4]{}

\definecolor{blue3}{RGB}{31, 119, 180}
\definecolor{red3}{RGB}{	214, 39, 40}
\definecolor{orange3}{RGB}{255, 127, 14}
\definecolor{green3}{RGB}{44, 160, 44}

\begin{document}

\begin{titlepage}
\setcounter{page}{1} \baselineskip=15.5pt 
\thispagestyle{empty}

\begin{center}
{\fontsize{18}{18} \bf TASI Lectures on \\[5pt]
Cosmic Signals of Fundamental Physics}
\end{center}

\vskip 20pt
\begin{center}
\noindent
{\fontsize{12}{18}\selectfont  Daniel Green }
\end{center}

\begin{center}
\vskip 4pt
\textit{ Department of Physics, University of California San Diego, La Jolla, CA 92093, USA}

\end{center}

\vspace{0.4cm}
 \begin{center}{\bf Abstract}
  \end{center}  
 \noindent The history of the Universe and the forces that shaped it are encoded in maps of the cosmos.  From understanding these maps, we gain insights into nature that are inaccessible by other means.  Unfortunately, the connection between fundamental physics and cosmic observables is often left to experts (and/or computers), making the general lessons from data obscure to many particle theorists.  Fortunately, the same basic principles that govern the interactions of particles, like locality and causality, also control the evolution of the Universe as a whole and the manifestation of new physics in data. By focusing on these principles, we can understand more intuitively how the next generation of cosmic surveys will inform our understanding of fundamental physics.  In these lectures, we will explore this relationship between theory and data through three examples: light relics ($\Neff$) and the cosmic microwave background (CMB), neutrino mass and gravitational lensing of the CMB, and primordial non-Gaussianity and the distribution of galaxies.  We will discuss both the theoretical underpinnings of these signals and the real-world obstacles to making the measurements. 
\end{titlepage}

\restoregeometry

\newpage
\setcounter{tocdepth}{2}
\tableofcontents

\newpage

\section{Introduction}

Cosmology is the greatest scientific discipline.\footnote{I have taken some liberty of injecting some hyperbole and personal opinion into these lectures.} It spans all of the length scales in physics, from the Planck scale to our cosmological horizon.  It also informs our understanding of the origin, history, and fate of our Universe; it even hints at the possibility of universes beyond ours.  Yet, cosmology doesn't merely pertain to existential questions, it answers them through information encoded in maps of the Universe that are accessible with current or near-term technology. One ongoing challenge for the modern cosmologist is decoding these maps to isolate the answers to the questions that are of the most pressing interest.

Traditional presentations of cosmology often tell only a piece of this story: the fundamental theory that allows us to ask deep questions about nature is treated as an independent subject from the phenomenology of the maps that are used to answer them.  Of course, both presentations acknowledge their dependence on the other, but usually through a handful of parameters that serve as a Rosetta Stone to relate theory and observations.  Few practitioners hold in their mind both the theoretical underpinnings and the complexity of the cosmological data as being two components of a whole.  Yet, the fundamental principles that shape the evolution of the Universe are the same across epochs, and inform how we think about inflation, galaxy formation, data analysis, and more.

The purpose of these lectures is to give you a flavor of this connection between theory and data.   Of course, the downside is that we will not have time to delve as deeply into each of our subjects as one might like.  Fortunately, there are excellent textbooks~\cite{Dodelson:2003ft,Weinberg:2008zzc,Baumann:2022mni}, TASI lectures~\cite{Baumann:2009ds,Silverstein:2016ggb,Cline:2018fuq,Baumann:2018muz}, and white papers~\cite{Dvorkin:2019jgs,Flauger:2022hie,Baumann:2022jpr,Dvorkin:2022jyg,Achucarro:2022qrl,Green:2022hhj} on many of these topics, whether it is inflation, thermal history, the cosmic microwave background (CMB), or large scale structure (LSS). In contrast, there is comparatively less material that synthesizes these perspectives into a single subject.  Our goal will be to learn about the connections between these topics, particularly with an eye towards testing fundamental physics with cosmological data.

\subsection*{Notation}

I will assume some basic familiarity with general relativity, cosmological spacetimes, and thermodynamics.    We will always be working in flat Friedmann-Robertson-Walker (FRW) background,
\beq
ds^2 = -dt^2 + a(t)^2 dx^2 \qquad H(t) = \frac{\dot a(t)}{a(t)} \ ,
\eeq
where we will define $\dot F(t) \equiv \frac{d}{dt} F(t)$. We will set $c=\hbar = k_B =1$ throughout, and define $\Mpl$ as the reduced Planck mass, $\Mpl^2 \equiv 1/(8 \pi G)$ where $G$ is Newton's gravitational constant.  With these choices, the behavior of the FRW solution is governed by the Friedmann equation,
\beq
3 \Mpl^2 H^2 = \sum_i \bar \rho_i(t) \ ,
\eeq
where $\rho_i(\x, t)$ are the energy densities of various species, and we define the homogenous density $\bar \rho$ and fluctuations $\delta \rho$ via $\rho_i(\x,t)= \bar \rho_i(t) +\delta \rho_i(\x,t)$. It is useful to recall that matter, radiation, and vacuum energy (dark energy) redshift like $a^{-3}$, $a^{-4}$, and $a^0$ respectively. Applying the Friedmann equation, the expansion rate scales as $H(t) \propto a^{-3/2},\, a^{-2}$, and $a^0$ during the respective epochs in the standard $\Lambda$CDM cosmology where each of these energy densities dominates the energy budget of the Universe.  We will also sometimes define
\beq
\Omega_i H_0^2  = \bar \rho_i(t_{\rm today}) \qquad \omega_i = \Omega_i h^2
\eeq
where $h = H_0 /(100 \, {\rm km/s/Mpc})$ is introduced so that $\omega_i$ behaves like a density as we change cosmological parameters while still carrying the units of $\Omega_i$.

At times, we will discuss perturbations of the metric and matter distribution using conformal Newtonian gauge such that (following~\cite{Baumann:2015rya})
\beq
\mathrm{d} s^{2}=a^{2}(\tau)\left[(-1-2 \Phi) \mathrm{d} \tau^{2}+(1-2 \Psi) \delta_{i j} \mathrm{~d} x^{i} \mathrm{~d} x^{j}\right] \ ,
\eeq
and work with the Weyl potential, $\Phi_+ \equiv \Phi+\Psi$, where $dt = a(\tau) d\tau$ define the conformal time variable $\tau$ (sometimes called $\eta$ instead). To clarify when we are using $\tau$, we will define 
\beq
X' \equiv \frac{\partial}{\partial \tau } X \qquad  \H(\tau) = \frac{a'(\tau)}{a(\tau)} = H(\tau) a(\tau) \ .
\eeq
Conformal time is particularly useful during inflation where $H$ is approximately constant and $a(\tau) = -1/(H \tau)$.  To avoid potential confusion with conformal time, we will define the optical depth to reionization as $\bar \tau_{\rm optical}$.  

Throughout these lectures, we will discuss fields in both position and Fourier space for the spacial coordinates $\x$.  We will define the Fourier transform of a field $\phi(\x,t)$ as $\phi(\k,t)$ so that 
\beq
\phi(\k,t) = \int d^3 x e^{i \k \cdot \x} \phi(\x,t) \qquad \phi(\x,t) = \int \frac{d^3 k}{(2\pi)^3} e^{-i \k \cdot \x} \phi(\k,t) \ .
\eeq
We will often use the notation $k = |\vec k|$ to denote the length of a vector, and $\hat k = \k / k$ is the unit vector.

It is also useful to remember that the redshift of an object, $z$, is related to the amount of expansion between the time the light was emitted ($t$) and today ($t_{0}$) by
\beq
\frac{a(t_0)}{a(t)} = 1+z(t) \ .
\eeq
If we set $a_0\equiv a(t_0)= 1$ for convenience, then we can define $ a(t) = 1/(1+z(t))$ with $z=0$ being the same as $t_{0}$, as it should.  While we could, in principle, discuss most of theoretical cosmology without ever mentioning the redshift directly, it is useful for making contact with observations where the redshift is what you measure.  

Finally, figures showing the power spectra were calculated using the CLASS Boltzmann code~\cite{Blas:2011rf}, unless otherwise stated.

\newpage

\section{Coins and Cosmology} \label{sec:coins}

In most disciplines of physics, there is the idea of signal and noise, and our task is to isolate the signals.  Cosmological maps, in contrast, are literally maps of noise.  This is not even hyperbole, the fluctuations of interest in these maps are simply random sound waves that propagated though the Universe.  To understand the role of fundamental physics in the Universe, we will also have to grapple with the meaning of the noise in our maps.  

\subsection{Bayesian Inference}

The challenge of inferring cosmological parameters from maps of the Universe can be understood as a generalization of the problem of determining if a coin is fair.  Suppose we have a coin where a flip will return heads with probability $p$ and tails with probability $1-p$. Of course, we don't actually know what $p$ is and our goal is to determine it by flipping the coin $N$ times.  The data we get from performing this experiment is that after $N$ flips we got $n$ heads and $N-n$ tails (we will take it as a given that the order of the heads and tails is not meaningful).  If we were told ahead of time what the correct value of $p$ was, we could calculate the probability of this outcome.  However, that is not the problem we are faced with.  Instead, given this outcome, we want to know what values of $p$ are consistent with our experiment.  

The most straightforward way to understand this problem is to use {\it Bayes' Theorem}.  The theorem is stated as follows
\vskip 7pt
\noindent {\bf Bayes' Theorem:}  {\it Given a data set, $D$, and a hypothesis, $H$ (a model), that would explain the data, the probability of the model given the data, $P(H | D)$ (the posterior), is 
\beq
P(H | D) = \frac{P(D | H) P(H)}{P(D)} \ ,
\eeq
where $P(D | H)$ is the probability of the data given the model (also known as the likelihood), $P(D)$ is the probability of the data, and $P(H)$ is the probability of the hypothesis.}
\vskip 7pt
The likelihood, $P(D | H)$, is something you should be able calculate given the model's assumptions. The prior, $P(H)$, is what we think the probability is that $H$ is correct before we see the data (it is helpful to imagine that $H$ is part of a family of hypotheses that we might not view as equal ahead of time).  Usually, it is a good idea to make the prior pretty weak, like $P(H) =$ constant, so that $P(H|D)$ is determine by the data and not our prior beliefs.  However, $P(H)$ may also be used to include information from other experiments and therefore it may play an important role.  Finally, the probability of the data (also known as the evidence), $P(D)$, is simply the probability that you would observe the specific data set given all the possible hypotheses you will consider.  Concretely, we really want $P(H | D)$ to be a probability distribution for the parameters of a model, so it should be true that 
\beq
\int dH P(H | D) = 1  \ .
\eeq
Notice that $P(D)$ does not depend on $H$ so once we know the data, $D$, $P(D)$ is simply a constant.  We can therefore define
\beq
P(D) = \int dH P(D | H) P(H)
\eeq
so that we have a normalized probability distribution.

\begin{figure}[h!]
\centering
\includegraphics[width=3.5in]{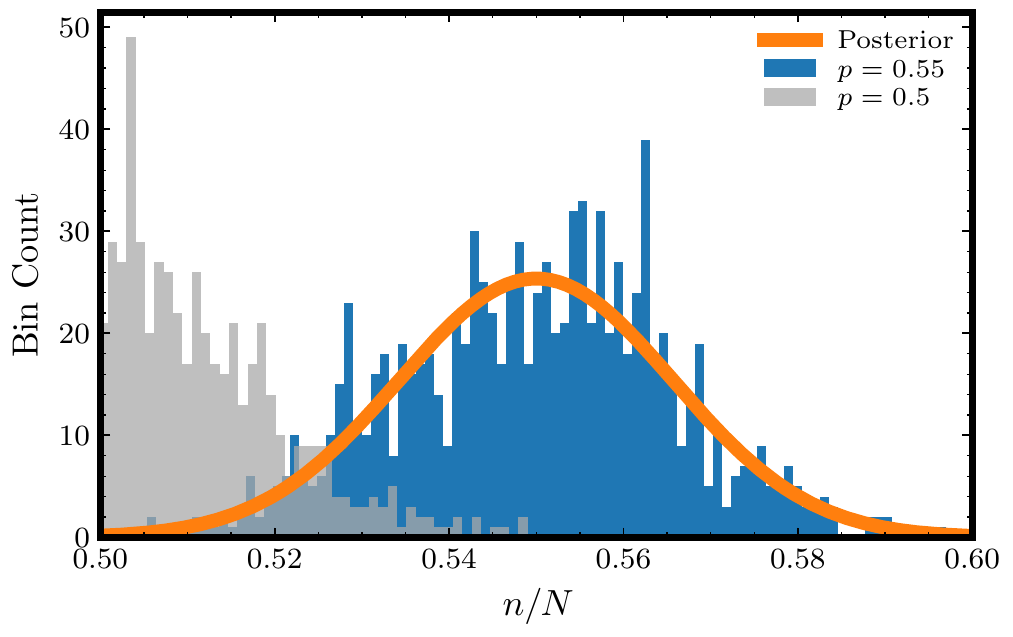}
\caption{The orange line is the posterior for $p$ from $N=1000$ and $n=550$.  This curve has little support for an fair coin, $p=0.5$.  Using (\ref{eq:p_sigma}) we have $\sigma = 0.016$, which would exclude $p=0.5$ at approximately $3\sigma$.  The blue histogram show the distributions of outcomes for $n/N$ from 1000 simulations, each with $N=1000$ flips and $p=0.55$. The grey histogram is the same distribution of outcomes for 1000 simulations each using $N=1000$ flips, but now using fair coin, $p=0.5$.  We see that our Bayesian inference (orange curve) is a very accurate representation of the range of possibilities that come from simulating the data with $p=0.55$ (blue histogram).}
\label{fig:flips}
\end{figure}

Bayes' theorem is much easier to understand when we apply it to examples, so let's return to the problem of coin flipping: our hypothesis is that the coin has probability $p$ of heads and $1-p$ of tails.  We assume we know nothing about $p$ ahead of time, so that $P(H = p) = $ constant.  We have in hand $N$ flips of a coin returning $n$ heads and $N-n$ tails.  The probability of getting this data given our $p$ is
\beq
P(n | p ) = p^n (1-p)^{N-n}  \binom{N}{n} = p^n (1-p)^{N-n}  \frac{N!}{n! (N-n)!} \ .
\eeq
The probability of the data is simply the integral over all $p$:
\beq
P(n ) =\int_0^1 dp \, p^n (1-p)^{N-n}  \frac{N!}{n! (N-n)!} \ ,
\eeq
so that the posterior for $p$ is given by
\beq
P(p | n) = \frac{ p^n (1-p)^{N-n} }{\int_0^1 dp \,  p^n (1-p)^{N-n}} \ .
\eeq
This result is shown in Figure~\ref{fig:flips} for the choice $N=1000$ and $n=550$.  A priori, it might have seemed reasonable to get 550 heads from 1000 flips of a fair coin.  Not only does Bayes' theorem tells us the fair coin is unlikely, repeating the experiment of flipping 1000 coins fair over and over again shows that it very rare to get 550 heads from a fair coin but the distribution simulations when we take our coin to have $p=0.55$ matches the posterior we derived using Bayes' theorem. 

In order to build intuition, let us write $ n = \bar p N$ so that
\beq
p^n (1-p)^{N-n} = \exp \left( N (\bar p \log p + (1-\bar p) \log (1-p) )\right) \ .
\eeq
At large $N$ we expect this will become very sharply peaked as reflected by the narrow Gaussian in Figure~\ref{fig:flips}.  The maximum value of our posterior, $p_0$, is determined by 
\beq
\frac{d}{dp } p^n (1-p)^{N-n}|_{p_0}  = 0 \to  \frac{\bar p}{p_0} - \frac{(1-\bar p)}{1-p_0} = 0 \to p_0 = \bar p  \ .
\eeq
This seems reasonable, our best guess for $p$ given $n$ heads is that $p = n/N$.  We also want to know how certain we are of this possibility, so let's find the behavior around the maximum, $p=\bar p+ \delta p$, and expanding the argument of the exponential in $\delta p$,
\bea
P(p=\bar p +\delta p | n) &\approx& C \exp \left( - \frac{N}{2\bar p} \delta p^2 - \frac{N}{2 (1-\bar p)} \delta p^2 \right) = C \exp \left( - \frac{N}{2\bar p(1-\bar p) } \delta p^2 \right) \ ,
\eea
for some constant $C$.  From here, we notice our first important result: we have found that, around the maximum, our posterior is a Gaussian distribution with a variance
\beq\label{eq:p_sigma}
\sigma^2 = \frac{\bar p(1-\bar p)}{N} \ .
\eeq
This tells us that our the uncertainty in $p$ decreases as $1/\sqrt{N}$.  In other words, if we want to lower the uncertainty by a factor of 10, we need $100\times N$ flips the coin!  

The other thing to notice is that we have an overall factor of $N$ times a function only of $p$ and $\bar p$. If we include higher powers of $\delta p$, we will not generate any new factors of $N$ and we find
\beq
P(p=\bar p +\delta p | n)  = C\exp \left( - \frac{N}{2\bar p(1-\bar p) } \delta p^2 - \frac{N}{3} \left(\frac{1}{(1-\bar p)^2} - \frac{1}{\bar p^2} \right) \delta p^3 +\ldots \right) \ .
\eeq
First, we notice that  the posterior will become exponentially small when $\delta p \sim \sigma = {\cal O}(N^{-1/2})$.  At this point $N \delta p^2 ={\cal O}(1)$, which tells us that $N \delta p^3 = {\cal O}(N^{-1/2})$.  In other words, the higher-order terms in the Taylor expansion aren't very important in region that determines the probability of $p$, up to some exponentially small piece. In this precise sense, adding data (flipping the coin more times) not only decreases the variance in our knowledge of $p$, but it makes the posterior increasingly Gaussian (the central limit theorem in action).  For the very same reason, this result didn't really depend on the fact that it was a coin but would essentially be true if we draw $N$ independent numbers for a random distribution.

Notice that we could have extracted the errors, assuming that it reaches a Gaussian distribution, by calculating
\beq
\frac{1}{\sigma^2} = - \frac{\partial^2}{\partial p^2} \log P(p | n)|_{p =\bar p} \ ,
\eeq
where we recall the first derivative will vanish, by definition, at $p=\bar p$ (the maximum of the probability distribution).  If we were to extend this to models to include more parameters (e.g. dice with probabilities for each of its sides, $p_i$), this expression becomes a matrix where the variances are determined by inverting the matrix
\beq\label{eq:fisher_def}
F_{ij} =\frac{\partial^2}{\partial p_i \partial p_j} \log P(\{p_i \} | n)|_{p_i =\bar p_i} \to \sigma_i^2 = [F^{-1} ]_{ii} \ .
\eeq
We have discovered the Fisher information matrix, $F_{ij}$, which one can also derive using more formal statistical techniques.  But, fundamentally, all the intuition can be reduced back to coin flipping, or any other calculable example you like.

\vskip 5pt
\noindent {\bf Takeway:} \hskip 5pt The way we learn properties of a probability distribution is by sampling the distribution many times.  Given a model for the distribution with some set of parameters, the posterior distribution for those parameters will approach a Gaussian with variances $\sigma^2 \propto 1/N$ when $N$ is sufficiently large. For more information on the topics in this section, see~\cite{Dodelson:2003ft,Heavens:2009nx}.

\subsection{Cosmic Microwave Background}

Much of our understanding of the history of the Universe is informed by measurements of the cosmic microwave background (CMB).  To first approximation, the CMB is a blackbody distribution of photons with a universal temperature of 2.7~K in every direction.  It is a relic from the time when the Universe first became neutral, roughly 380~000 years after the Big Bang, often called the epoch of recombination.\footnote{Like many, but not all, uses of `re' in cosmology, this is the first time the protons and electrons have combined to form hydrogen so it might be more accurate to call it ``combination".} Prior to recombination, the temperature of the Universe was high enough to prevent hydrogen from forming, $T \gg 1$ eV, and the Universe formed a hot plasma of electrons and protons.  For the purposes of this section, we are still late enough in the history of the Universe where dark matter and neutrinos  were only coupled gravitationally to other particles, as we will discuss later.  Photons and free electrons scatter efficiently and the combined system of photons, protons and electrons (including helium which had formed earlier) behaved as a single fluid, which cosmologists often give the confusing name ``the photon-baryon fluid".\footnote{Protons and electrons are ``baryons" presumably in the same sense that everything heavier than helium is called a ``metal" by astronomers.}

Baryons and electrons are vastly outnumbered by photons, by roughly a factor of $10^{10}$ (see e.g.~\cite{Cline:2018fuq} for discussion and the relation to baryogenesis).  This fact is important for two reasons: first, it explains why the Universe remained ionized to temperatures well below the binding energy of hydrogen (13.6 eV).  In order to ionize hydrogen, we only need $1$ in $10^{10}$ photons to have energies above $13.6$ eV.  If we simply estimate the temperature of recombination, $T_\star$ by the point when $e^{-13.6 \, {\rm eV} /T_\star} = 10^{-10}$ we find $T_\star \sim 0.5$ eV which is a close to the more precise value of $T_\star = 0.2$ eV. The second important consequence is that the energy density of the plasma is dominated by the photons, which are relativistic.  As a result, sound waves traveling through the photon-baryon fluid will have a large sound speed,
\beq
c_s^2 = \frac{1}{3(1+R_b)} \approx \frac{1}{3} \ ,
\eeq
where $R_b =\omega_b/\omega_\gamma\propto a(t)$ is the ratio of energy in baryons to photons. During the radiation era $R_b \ll 1$, and grows to $R_b(t_\star) = 0.6$. at recombinaton. The baryons are heavy and their inertia is responsible for slowing down the sound waves in this fluid.  All of the pressure in the fluid is supplied by photons which are the only reason there are sound waves at all.  However, since photons don't scatter with themselves, the presence of free electrons is essential for keeping the fluid together. Even though the electrons dominate the scattering with photons, the Coulomb force between electrons and protons brings the nuclei along for the ride and meaningfully alters the inertia of the waves.

Once neutral hydrogen forms, the photons are no longer bound in place by scattering and the baryons lose their source of pressure.  The photons travel from that point forward without scattering again, to first approximation.  As a result, the phase space distribution of the photons goes effectively unchanged, up to the expansion of the Universe which rescales the wavelengths of the photons and the temperature of the distribution in unison.  A small fraction of photons will scatter with electrons after the Universe is reionized\footnote{This is a correct use of `re'.} by stars or will be gravitationally lensed by the matter, but otherwise we are left with a relic of the recombination era.\footnote{One might wonder why the CMB is not returned to the form of a fluid after reionization?  The reason is that recombination occurs at $z\sim 1100$ and reionization happened around $z\sim 6$.  The density of electrons is therefore $(1+6)^3/(1+1100)^3 = 10^{-7}$ times smaller at $z=6$, which makes scattering during reionization rare.}

\begin{figure}[h!]
\begin{center}
\includegraphics[width=4.5in]{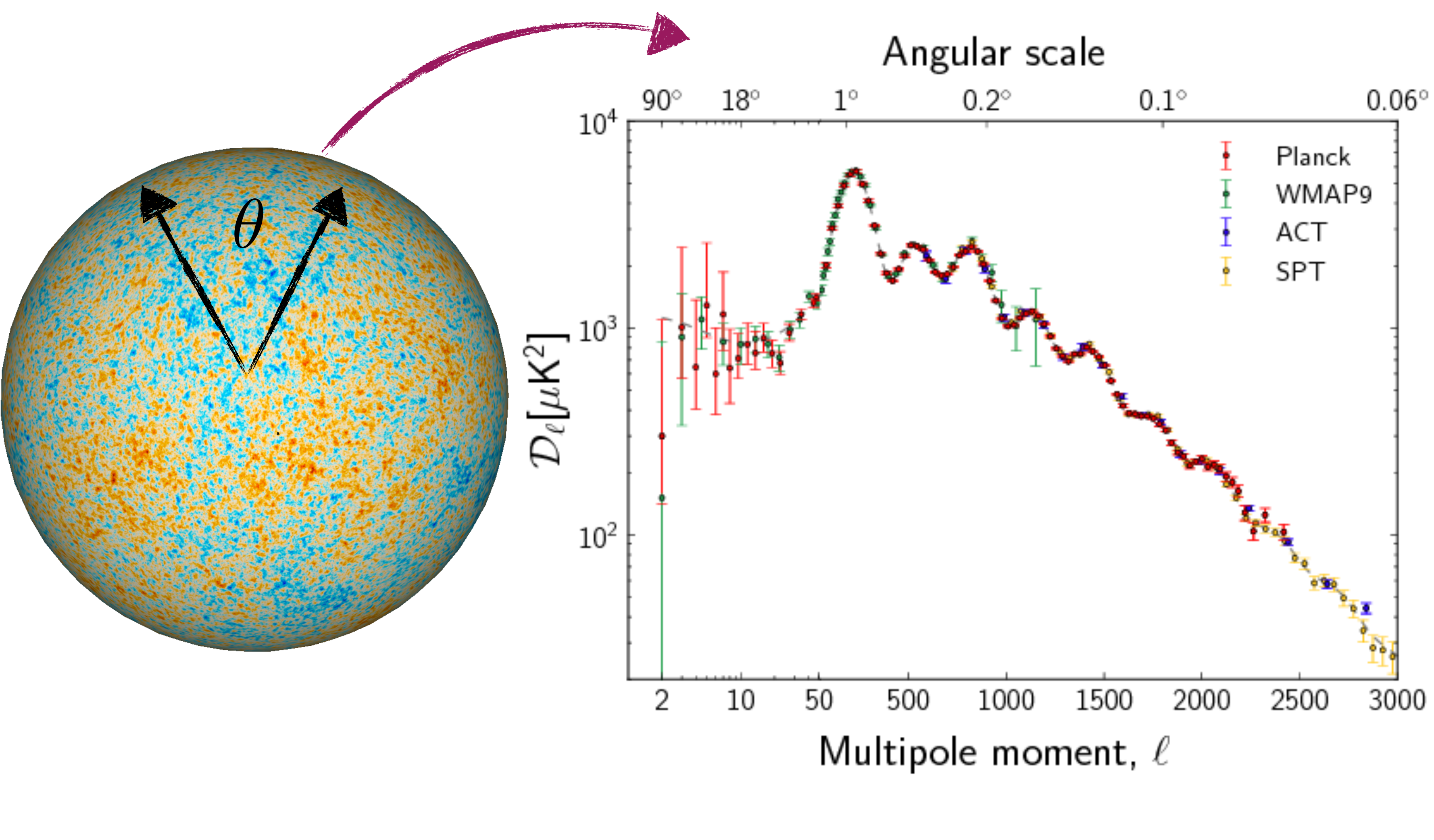}
\end{center}
\caption{{\it Left:} Planck map for the temperatures fluctuations across the sky, via \href{http://thecmb.org}{thecmb.org}. {\it Right:}  Power spectrum of fluctuations of the CMB temperature as a function of the angular size of the fluctuations or, equivalently, $\ell$ of the associated spherical harmonics. }
\label{fig:cmb_power}
\end{figure}

The image of the CMB temperature from Planck (Figure~\ref{fig:cmb_power}) is presumably well-known to most readers and looks like a sphere covered in spots.  These spots on the CMB sky are variations in temperature of order $10^{-5}$ from point to point.  What this map represents is an image of the sound waves that were propagating in the plasma at that time of recombination. Decomposing the sound waves according to their wavevector, $\vec k$, then we would expect the variation in temperature (or energy density) due to the wave will follow the solutions of the wave equation\footnote{What we observe is actually a combination of the local energy density (including the gravitational redshift) and the velocity of the fluid along the line-of-sight (Doppler shift). We will be dropping the Doppler term for now, as it is somewhat smaller and including it doesn't alter the overall conclusion. }, namely
\beq
\frac{\delta T}{T}(\k) = A_{\k} \cos(c_s k \tau_\star) + B_{\vec k} \sin(c_s k \tau_\star) \ ,
\eeq
where $\tau_\star$ is the (conformal) time at recombination, and $A_\k$ and $B_\k$ are numbers obeying the reality condition $A^*_\k =A_{-\k}$ and $B^*_\k =B_{-\k}$.  When we look at the CMB, there is no obvious discernible wave-like pattern, it just looks like a bunch of random spots (although maybe you can see that many of the spots seem similar in size, which we will explain later).  This is a reflection of the fact that the amplitudes $A_\k$ and $B_\k$ are random variables with
\beq
\langle A_\k \rangle = \langle B_\k \rangle = 0 \ .
\eeq
So far, this isn't really surprising - unless there was some coherence source of sound that picked out some cosmologically enormous scale, we would have expected the sound to be we generated by something random on small scales like thermal fluctuations, in which case we might expect that $A_\k$ and $B_\k$ are drawn from random distributions~\cite{Dodelson:2003ip}.  Since the Universe is rotationally invariant, we would also assume that the distribution should be independent of the direction of the $\k$ vector and because it is translationally invariant, the statistics will look like momentum (defined by $\k$) is conserved. If we calculate the two-point statistics of the temperature fluctuations, we should find
\begin{align}
\langle \delta T(\k) \delta T (\k') \rangle =  \Bigg[ P_A(k) \cos^2(c_s k \tau_\star) & + 2 C_{AB}(k) \sin(c_s k \tau_\star) \cos(c_s k \tau_\star) \nonumber \\
&+ P_B(k) \sin^2(c_s k \tau_\star) \Bigg] (2\pi)^3 \delta(\k +\k')
\end{align}
where 
\beq
P_A(k)  = \langle A(\k) A(\k') \rangle' \ , \qquad C_{AB}(k)  = \langle A(\k) B(\k') \rangle' \ ,\qquad P_B(k)  = \langle B(\k) B(\k') \rangle' \ ,
\eeq 
and $\langle .. \rangle$ is the statistical average over these random amplitudes and we defined $\langle .. \rangle = \langle .. \rangle' (2\pi)^3 \delta(\k+\k')$.

Now, if we think about sound waves in a room created by a random source, you might reasonably assume that the amplitude was drawn from some smooth distribution and that the phase is drawn from a uniform distribution in $[0,2\pi)$.  Except for things like lasers, making waves in phase is usually very difficult.  Making the assumption of a uniform phase distribution implies that $C_{AB}(k)= 0$ and $P_A(k) = P_B(k)\equiv P_{\rm source}(k)$.  While this may seem like a reasonable guess, when we plug it into the temperature power spectrum gives us
\beq\label{eq:CMB_sound}
\langle \delta T(\k) \delta T (\k') \rangle  \stackrel{?}{=} P_{\rm source}(k) (2\pi)^3 \delta(\k +\k') \ .
\eeq
We will see that this isn't correct.  The first sign of trouble is that this answer does not depend on parameters like the sound speed, $c_s$, as all the physics of the propagation of the sound waves averaged out (at least to first approximation).

So what do we know about this distribution from data? Well, if we stick with our assumption that the answer doesn't depend on direction, then we can take every $\delta T(\k)$ with the same $k$ but different direction $\hat k$ and take their average.  If this is a valid assumption, then determine the amplitude of $\langle \delta T(\k) \delta T (\k') \rangle$ is literally the same problem as drawing $N$ samples from a probability distribution and using it to infer the variance of the distribution (i.e.~it is literally just our coin flipping problem).  In detail, the number of samples for a given $k$ is not our choice but is determined by the spherical geometry of the sky (in the case of the CMB).  If we were sampling the entire sky, we can decompose the sphere into spherical harmonics, 
\beq
T(\hat n) = \sum_{\ell=0}^{\infty} \sum_{m=-\ell}^{\ell} a_{\ell, m} Y_{\ell,m}(\theta, \varphi) \ , 
\eeq
where $\hat n$ is a direction on the sky determined by $(\theta, \varphi)$, On small angular scales (large $\ell$), the decomposition into spherical harmonics will take the form of a Fourier transform where $k  \approx \ell /(\tau_0-\tau_\star)$ (recalling that $\tau_0-\tau_\star$ is the distance to the last scattering surface), $m$ is the orientation of $\k$, and $a_{\ell,m} \approx \delta T(\k)/T$. The main takeway is that the CMB is like our coin flipping problem where for each $\ell \leftrightarrow k$, each $m$ corresponds to a single flip (draw from the random distribution), such that we have $N=2\ell +1$ total samples for each $\ell$.

The Planck map of $\delta T(\hat n)$ across the sky and the associated power spectrum are shown in Figure~\ref{fig:cmb_power}, where 
\beq
C^{\rm TT}_{\ell} = \frac{1}{2\ell +1} \sum_{m=-\ell}^{\ell} a_{\ell,m} a_{\ell,-m} \qquad D_\ell \equiv \ell(\ell+1)C^{\rm TT}_{\ell}  / (2\pi) \ .
\eeq
For $\ell < 1500$, the pattern of the error bars follows exactly the expectation from coin flipping: the error bars are set by the number of modes and scale as $1/\sqrt{2\ell +1}$ and are therefore largest at low~$\ell$.  This contribution to the error of the observed CMB power spectrum is called {\it cosmic variance}: it is the name for the minimum amount of error we need to include because we are just measuring a finite numbers of samples drawn from a probability distribution.  For $\ell > 1500$, the error begins to increase again due to the instrumental noise of the Planck detectors.  We will return to discuss this in detail in Section~\ref{sec:Neff}.  

We expect that the shape of the power spectrum should follow\footnote{ Factors of $\ell(\ell+1)$ and $k^3$ are needed account for the difference between a two-dimensional transform on the sky and three-dimensional Fourier transform in space.} the temperature power spectrum, $\ell(\ell+1)C_\ell \approx k^3P(k)|_{k=\ell/(\tau-\tau_\star)}$.  In Figure~\ref{fig:cmb_power}, we clearly notice an oscillatory pattern that is certainly much larger than the errors.  This is completely inconsistent with our assumptions that lead to Equation~(\ref{eq:CMB_sound}), unless $P_{\rm source}(k)$ itself contains these oscillations.  However, such an assumption is already a problem, as it would seem to break the idea that the source of noise is local; if this were the case, we would expect some kind of polynomial in $k$ so that it Fourier transforms to something localized, perhaps multiplying an overall power-law if the source has frequency dependence. 

If we instead give up on the idea that the phase is random~\cite{Dodelson:2003ip}, then the CMB observations match exactly what we would expect if the amplitudes of all sine waves was zero so that
\beq\label{eq:power_spec}
P_A(k) = \frac{A_s}{k^{3}} k^{n_s-1} \qquad P_B(k) = C_{AB} (k) = 0 \ .
\eeq
This has the advantage of explaining the frequency in terms of physical parameters like $c_s$ but leaves us with a serious puzzle of how these sound waves were created in phase.  To the accuracy we have measured\footnote{In detail, the actual CMB temperature receives a smaller contribution from the Doppler shift that is proportional to the velocity of the fluid at last scatter, $\vc \cdot \hat n \propto A_\k \sin(c_s k \tau_\star)$.  This is not of the same form as a phase shift of the density itself and thus is distinguishable from $B\neq0$.  More importantly, our puzzle remains even if we have $0<P_B(k)\ll P_A(k)$.} in the CMB, the phases are indeed identical and it is hard to imagine how a purely local process during recombination, such as thermal fluctuations, could give rise to such large scale coherence.
\vskip 5pt
The puzzle we have just encountered is just another incarnation of {\it the horizon problem} (see e.g.~\cite{Baumann:2009ds} for discussion). The fluctuations in CMB do not arise from local physics around the time of recombination.  Instead, there is a very delicate pattern of correlations between sound waves traveling in different directions that require something more.  This puzzle is resolved by inflation, a period of exponential expansion, but what the data from the CMB is really telling us (more model independently) is that the physics responsible for the origin of these sound waves, and ultimately the origin of all structure in the Universe, must have come from an epoch {\it before the hot Big Bang}.  

The way it works is as follows: we described the observed fluctuations as sound waves because when their wavelengths are small compared to the scale of the spacetime curvarture, they behave like density fluctuations of a plasma in flat space and follow the solutions of a wave equation.  However, if we postulate that these waves were not actually created during the era of recombination, but existed long before, then if we go back far enough in time, their wavelength was larger than the curvature scale of the Universe.  Concretely, the modes we see at recombination were definitely smaller than the curvature scale or equivalently
\beq
k_{\rm physical}(t_\star) = \frac{k}{a(t_\star)}   \gg H(t_\star) \ .
\eeq
Of course, because the Universe expands, $a(t)$ is monotonically increasing and $k_{\rm physical}$ is monotonically decreasing.  Therefore, we must have
\beq
 k_{\rm physical}\left(t_\star \right) \,  \ll  \,k_{\rm physical}\left(t \ll t_\star\right) \ ,
\eeq
or that the physical wavenumber was larger in the past.  However, the curvature scale of the Universe also changes as the matter and radiation densities redshift so the comparison of $k_{\rm physical}(t)$ and $H(t)$ is non-trivial. Well before recombination, the evolution of the Universe was dominated by the radiation. The Friedmann equation tells us
\beq
3 \Mpl^2  H^2 = \rho(t) \ ,
\eeq
and the radiation redshifts like $\rho \propto a^{-4}$. Therefore, during the radiation era $H(t) \propto a^{-2}$, which tells us that the time evolution of $H(t)$ from the dilution of the energy in radiation is a larger effect than the redshifting of the wavelength of the sound wave. As a result, if we go far enough to the past (while maintaining radiation domination), a physical wavenumber was much smaller than the Hubble scale (curvature scale of the Universe), assuming it was created long before,
\beq
k_{\rm physical}(t \ll t_{\star} ) \ll H(t \ll t_{\star} ) \ .
\eeq  
In this regime, the solutions do not oscillate as their evolution is dominated by the curvature of the Universe rather than the pressure of the fluid. It turns out the solutions in this regime are power laws always given by~\cite{Weinberg:2003sw}
\beq
\delta T(\k,t) \propto c_\k + a^{-3} d_\k \ ,
\eeq
where $c_\k$ and $d_\k$ are constants.  Of course, these solutions must match onto the oscillating wave solutions that describe the sound waves at recombination (i.e.~they set the values of $A_\k$ and $B_\k$).  However, when we match $c,d \to A_\k, B_\k$, some linear combination will be suppressed by $a^{-3}$, i.e.~the inverse of the volume element of the Universe, $\sqrt{g} =a^3$.  At this point, you can presumably guess that this is the sine wave, $B_\k \sin(c_s k \tau)$, and $B \propto 1/ a^3(t) \to 0$.  Therefore, if we can find a mechanism for creating these very long wavelength fluctuations (much larger than the particle horizon during the hot Big Bang) before the era of recombination, then we can explain the phase coherence of the CMB through the time evolution of the modes themselves.

One of the benefits of phase coherence is that the fluctuations in the CMB can be understood to factorize into the statistical fluctuations from inflation ($A_\k$) and the evolution of the sound waves in this plasma ${\cal T}(k)$.  To first approximation, this takes the form
\beq\label{eq:factorization}
\delta T(\k) = {\cal T}(k) \zeta(\k) \approx \cos(c_s k \tau_{\star} )\zeta(\k) 
\eeq
where ${\cal T}(k)$ is a transfer function that depends only on $k\equiv |\k|$, and we made the replacement $A_\k \to \zeta(\k)$, the comoving curvature perturbation produced during inflation.  Correlation functions of temperature are then just linearly related to inflationary correlation functions by
\beq
\langle \delta T(\k_1) .. \delta T(\k_n) \rangle = {\cal T}(k_1) .. {\cal T}(k_n) \langle \delta \zeta(\k_1) .. \delta \zeta(\k_n) \rangle \ .
\eeq
This is important because it tells us that random part of the fluctuation is due to inflation and the deterministic part is the plasma in the evolution up to recombination.  This is roughly why we can measure properties of inflation and the plasma simultaneously.

\subsection{Large Scale Structure}\label{sec:intro_lss}

At the time of recombination, $z \approx 1100$, most of the energy in the Universe is contained in dark matter, not in photons and baryons.  The moment when the two energy densities were equal, known as matter-radiation equality, occurs at $z\approx 3400$.  Despite this fact, the dominant role of the dark matter does not translate into a large impact on the appearance of the CMB, although its presence is unmistakable if you know where to look.\footnote{One specific imprint of dark matter is the offset in the relative heights of the even and odd peaks~\cite{Dodelson:2003ft}, particularly the second, third, and fourth, temperature peaks.}

The need for dark matter becomes completely unavoidable when you start thinking instead of the Universe at lower redshifts.  The Universe at present time is filled with all kinds of nonlinear structures, most obviously planets, stars, galaxies, and galaxy clusters.  For these structures to have formed, we need large density fluctuations so that linear evolution fails, which is usually characterized by a large density contrast
\beq
\delta_m (\x,t) \equiv \frac{\delta \rho_m(\x,t) }{\bar \rho_m(t)} > 1 \ .  
\eeq
We assume that this must have arisen from gravitational collapse, as gravity is the only force we know that operates on these enormous scales. Since $\delta_m \ll 1$ at recombination, most of this growth must occur where we can neglect nonlinear effects (i.e.~it has to grow enough under linear evolution to become nonlinear).  In a matter-dominated universe, the matter fluctuations on small scales, $k \gg aH$, evolve according to Newtonian gravity as~\cite{Baumann:2022mni}
\beq
\begin{aligned}
\dot \delta_m- \frac{k^2}{a} u_m &=0 \\
\dot u_m+\H u_m &=\frac{1}{a} \Phi \\
k^{2} \Phi &=\frac{3}{2} H^{2} a^2 \delta_m \ ,
\end{aligned}
\eeq
where the first two equations describe the local conservation of mass and momentum and the third follows from Einstein's equation at $k\gg aH$, or (equivalently) Newtonian gravity. Here we define the (Eulerian) velocity of the matter of the matter fluid, $\vc_m$, in terms of the velocity potential $u$, via $\vec v_m = \vec \nabla u_m$. Combining these equations, the matter overdensities evolve according to
\beq\label{eq:density_fluc}
\ddot \delta_m + 2 H \dot \delta _m- \frac{3}{2} H^2 \delta_m = 0 \ .
\eeq
From the Friedmann equation, we have 
\beq
3\Mpl^2 H^2 = 3 \Mpl^2 \frac{\dot a^2}{a^2} = \rho_m = \omega_m a^{-3} \ ,
\eeq
which can be integrated to find
\beq
a(t) \propto t^{2/3} \qquad H = \frac{2}{3 t} \ .
\eeq
Plugging this back, we now find the evolution in a matter-dominated universe is
\beq\label{eq:density_fluc2}
\ddot \delta_m + \frac{4}{3t} \dot \delta _m- \frac{2}{3t^2} \delta_m = 0 \ .
\eeq
Making the power-law ansatz, $\delta  = t^\gamma$ where $\gamma$ is a constant, yields the result
\beq
\gamma^2 +\frac{1}{3} \gamma - \frac{2}{3}  = 0 \to \gamma=-1, 2/3
\eeq
or 
\beq
\delta_m(\k,t) = a(t) \delta_+(\k) + a(t)^{-3/2} \delta_-(\k) \ ,
\eeq
where $\delta_{\pm}(\k)$ are some initial conditions set after the baryons have decoupled from the photons~\cite{Green:2020fjb}. The key takeway from this equation is that the growing solution\footnote{The time-dependence of the growing solution is often given the name of the growth function, $D(a)$, which in this case is $D(a)=a(t)$.} for the matter density contrast, $\delta_m$, scales as $\delta_m(\k, t) \propto a(t)$ (and $k^2 \Phi \propto a^2 \bar \rho_m \delta_m \propto a^{-1} \delta_m $ is a constant).  Clearly, if we wait long enough, the gravitational collapse of the matter will produce nonlinear fluctuations.  

This simple picture of structure formation {\bf cannot} produce nonlinear structure if the matter in the Universe was made only of baryons.  The fluctuations we observe in the CMB have $\delta_{\gamma,b}(a_\star) \sim 10^{-5}$ at $a_\star = 10^{-3}$ ($z_\star=1100$); gravitational collapse of the baryons would scale as $\delta_b \propto a$ and the size of the matter fluctuations today would only reach
\beq
\delta_b(t_0) = \frac{a_0}{a_\star} \delta_b(a_\star) \approx (1+z_\star) 10^{-5} \approx 10^{-2} \ .
\eeq
We see that there simply wasn't enough time for matter to become nonlinear if structure forms from the very same fluctuations we observe in the CMB.\footnote{This fact was historically important as people had believed $\delta T / T$ must be no smaller than $10^{-3}$ and therefore caused some apprehension when they were not found at that level~\cite{Peebles:1982ff,Peebles:2017bzw}.}

\subsection{The Need for Dark Matter}

The resolution to our paradox in $\Lambda$CDM cosmology is that dark matter fluctuations start growing under gravitational collapse once they cross the horizon, $k = H(t) a(t)$, and they do not have to wait until recombination (the collapse of the baryons is prevented by the pressure of the photons which is why this extra matter must be dark and non-baryonic).  The modes that entered earlier have had more time to grow and thus are able to become nonlinear by $z=1$ without postulating a new force beyond gravity.  Furthermore, since we don't directly observe the dark matter, there is no contradiction with the amplitude of temperature fluctuations of the CMB. The consequence is that the modes that become nonlinear must be the ones with higher $k$, as they crossed the horizon first. Given that the primordial metric fluctuations have an amplitude of about $10^{-5}$ even the modes that crossed the horizon at matter-radiation equality are still linear today.  The truly nonlinear modes must have entered the horizon deep in the radiation era.  Because the Universe at that time was dominated by the energy in photons, which do not collapse under the force of gravity from the dark matter, the growth was substantially slower, $\delta_m(\k,t) \propto \log a(t)$.  We can understand this simply if we note that $k^2 \Phi \propto H^2 a^2\delta_\gamma$ during radiation domination.  Since the radiation density simply oscillates, $\delta_\gamma \propto \cos(c_s k \tau)$, but its amplitude does not grow, we have $\Phi \propto a^{-2}$ during matter domination and therefore, according to Equation~(\ref{eq:density_fluc}), the collapse of the matter follows from $\ddot \delta +2 H \dot \delta \approx 0$ with $a(t) \propto t^{1/2}$, $H(t) = 1/(2t)$ during radiation domination. 

\begin{figure}[h!]
\begin{center}
\includegraphics[width=6.in]{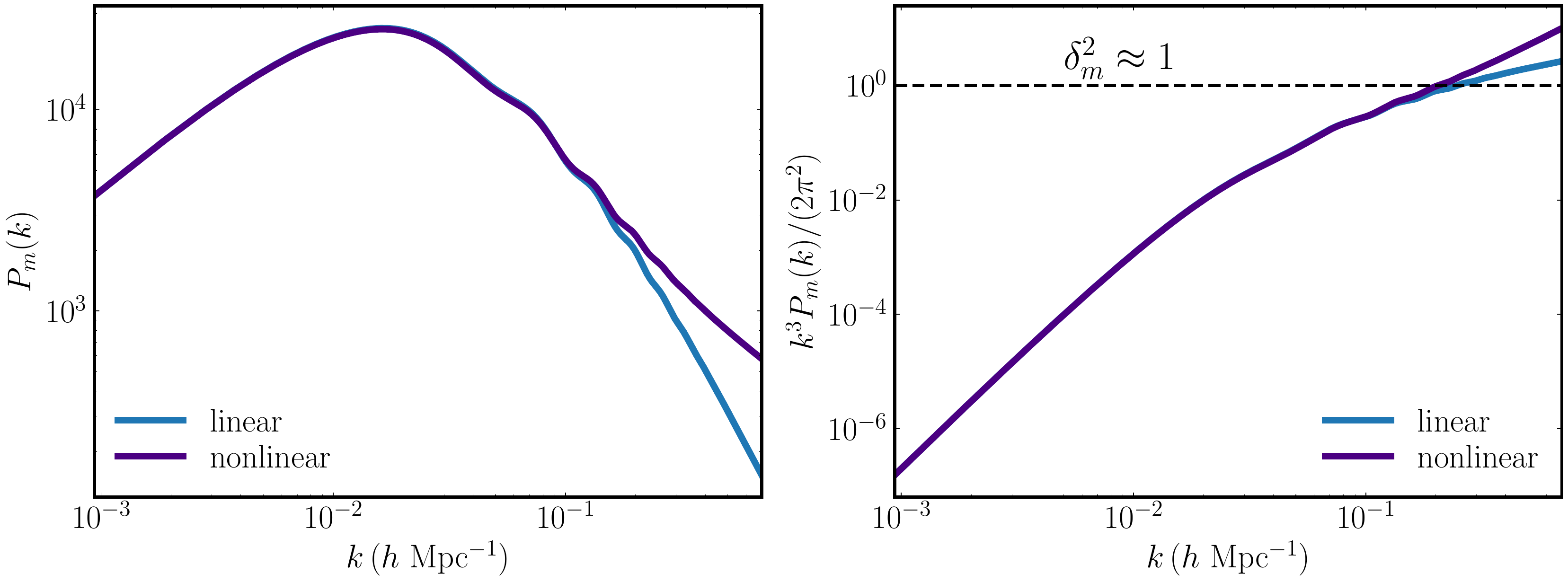}
\end{center}
\caption{ {\it Left:} The linear and nonlinear matter power spectra at $z=0$. The peak of this curve corresponds to modes entering the horizon at matter-radiation equality, $z\approx 3400$  {\it Right:} Comparison of $k^3 P_m(k)/(2\pi^2) \approx \delta_m^2$ at $z=0$ for both the linear and nonlinear matter power spectrum with the rough scale of nonlinearity $\delta_m = 1$. Only the modes growing logarithmically during the radiation era became nonlinear by today.}
\label{fig:matter_power}
\end{figure}

The net effect is that the power spectrum of matter fluctuations 
\beq
\langle \delta_m(\k) \delta_m(\k') \rangle = P_m(k) (2\pi)^3 \delta(\k+\k') \ ,
\eeq
has a somewhat surprising shape, shown in the left panel of Figure~\ref{fig:matter_power}.  At low-$k$, it scales linearly with $k$, which is a reflection of the fact that the Newtonian potential is constant in the matter era, $\delta_m \propto k^2 \Phi$, and that the metric fluctuations are scale invariant $P_\Phi \propto k^{-3}$ (i.e. $P_m(k\ll k_{\rm eq}) \approx k^4P_\Phi \sim A_s k$).  During the radiation era, the Newtonian potential decays with time because it is dominated by the sound waves which don't cluster.  The logarithmic growth of $\delta_m(k) \propto \log(k /(aH))$ we found during radiation domination then implies that $P_m(k\gg k_{\rm eq} ) \propto (\log k)^2/k^3$.  The nonlinear scale, where $\delta^2 \sim 1$, is then determined by $k^3 P_m(k) = {\cal O}(1)$. Figure~\ref{fig:matter_power} confirms this intuition and shows it is the small scales modes that give rise to non-linear structure.

The baryons are, of course, still a part of the matter content of the Universe and have an important role to play in structure formation. Before recombination the baryons are just one component of the photon-baryon fluid and, as a result, when recombination ends the fluctuations in baryons are still look like sound waves, $\delta_b(\k)\propto \zeta(\k) \cos(c_s k \tau_\star)$. Once the photons decouple, the waves no longer have pressure support and so the baryons eventually stop moving.  At this point, there is no difference between the baryons and the dark matter (on cosmological scales) and so they both grow together under the force of gravity.  Yet, the initial distribution of baryons looks totally different than the dark matter, so the overall shape combines the initial positions of the baryons and dark matter.  The result is that the oscillatory feature of the baryons, called the baryon acoustic oscillations (BAO), survives in the power spectrum~\cite{Eisenstein:1997ik} as,
\beq\label{eq:bao}
P_m(k,\omega_b) \approx \left(1+A_{\rm BAO} c_s k \sin(c_s k \tau_\star)\right) P_m (k,\omega_b=0) \ ,
\eeq
where $\cos(c_s k \tau) \to c_s k \sin(c_s k \tau_\star)$ is due to the velocity overshoot~\cite{1970Ap&SS...7....3S,Press:1980is,Green:2020fjb}, which is the result of the non-zero velocity the baryons have at recombination (they don't freeze in place instantly at recombination, they slow down and stop). From Figure~\ref{fig:matter_power}, it is easy to see that the amplitude of the BAO feature is much smaller than the rest of the matter power spectrum.  This occurs because the baryon fluctuations don't start growing until $z=1100$ while the dark matter fluctuations were growing linear starting from matter-radiation equality and logarithmically from the time of horizon crossing to equality.

The BAO itself contains important lessons for our understanding of structure formation. First, the small amplitude of the BAO is a confirmation of the intuition that the amplitude of the CMB fluctuations is not large enough to explain nonlinear structure; baryons alone would have been linear at $z=0$. Second, it is important to notice that $C^{\rm TT}_{\ell}$ reflects the location of baryons (and photons) in space at $z=1100$, while $P_m(k)$ maps the location of matter at $z=0-10$.  This presents a serious problem for anyone that doubts the existence of dark matter~\cite{Pardo:2020epc}: if you want the Standard Model (SM) matter to do all the work of structure formation, then you have to explain how the baryons moved over distances comparable to the entire observable Universe between $z=1100$ and $z\approx 10$ so that their distribution in space will match measurements of both $C^{\rm TT}_{\ell}$ and $P_m(k)$.  Introducing new physics to accomplish this goal will require far more drastic and baroque changes to the laws of nature than just adding cold dark matter.

The BAO is itself a gift to cosmologists for other reasons~\cite{Seo:2003pu,Eisenstein:2006nj,Weinberg:2013agg}.  The scale $c_s \tau_\star= r_s$ is the sound horizon at recombination and it determines the shape of the CMB.  This means that if we can measure the BAO feature at lower redshifts, then it forms a standard ruler: i.e. we know the physical size of $r_s$ and we measure the redshift $z$ and angular size of the feature on the sky.\footnote{This is basically just the fact we know from flat space that if some small object has angular size $\theta$ and has a true physical size $r_s$ then it is at a distance $d \approx r_s/\theta$.  In an expanding universe, this is replaced with the angular diameter distance which has a non-trivial dependence on the expansion history~\cite{Dodelson:2003ft,Weinberg:2008zzc,Baumann:2022mni}.}  From this information we can determine $H(z)$. Current galaxy surveys, like BOSS, provide the most precise measurements of the late time expansion precisely from this effect~\cite{deMattia:2020fkb}. 

\newpage

\section{Light Relics and the CMB} \label{sec:Neff}

Ironically, the entire post-recombination era (i.e.~the era we can directly observe) has been dominated by components of the energy that we cannot see directly, dark matter and dark energy.  Naturally, we wonder if there could also be some form of dark radiation to complete the trinity of the dark world. Like the other two, dark radiation would also leave a gravitational imprint that we might hope to uncover.  However, being radiation, it redshifts like $a^{-4}(t)$ and we know that this effect cannot be large enough to directly impact gravitational evolution of the late Universe.

Fortunately, there is a source of ``dark radiation" in the Standard Model, namely neutrinos, that has given us a good reason to develop our understanding of what dark radiation might look like and where to look for it.  Therefore, we will first make a detour to discuss the cosmic neutrino background and its signatures.  See also~\cite{Baumann:2018muz} for an introduction to many of these topics.

\subsection{Cosmic Neutrinos}

Neutrinos interact with the rest of the Standard Model through the weak force.  Neutrinos are therefore produced or scatter at $E \ll M_W$ with an amplitude proportional to $G_{\rm F} = \sqrt{2} g^2/(8 M_W^2)$, where $M_W$ is the $W$-boson mass and $g$ is $SU(2)$ weak coupling constant.  By dimensional analysis, the production rate (which carries units of energy) of neutrinos scales as $\Gamma_\nu \approx G_F^2 T^5$. Neutrinos will stay in equilibrium with the SM as long as $\Gamma_\nu(t) \gg H(t)$ and thus are thermalized at high energy temperatures.  Remembering that the temperature will redshift like $1/a(t)$, and using $\rho_\gamma \propto T^4$ and $3 \Mpl^2 H^2 = \rho_\gamma$ in the very early Universe, neutrinos will begin to decouple at a temperature $T_{\rm dec}$, when
\beq
\Gamma_\nu \approx G_{\rm F}^2 T_{\rm dec}^5 = H(T) = \sqrt{g_\star \frac{\pi^2}{90}}  \frac{T^2}{\Mpl} \to T^3_{\rm dec} \approx \frac{M_W^4}{M_{\rm pl}} \approx (1 \, {\rm MeV})^3 \ ,
\eeq 
where $g_\star$ is the effective number of degrees of freedom defined in Appendix~\ref{app:thermo} along with some other useful thermodynamic quantities.  Numerical calculations show that decoupling begins somewhere between 1-10 MeV, which is consistent with this very rough estimate.  See e.g.~\cite{Cyburt:2015mya,Fields:2019pfx} for reviews.

After the neutrinos decouple, their phase space distribution stays more or less the same so that the neutrinos maintain a thermal distribution with an apparent temperature $T_\nu \propto a^{-1}$. As long as temperatures evolve only according to expansion, the temperature of the photons and neutrinos would still match at later times $T_\gamma = T_\nu$ even though they are no longer in equilibrium.  However, there is a change from this simple redshifting behavior, not due to the neutrinos but inside the Standard Model.  At $T_\gamma > 1$ MeV, electrons and positrons are still relativistic and make up an important component of the plasma.  However, when the temperature drops below their mass, $m_e = 511$ keV, the number densities of the positrons and electrons become Boltzmann suppressed.  Yet, because they remain in thermal equilibrium with the photons, the comoving entropy is conserved and therefore the entropy that was once in $e^+ e^-$ must be converted to the photons.  

Using these observations, we can calculate the increase to the temperature of the photons, relative to the neutrinos, using entropy conservation.   Consider a time $t_1$ where the temperature was large so that $T \gg m_e$.  At this time, the entropy of the Universe was determined by the photons, electrons, positrons and neutrinos, all with the same temperatures as the neutrinos, $T_\nu(t_1)$, 
\beq
S = \frac{\pi^2}{45} g_\star(t_1) (T_\nu(t_1) a)^3  \qquad g_\star(t_1) = n^{(s)}_\gamma + \frac{7}{8} \left(n^{(s)}_\nu + n^{(s)}_{e^+}+ n^{(s)}_{e^-} \right) \ .
\eeq
All three particles have 2 helicity states and, in addition, we have 3 species of neutrinos so that\footnote{Since only the left-handed neutrinos are required to interact with the Standard Model, the number of thermalized degrees of freedom often does not depend on whether we have Majorana or Dirac neutrinos.} $n^{(s)}_\nu = 6$ and all others are 2, for a total of $g_\star (t_1)= 43/4$.  We then compare this to the entropy calculated at $t_2$ when $T \ll m_e$.  The electrons and positrons are Boltzmann supressed but the temperature of the neutrinos and photons need not be the same,
\beq\label{eq:Tnu_entropy}
S = \frac{\pi^2}{45}   \frac{7}{8} n^{(s)}_\nu  (T_\nu(t_2) a)^3  +\frac{\pi^2}{45}  n^{(s)}_\gamma  (T_\gamma(t_2)  a)^3 \ .
\eeq
By entropy conservation, we can equate both expressions for $S$.  Furthermore, assuming the neutrinos are decoupled, we have $T_\nu(t) a(t) =$ constant so that $T_\nu(t_1)a(t_1) = T_\nu (t_2)a(t_2)$ and therefore  
\beq
2 (T_\gamma(t_2) a)^3 = \left(g_\star(t_1)  -\frac{7}{8} n^{(s)}_\nu  \right) (T_\nu(t_2) a)^3 \ .
\eeq
Notice that $n^{(s)}_\nu$ actually cancels out in this expression so that the result can be expressed as
\beq\label{eq:Tnu}
\left(\frac{T_\nu}{T_\gamma}\right)^3 = \frac{g_\gamma}{g_\gamma+ g_{e^\pm}} = \frac{4}{11}  \ .
\eeq
This famous result implies that there is a cosmic background of neutrinos with a temperature of 1.9 K filling the Universe today.

Given the temperature of the neutrinos, we can determine their total energy density while they are relativistic, $\rho_\nu \propto T_\nu^4$.  Although the energy density in neutrinos is what ultimately affects observations, it has become conventional to express $\rho_\nu$ in terms of the parameter $\Neff$, defined by
\beq
\rho_{\rm \nu}=\frac{7}{8}\left(\frac{4}{11}\right)^{4 / 3} N_{\mathrm{eff}} \, \rho_{\gamma} \ ,
\eeq
so that $\rho_{\rm r} = \rho_\gamma (1+\frac{7}{8}\left(\frac{4}{11}\right)^{4 / 3} N_{\mathrm{eff}})$ is the total energy density that redshifts like radiation, $\rho_{\rm r} \propto a^{-4}$, around the time of recombination.

In practice, the calculation of $T_\nu$ and hence $\rho_\nu$ is corrected by several effects.  First of all, the neutrinos are not completely decoupled from the Standard Model at $T \approx m_e$ and there is enough $e^+ e^- \to \nu \nu$ (for example) to correct the neutrino temperature. Secondly, in writing the entropy of the Standard Model, we assumed it was a dilute gas.  This is a decent approximation at $T\sim 1$ eV, but QED will introduce corrections suppressed by the fine-structure constant.  A more complete calculation needs to include these and other effects, giving a small increase to the neutrino temperature, which can be interpreted as the statement that $\Neff = 3.044$ in the Standard Model~\cite{EscuderoAbenza:2020cmq,Akita:2020szl,Froustey:2020mcq,Bennett:2020zkv} (where the current {\it theoretical} uncertainty is in the 4th decimal place).

\subsection{Light Thermal Relics}

Beyond the Standard Model (BSM) physics is awash with new light particles such as axions, sterile neutrinos, dark sectors, mediators of new forces, etc~\cite{Asadi:2022njl}.  We can abstractly consider all of these possibilities~\cite{Brust:2013ova} by calling this new particle or particles $\phi$, and assigning to it an effective number of degrees of freedom $g_\phi$.  In this way, the gravitational influence of a complex hidden sector is parametrized in terms of its energy density through $g_\phi$ and its temperature, $T_\phi$.  To explain the absence of these particles in the lab, one might imagine that $\phi$ is coupled to the Standard Model through some irrelevant operator so that we naturally avoid experimental constraints. To be concrete, let us suppose it couples through an operator of dimension $\Delta = 4+n > 4$ so that, by dimensional analysis, the thermal production rate of $\phi$ follows from
\beq
{\cal L}_{\rm int} \supset \frac{\lambda}{\Lambda^{n}} {\cal O}_\phi {\cal O}_{\rm SM} \to \Gamma_\phi \approx \frac{\lambda^2}{\Lambda^{2n}} T^{1+ 2 n} \ ,
\eeq
where ${\cal O}_\phi$ and ${\cal O}_{\rm SM}$ are operators made from hidden sector and SM fields respectively. When $n > 1/2$, the production rate become less important than expansion ($\Gamma_\phi <H$) at low temperatures but is potentially in equilibrium ($\Gamma_\phi \gg H$) at high temperature, just like the neutrinos.  Following the neutrino playbook, we can define a freeze-out (decoupling) temperature, $T_{\rm F}$, by
\beq
\Gamma_\phi (T_{\rm F})\approx \frac{\lambda^2}{\Lambda^{2n}} T_{\rm F}^{1+ 2 n} = H(T_{\rm F}) \approx \frac{T_{\rm F}^2}{\Mpl} \quad \to \quad T_{\rm F}^{2 n -1} \approx \frac{\Lambda^{2n}}{\Mpl} \label{eq:TF_theory}
\eeq
Notice that for $\Lambda \ll \Mpl$ and moderate $n$, decoupling happens within the regime of control of effective field theory (EFT) namely $T_{\rm F} \ll \Lambda$. Unless reheating occurred at sufficiently low temperature,  $T_{\rm reheat} < T_{\rm F}$, we would expect a thermal background of $\phi$ to arise in any such model.  Inverting this logic, if we did not see such a background, it would requires $\Lambda^{2n} \gg T^{2n-1}_{\rm reheat} \Mpl$. For smaller values of $n$, the appearance of $\Mpl$ in this expression would imply a particularly stringent constraint on $\Lambda$.  Even if we only assume reheating above the weak scale $T_{\rm reheat} = 1$ TeV, for $n=1$ we would get $\Lambda > 10^{10}$ GeV.  For high-scale reheating, $T_{\rm reheat} \gg 1$ TeV, we will get very strong constraints on a wide range of couplings.

\begin{figure}[h!]
\centering
\includegraphics[width=4.5in]{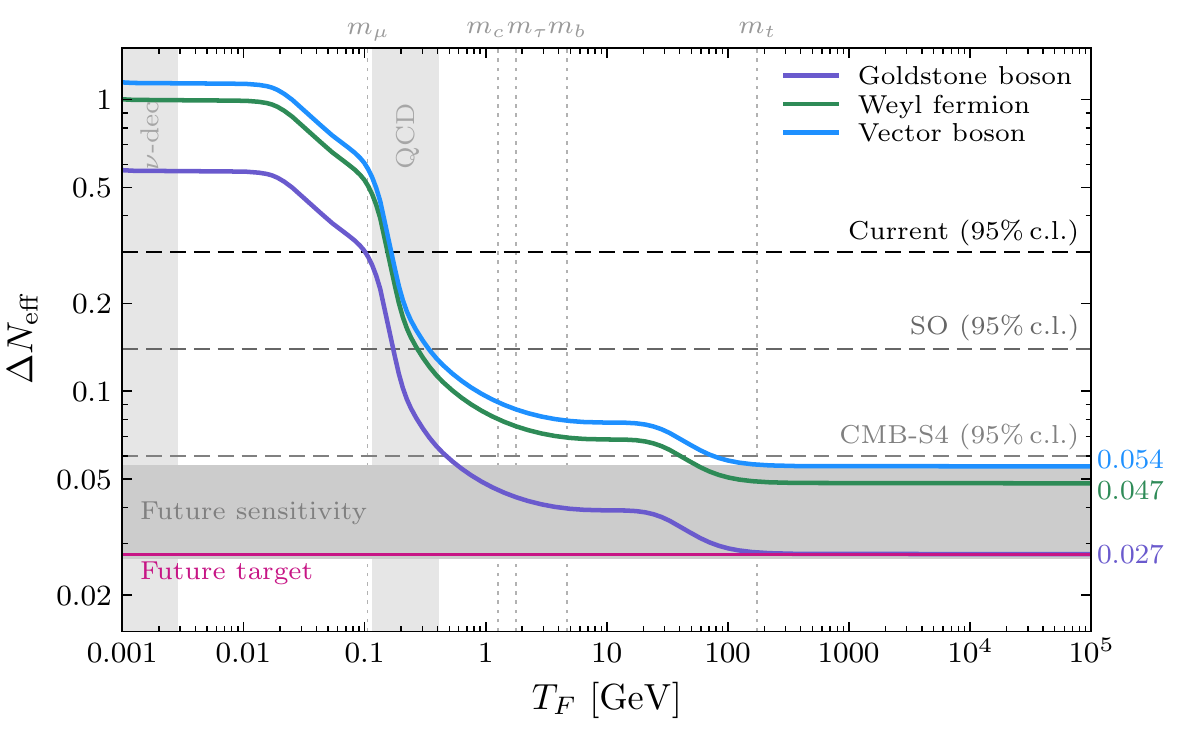}
\caption{Contributions to $\Delta \Neff$ as a function of $T_{\rm F}$.  Reproduced from \cite{Wallisch:2018rzj,Dvorkin:2022jyg}.}
\label{fig:DeltaNeff}
\end{figure}

Like for cosmic neutrinos, cosmological constraints will depend on the energy density of $\phi$ around the time of recombination.  We can determine their temperature, relative to the temperature of the neutrinos, by repeating the conservation of entropy argument in Equation~(\ref{eq:Tnu_entropy}).  Comparing the entropy at the temperature $T_F$ to the entropy at $T \approx 10$ MeV, before neutrino decoupling, we get
\beq
\left(\frac{T_\phi}{T_\nu}\right)^3 = \frac{g_\gamma+ g_{e^\pm}+g_\nu}{g_\star(T_{\rm F})} = \frac{43}{4} \frac{1}{g_\star(T_{\rm F})}  \ .
\eeq
Significantly, this cosmic-$\phi$ background is just another source of free-streaming\footnote{Here we are making the additional assumption that interactions with the SM and self-interactions of $\phi$ (or other non-SM particles) are controlled by the same scale $\Lambda$, or higher.  } (non-interacting) radiation that is coupled to the Standard Model via gravity. The energy density in $\phi$, $\rho_\phi = g_\phi T_\phi^4 (\pi^2/30)$, is therefore no different from an increase to $\rho_\nu$ and is therefore included as an additional contribution to $\Neff = 3.044 +\Delta \Neff$, where
\beq
\Delta \Neff = g_\phi \frac{4}{7} \left(\frac{43}{4} \frac{1}{g_\star(T_{\rm F})}  \right)^{4/3} \ .
\eeq 
Note that $g_\phi = 7/4$ and $T_F \approx 1$~MeV corresponds to $\Delta \Neff =1$, as it should.  Importantly, since $T_\phi$ is a universal function of $T_{\rm F}$, the predictions for $\Delta \Neff$ for any specific model is determined by the number of internal degrees of freedom of $\phi$ and the number of degrees of freedom of the Standard Model at $T \approx T_{\rm F}$.  These universal curves are shown in Figure~\ref{fig:DeltaNeff} for $g_\phi = 1,7/4$ and $2$ corresponding to a real scalar (Goldstone boson), Weyl fermion, and massless vector/complex scalar respectively.  

There are two noteworthy features of these curves.  First, all the curves asymptote to $\Delta \Neff = 0.027 g_\phi$ for $T_F \gg 100$ GeV .  The fact these curves reach minimal values of $\Delta \Neff$ is a consequence of the finite number of degrees of freedom in the Standard Model so that at high temperatures $g_\star(T \gg m_t) = 106.75$.  Unless we double the particle content of the Standard Model at the weak scale,\footnote{Weak-scale SUSY would clearly be relevant to $\Neff$ for $T_{\rm F} > 100$ GeV while a single WIMP would be negligible.} observational sensitivity to these asymptotic values probes vast regions of parameter space where these light particles would have thermalized.  Second, there is a major change in $\Delta \Neff$ for freeze-out before and after the QCD phase transition, due to the large change in $g_\star(T)$ of the Standard Model.  For this reason, current observations largely tell us about physics after the QCD phase transition, but this is about to change with coming observations.

Figure~\ref{fig:DeltaNeff} also show the current and future constraints on $\Delta \Neff$ that arise from the CMB + BAO.  A concrete future survey of interest is CMB-S4 which is expected to excluded $\Delta \Neff < 0.06$ at 95\% confidence~\cite{Abazajian:2019eic}. We can see that these constraints are reaching very interesting values of $\Neff$, excluding $T_{\rm F} < {\cal O}(100 \, {\rm GeV})$ for particles with spin and excluding the thermalization of any dark sector with $g_\phi > 2.2$.  Translating these into constraints on beyond the Standard Model physics is competitive with, and often more sensitive than, lab-based and astrophyiscal constraints~\cite{Baumann:2016wac,Green:2017ybv,Knapen:2017xzo,DEramo:2021lgb,Green:2021hjh}.  Naturally, we should understand where this constraint is coming from and what else it might tell us about the Universe.

\subsection{Implications for the CMB}

\begin{figure}[h!]
\centering
\includegraphics[width=4.5in]{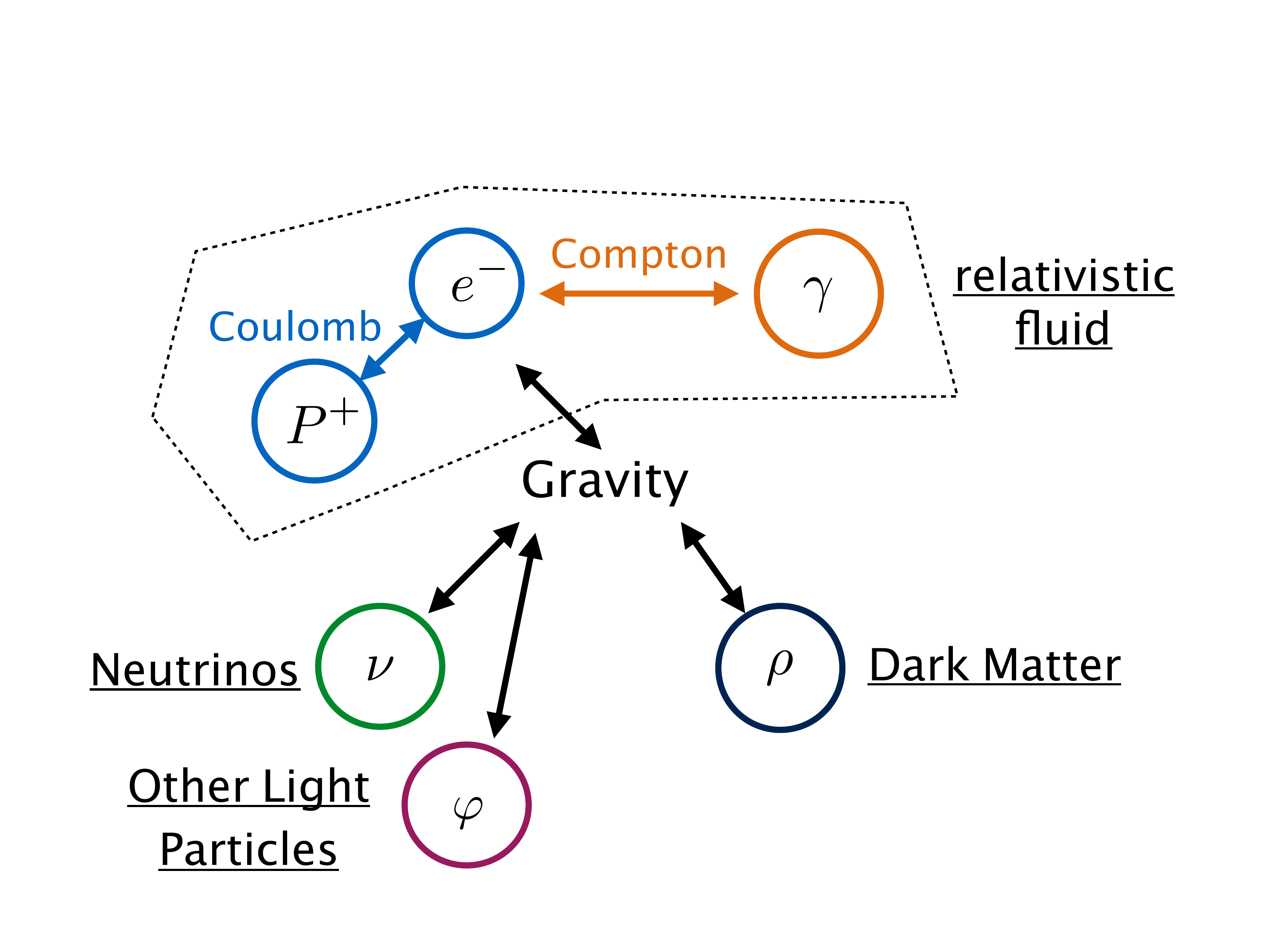}
\caption{Pre-recombination matter content and forces}
\label{fig:forces}
\end{figure}

The measurement of $\Neff$ is driven by the physics of the plasma that fills the Universe prior to recombination.  Recall that this plasma, and its associated sound waves, are made up of photons, electrons, and protons.  The signals of cosmic neutrinos (and dark matter) we measure in the CMB arise only from the gravitational influence of these additional sources of energy density on the sound waves.  The relevant forces at before recombination are illustrated in Figure~\ref{fig:forces}.

The goal of these lectures is not to derive every formula from first principles, but to give you an idea of where the results come from.  To understand the CMB in full detail, one starts with a distribution function for each species $f_i(\x, \p,t)$.  In short, this tells us the number of particles with momentum $\p$ and energy $E(p)$ at location $\x$ at time $t$.  From the forces acting between all of these particles, the evolution of $f_i$ is just the repeated application of Newton's laws (or their relativistic counter-parts).  However, when the number density of these particles is sufficiently large, it can be a very good approximation to define quantities like the energy and average local velocity at $\x$ and $t$, 
\beq
\rho_i(\x,t) = \int \frac{d^3 p}{(2\pi)^3} \,  E(p)  \, f_i(\x, \p,t) \qquad  \vec v_i(\x,t)  =\int \frac{d^3 p}{(2\pi)^3} \, \frac{\vec p}{m} \, f_i(\x, \p, t) \ ,
\eeq 
rather than the full distribution function.  Working with the metric
\beq
\mathrm{d} s^{2}=a^{2}(\tau)\left[(-1-2 \Phi) \mathrm{d} \tau^{2}+(1-2 \Psi) \delta_{i j} \mathrm{~d} x^{i} \mathrm{~d} x^{j}\right] , 
\eeq
the conservation of energy and momentum completely dictate the form of some of the equations for time evolution for each decoupled sector $i$, namely~\cite{Bashinsky:2003tk,Baumann:2015rya}
\beq
\begin{array}{l}
\delta'_{i}=\nabla^{2} u_{i}+3 \Psi'  \ , \\
u'_{i}=c_{i}^{2} \delta_{i}-\chi_{i} u_{i}+\nabla^{2} \sigma_{i}+\Phi \ ,
\end{array}
\eeq
where $c_i^2 = \frac{dP_i}{d\rho_i}$, $\chi_{i} \equiv \mathcal{H}\left(1-3 c_{i}^{2}\right)$, and $\vec \nabla u_i = \vec v_i$. The quantity $\sigma_i$ is the anisotropic stress, which is a determined by the higher moments of the distribution function, while $\Psi$ and $\Phi$ are determined by Einstein's equations.

Defining $d_\gamma\equiv \delta_\gamma - 3\Psi$ and taking $c_\gamma^2 \approx 1/3$, the photon-baryon fluid is described by
\beq\label{eq:linear_density}
d''_\gamma - c_\gamma^2 \nabla^2 d_\gamma = \nabla^{4} \sigma_{\gamma}+\nabla^2 \Phi_+ \ ,
\eeq
where $\Phi_+ = \Phi+\Psi$. The homogeneous equation is the wave equation, as advertised, and imposing our initial conditions from Equation~(\ref{eq:power_spec}) has solutions
\beq
d_\gamma(\k , \tau) = A_\k \cos(c_s k \tau) \ .
\eeq 
To first approximation, what we observe is this sound wave at the time of recombination, $\tau = \tau_{\star}$.  After recombination, the photons free steam so that we observe them more or less unchanged, up to gravitational redshift from climbing out of their potential wells, and appear on the sky as spherical harmonics with $\ell \approx k /(\tau_0- \tau_\star)$.  Very roughly~\cite{Pan:2016zla}, the $n$th peak of the CMB temperature power spectrum ($C_\ell^{\rm TT}$), $\ell_n$, occurs at the $n$-th maxima of $\cos(\ell \tau_\star/(\tau_0-\tau_\star))^2$, or  
\beq
\frac{\ell_{n} \tau_\star}{\tau_0-\tau_\star} =  n \pi \ .
\eeq
The values of $\ell_{n}$ are very well-measured\footnote{Very roughly, there are $100(2\ell_n+1)$ modes around the $n$-th peak, so we would expect from our mode counting that we can measure these peaks at $10^{-2}-10^{-3}$ accuracy, for $\ell_n ={\cal O}(10^3)$, in agreement with the actual measurements.} in the CMB~\cite{Planck:2015fie} and are not dependent on our cosmological interpretation of the data: this is just a fact about angular scales of fluctuations we see in the sky.  However, from our theory, we know that these locations depends on $\tau_\star$ and $\tau_0$, which are defined by 
\beq
\tau= \int_0^{t}  \frac{dt'}{a(t')} = \int_0^{a} \frac{d a'}{H(a') a'{}^2}  \ .
\eeq
During radiation domination, 
\beq
H(a)  \propto \left(1+\frac{7}{8}\left(\frac{4}{11}\right)^{4 / 3} N_{\mathrm{eff}}\right)^{1/2} \ ,
\eeq
while at late times, $H(a) \propto H_0$.  If we vary $\Neff$ while holding $H_0$ fixed, then the locations of the maxima will change because of the change to $\tau_\star$, with $\tau_0 -\tau_\star \approx \tau_0$ held fixed,
\beq
\ell_{n}(\Neff) \propto \left(1+\frac{7}{8}\left(\frac{4}{11}\right)^{4 / 3} N_{\mathrm{eff}}\right)^{1/2} \times \tau_0\ .
\eeq
Since the locations of the peaks are known, namely $\ell_n$ are fixed, the better way to understand the impact of $\Neff$ is to also change $H_0 \propto  \left(1+\frac{7}{8}\left(\frac{4}{11}\right)^{4 / 3} N_{\mathrm{eff}}\right)^{1/2}$, or equivalently $\tau_0^{-1}$, so that we hold the angular locations of the acoustic peaks fixed~\cite{Hou:2011ec}, usually parameterized by the angular scale of the first peak, $\theta_\star$. 

The true impact on $\Neff$ is therefore encoded in the corrections to the homogeneous solutions arising from the non-perfect fluid behavior through $\sigma$, and the gravitational influence of the density perturbations through $\Phi$. Solving these equations in detail will be too large a detour, but we can summarize the most important effects.
\vskip 10pt
\noindent {\bf Diffusion (Silk) Damping}

\begin{figure}[h!]
\centering
\includegraphics[width=6in]{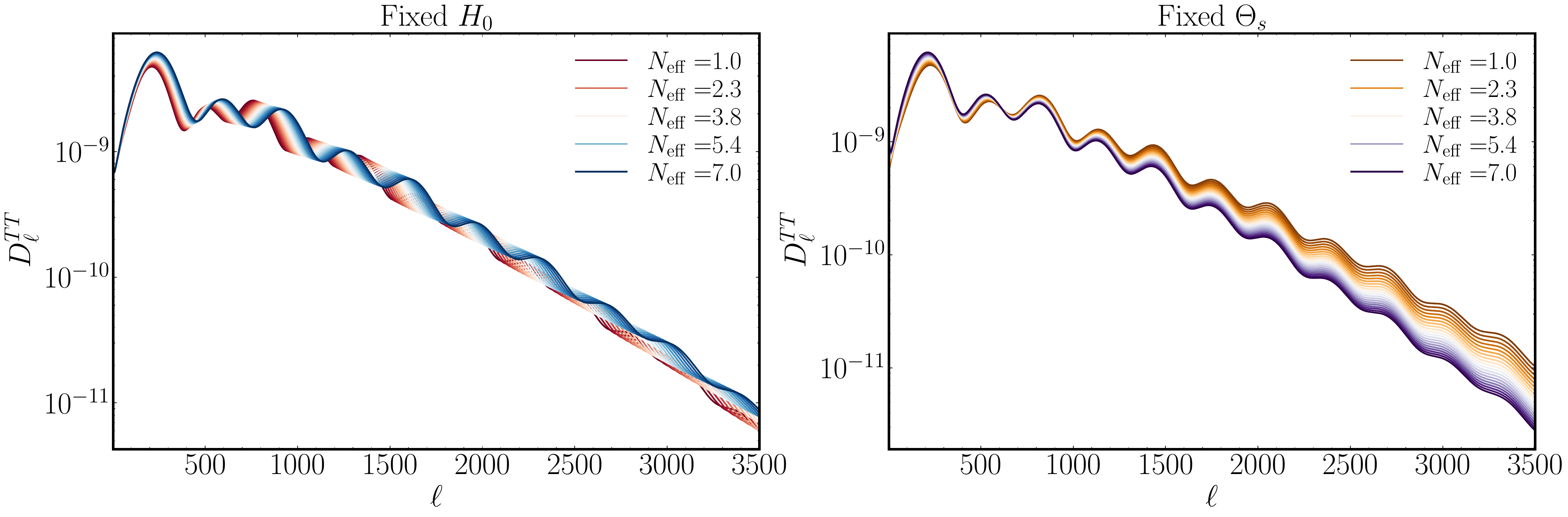}
\caption{The effect of varying $\Neff$ on $C_{\ell}^{\rm TT}$. {\it Left:} Changing $\Neff$ holding $H_0$ fixed. {\it Right:} Changing $\Neff$ holding $\theta_\star$ fixed. }
\label{fig:damping}
\end{figure}

\noindent The most important effect is the damping of the sound waves from the diffusion of photons.  In short, because the photons move a finite distance between scattering events, they don't stay perfectly coupled to the density perturbations of the baryons and slowly dissipate the energy in the waves. The result is a damping factor~\cite{Zaldarriaga:1995gi}
\beq
d_\gamma \approx A_\k \cos(c_s k \tau) e^{- k^2 /k_d^2} \qquad \frac{1}{k_d^2}= \int \frac{da}{n_e \sigma_T a^3 H(a)}\frac{1}{6(1+R_b)^2} \left(R_b^2 + \frac{16}{15} (1+R_b) \right) \ , 
\eeq
where $\sigma_T$ is the Thomson cross-section and $n_e$ is the density of free electrons. Notice that this Gaussian suppression is just the Fourier transform of probability distribution for the distance the photons travel during random walk, where the variance, $\sigma_X^2$, is given by
\beq
k_d^{-2} =  \sigma_X^2\propto  \frac{t}{n_e \sigma_{T}} \propto   \frac{1}{H(a_\star) n_e \sigma_{T} } \ .
\eeq
Using $k \to \ell / (\tau_0 -\tau_\star)$ we see that the damping in $\ell$ will take the form
\beq\label{eq:dampling}
\delta T(\ell) \approx A_{\ell, m} \cos\left(c_s \frac{\ell \tau_\star}{\tau_0-\tau_\star}\right) e^{-\ell^2 \sigma_X^2/ (\tau_0-\tau_\star)^2} \ .
\eeq
The main point to notice is that the behavior of the damping as we change $\Neff$ will be different if we hold $H_0$ or $\theta_\star$ fixed~\cite{Hou:2011ec}.  Concretely, the damping at high-$\ell$ is controlled by
\beq
\frac{\sigma_X^2}{(\tau_0- \tau_\star)^2} \propto \frac{H_0^2}{H(a_\star)}  \ .
\eeq
If we vary $\Neff$ holding $H_0$ fixed, it means $\frac{\sigma_X^2}{(\tau_0- \tau_\star)^2}  \propto \left(1+\frac{7}{8}\left(\frac{4}{11}\right)^{4 / 3} N_{\mathrm{eff}}\right)^{-1/2}$.  If we additionally vary $H_0$ with $\Neff$ so that we hold $\theta_\star$ fixed, then the damping scale evolves as
\beq
\frac{\sigma_X^2}{(\tau_0- \tau_\star)^2} \propto \frac{H_0^2}{H(a_\star)} \to \left(1+\frac{7}{8}\left(\frac{4}{11}\right)^{4 / 3} N_{\mathrm{eff}}\right)^{1/2} \ .
\eeq
Notice that the $\Neff$ dependence at fixed $\theta_\star$ has the opposite behavior we would expect for the damping tail is we had just calculated $k_d^{-2}$ and held $H_0$ fixed.  This difference can be seen clearly in Figure~\ref{fig:damping}.

It turns out that the change to the damping scale is the largest measurable effect that changing $\Neff$ has on the CMB and thus drives the constraints from current and future observations.  However, it is not particularly sensitive to the detailed physics of the light relics. First, it is just a measure of the overall size of $H(a)$ at recombination so it tells us nothing more than the overall energy density. The weak self-interactions of the neutrinos play no role.  Second, diffusion damping is sensitive to a number of other quantities, most notably the number density of free electrons, $n_e$.  This number is itself sensitive to BSM physics.  For example, $n_e$ is changed by the amount of helium, $n_e\propto (1-Y_p)$, where $Y_p$ is the primordial helium mass fraction. Electrons are bound to helium at recombination, so increasing the fraction of helium reduces the number of free electrons (holding $\rho_b$ fixed, which includes a contribution from helium). The amount of helium is sensitive to $\Neff$ at the time of Big Bang nucleosynthesis (BBN), but also any other changes to the physics during that epoch as well~\cite{Cyburt:2015mya}.  For these reasons, constraints on $\Neff$ change significantly if you allow the helium fraction (or any other parameter that changes the shape of the damping tail) to vary . That being said, even if you only want to consider a single light relic that contributes to $\Neff$ at all times after decoupling, we must change $Y_p$ in order to be consistent with $\Neff \neq 3.044$ during BBN (holding $\omega_b$ fixed) as
\beq
Y_{p} \approx 0.247+0.014 \times \Delta N_{\text {eff}} \ .
\eeq
This change to $Y_p(\Neff)$, also known is BBN consistency, is implemented by default in most, but not all, CMB codes.

\vskip 10pt
\noindent {\bf Gravitational Influence of Fluctuations}

\noindent We can understand a lot about the role of gravity in the recombination era by solving the inhomogeneous equations for the densities, Equation~(\ref{eq:linear_density}), while treating $\Phi_+$ as an external source. Fourier transforming $\x$ so that we have only an ODE in terms of $\tau$, we can find the inhomogeneous solution using Green's function,\beq\label{eq:green_phi}
d_\gamma^{\Phi}(\k,\tau) = \int d\tau' \frac{1}{c_s k} \sin\left(c_s k (\tau -\tau')\right) (-k^2 \Phi_+(\k,\tau')) \ .
\eeq
When $\Phi_+$ is determined by radiation, we can assume the role of gravity is negligible in the far future, namely $\Phi_+(\k, \tau\to \infty) \to 0$.  Since the source is no longer present in the far future, $d_\gamma^{\Phi}(\k,\tau \to \infty)$ is determined by the homogeneous solutions, $\sin(c_s k \tau)$ and $\cos(c_s k \tau)$. Inserting the homogeneous solution on the LHS, we can manipulate~(\ref{eq:green_phi}) to give
\begin{align}
d_\gamma^{\Phi}(\k,\tau \to \infty) \to A \cos(c_s k \tau) + B \sin(c_s k \tau) &=\int d\tau' \frac{1}{c_s k} \sin\left(c_s k (\tau -\tau')\right) (-k^2 \Phi_+(\k,\tau')) \ , \\
{\rm Im}\left((B + i A)e^{i c_s k \tau} \right) &=  {\rm Im} \, e^{i c_s k \tau} \int_0^\tau d\tau' e^{i c_s k \tau'} \Phi_+(\tau') \ .
\end{align}
It is useful to rewrite the integral on the RHS so that
\begin{align}
(B + i A) &=  \frac{k}{c_s}  \int_{-\infty}^\infty d\tau' e^{i c_s k \tau'} (\Phi^{(A)}_+(\tau') + \Phi^{(S)}_+(\tau') ) \ ,
\end{align}
where $\Phi_+^{(A)}(\tau<0) = - \Phi_+^{(A)}(|\tau|)$ and $\Phi_+^{(S)}(\tau<0) =  \Phi_+^{(S)}(|\tau|)$.  Defining these variables allows us to isolate the sine and cosine of the late-time solution with the real and imaginary parts of the integral,
\bea
A &=& \frac{k}{c_s }  \int_{-\infty}^\infty d\tau' e^{i c_s k \tau'} \Phi^{(A)}_+(\tau') \ ,\\
B &=& \frac{k}{c_s}  \int_{-\infty}^\infty d\tau' e^{i c_s k \tau'}  \Phi^{(S)}_+(\tau')  \ .
\eea
We notice two things about this solution: 
\begin{enumerate}
\item Radiation of any kind will generically produce $A \neq 0$, including just from the photons back-reacting on themselves.  This is sometimes called radiation-pumping as the radiation amplifies its own fluctuations (in phase).  This can be seen from the integral over $\tau'$ if we notice that $\Phi_+^{(A)}(\tau')$ is generally a non-analytic function at $\tau' =0$. The amplitude $A$ will also get an additional contribution from $\Neff$ but, as we will discuss, these effects will largely be degenerate with other cosmological parameters.
\item We will have $B = 0$ unless our source can move faster than $c_s \approx \frac{1}{\sqrt{3}}$.  This is the due to the effective role of causality in the fluid, as $B=0$ is required because information cannot travel faster than the sound speed of the fluid.  However, in the Standard Model, the neutrinos are not part of this fluid and travel near the speed of light so that $c_\nu > c_s$.  This gives us the result that $B \propto \Neff$ (at least to linear order), which follows from the integral because, for adiabatic fluctuations, $ \Phi^{(S)}_+(\tau') $ is an analytic function and therefore $B$ vanishes if we can close the contour in the upper-half plane.  This fails, for example, when $\Phi^{(S)} \propto \cos(c_\nu k \tau)$, with $c_\nu > c_s$.  
\end{enumerate}

Remarkably, this phase shift ($B\neq 0$) can be searched for directly in the data and was first detected at 5$\sigma$ TT data by~\cite{Follin:2015hya}.  It has subsequently been inferred from TTTEEE at $10\sigma$~\cite{Baumann:2015rya}.  The same phase shift survives in the BAO feature, Equation~(\ref{eq:bao}), in large scale structure~\cite{Baumann:2017lmt} and was been measured at $2.5\sigma$~\cite{Baumann:2019keh}. Taken all together, we have a fairly unambiguous detection of the cosmic neutrino background using its gravitational influence of free-streaming neutrinos on the baryonic matter in the Universe, with an energy density (temperature) consistent with theoretical predictions from BBN.

\subsection{Future CMB Experiments}

Now we come to the measurements of $\Neff$ via the CMB.  We have focused, so far, on the measurement of the CMB temperature ($C_\ell^{\rm TT}$), however, the coming generations of surveys will make a lot of their gains in sensitivity from the measurement of the polarization of the CMB~\cite{Chang:2022tzj}. This is essential in the measurement of gravitational waves from inflation, i.e.~primordial B-modes~\cite{Seljak:1996gy,Kamionkowski:1996ks}.  However, for generic BSM physics, the role of of the polarization measurement may not seem to be essential but is important for understanding the real-world opportunities and limitations of these surveys.

Thomson scattering polarizes the CMB for the same reason that light reflected off any object is polarized (and hence why good sunglasses use polarized lenses).  For unpolarized photons scattering off an electron from a specific incoming direction, the probability of scattering in different outgoing directions depends on the polarization of the outgoing photon.  As a result, scattered light from a localized source is polarized. However, the CMB is not a localized source but instead is, to first approximation, a uniform distribution of photons in every direction. For the CMB maps to exhibit a net polarization, we must have a residual effect after adding up the contributions of photons scattering from all possible incoming directions. If the incoming photons are unpolarized and equally likely to come with the same energy from any given direction, then there is no net polarization.  This can been seen in detailed calculations, but is fundamentally just a result of symmetry: if the initial distribution of photons has no preferred directions, it cannot produce a preferred polarization from scattering.  The same is even true for a local temperature dipole: polarization is not actually a vector (two polarization ``vectors" rotated by $\pi$ are equivalent) and cannot be proportional to a just a dipole.

In the end, generating a polarized CMB map requires that radiation at recombination has a non-zero local quadrupole moment. There is no such quadrupole generated in a perfect fluid, which is described only by a monopole (the local temperature $T$) and a dipole (the local velocity $\vec v$).  Fortunately, the plasma at recombination is not a perfect fluid and a quadrupole is generated by the finite mean-free path of the photons. However, the polarization field $P = Q + i U$, where $Q$ and $U$ are the Stokes parameters, is suppressed relative to the CMB temperature fluctuations by the mean-free path $n_e \sigma_T$, so that $|P| \approx T/(n_e \sigma_T)$.

Given this suppression, a natural question is why polarization is useful at all? Since the amplitude of the polarization signal is much smaller than that of the temperature fluctuations, we might imagine deeper CMB maps would have the most gains from the temperature maps which have much higher signal to noise. In addition, the physics responsible for the polarization is essentially the same, as it is just encoding the propagation of sound waves in the pre-recombination plasma (except, again, for the gravitational waves). 

In order to understand the value of the polarization maps, we need to discuss how we quantify the sensitivity of a CMB survey. To start, we need to understand the sources of noise in our maps.  As discussed in Section~\ref{sec:coins}, the maps of the CMB are themselves maps of noise and therefore there is a fundamental limitations on the amount of information. However, we also have two additional sources of noise we have to worry about, detector noise and point sources. If we Fourier transform\footnote{Note here that $\vl$ is a two dimensional Fourier of $\n$ on the flat sky so that $\vl \to \ell, m$ on the full sky.  } a map of the CMB (taking the flat-sky limit for simplicity), then a good approximation to what we see on a wide range of scales is
\beq
T_{\rm measured}(\vl) = B(\vl) (T_{\rm CMB}(\vl)+ T_{\rm PS}(\vl))+ N_{\rm detector}(\vl) \ ,
\eeq
where $T_{\rm PS}$ is the contribution from unresolved point sources between us and the CMB, $N_{\rm detector}$ is the white noise associated with the sensitivity of the CMB detectors, and $B(\vl)$ is a transfer function associated with the ``beam" of the telescope (i.e.~the limiting resolution).  In principle, we can determine the function $B(\vl)$ exactly (or close enough for this discussion) by calibration of the instrument, so that we can simply remove the effect by dividing through 
\beq
T_{\rm map}(\vl) = (B(\vl))^{-1} T_{\rm measured}(\vl) = T_{\rm CMB}(\vl)+ T_{\rm PS}(\vl) + (B(\vl))^{-1}  N_{\rm detector}(\vl) \ .
\eeq
The same formula will also hold for the polarization maps, except that the detector noise in polarization is usually larger than in temperature by a factor of $\sqrt{2}$, and that the the point-source contribution is suppressed by the small fraction of the emission from point sources that is polarized.

The sensitivity of these CMB experiments to a set of cosmological parameters $\Theta$ can be quantified using the Fisher matrix~\cite{Wu:2014hta}:
\beq\label{eq:fisher_CMB1}
F_{i j}=\sum_{\ell} \frac{2 \ell+1}{2} f_{\rm sky} \operatorname{Tr}\left(\boldsymbol{C}_{\ell}^{-1}(\Theta) \frac{\partial \boldsymbol{C}_{\ell}}{\partial \Theta_{i}} \boldsymbol{C}_{\ell}^{-1}(\Theta) \frac{\partial \boldsymbol{C}_{\ell}}{\partial \Theta_{j}}\right) \ ,
\eeq
where $f_{\rm sky}$ is the sky fraction of the survey, $\Theta_i$ is one of the cosmological parameters being measured, 
\beq
\mathbf{C}_{\ell} \equiv\left(\begin{array}{cc}
C_{\ell}^{T T}+N_{\ell}^{T T} & C_{\ell}^{T E}  \\
C_{\ell}^{T E} & C_{\ell}^{E E}+N_{\ell}^{E E} 
\end{array}\right)  \ ,
\eeq
and $N_\ell = C^{\rm PS}_\ell + B_\ell^{-2} C^{\rm detector}_\ell$ is the effective noise power spectrum including both point sources and detector noise.\footnote{One might ask why the CMB data usually follows $C_\ell^{TT}$, and not $C_\ell^{\rm TT} + N_\ell^{\rm TT}$, even when the detector noise is becoming large ($N_\ell^{\rm TT} \gtrsim C_\ell^{\rm TT}$).  The reason is that the reported $C_\ell^{TT}$ is usually obtained by cross-correlating maps measured at different times so that $\langle N_{\rm dectector}(t_1)N_{\rm dectector}(t_2)\rangle$ vanishes.  The same approach does not remove the point sources and those usually do show up if we plot $C_\ell^{TT}$ to high enough $\ell$.}  Here we replaced $P \to E$, where $E$ ($B$) is the parity even (odd) component of the polarization map, which are the only polarization configurations that arise in the primary\footnote{The primary CMB is the CMB without a number of effects that alter the appearance of the CMB after recombination (the secondaries), including CMB lensing that we will discuss in Section~\ref{sec:lensing}. } CMB in the absence of primordial gravitational waves (see~\cite{Zaldarriaga:2003bb} for review). The formula here is literally the same as Equation~(\ref{eq:fisher_def}) where we have to take into account that the amplitude of the fluctuations (the variance) depends on $\ell$ and the type of map.  We will focus on the case of $\Lambda$CDM+$\Neff$ where 
\beq
\Theta_i \in \{\theta_s,\omega_{\rm m}, \omega_{\rm b}, A_s, n_s, \bar \tau_{\rm optical}, \Neff \} \ ,
\eeq 
are the parameters we are trying to constrain simultaneously.  The noise, $N_\ell$, does not depend on the cosmological parameters (and does not contribute to the derivatives) and therefore the Fisher matrix can be understood as representing the $(S/N)^2$ for the individual parameters.

The sensitivity of most experiments to any specific cosmological parameter is limited both by the experimental noise and the degeneracy with other parameters.  The degeneracy with other parameters is easy to understand as follows: imagine the probability of getting heads from a coin $p$ is given by $p = A+B$ where $A$ and $B$ are the parameters we want to know.  Repeatedly flipping the coin will help us determine $p$, but we will not have enough information to determine $A$ or $B$.  We can see this at the level of the Fisher matrix as follows: if two parameters affect the observable in the same way, $\frac{d}{d\Theta_1} C_\ell \propto \frac{d}{d\Theta_2} C_\ell$, then the Fisher matrix is not of full rank and we won't be able to invert it.  This is the same as saying that $\sigma(\Theta_{1,2} ) \to \infty$ when $\Theta_1$ and $\Theta_2$ have the same effect on the CMB (i.e.~are degenerate).

As a general rule, parameters that control the changes to smooth functions are often degenerate.  The basic reason is that if a function is well approximated by a low-order polynomial, then all the fundamental parameters are effectively controlling these few coefficients of the polynomial. As a result, if we have lots of free parameters that control a smooth function (or a smooth envelope of our oscillatory function), then odds are that several linear combinations of them will cause very similar changes to the function.  We saw this already when discussing the CMB damping tail.  In contrast, the oscillatory part of the shape is usually much more robust: the location of each peak and trough carries independent cosmological information.

\begin{figure}[h!]
\centering
\includegraphics[width=6in]{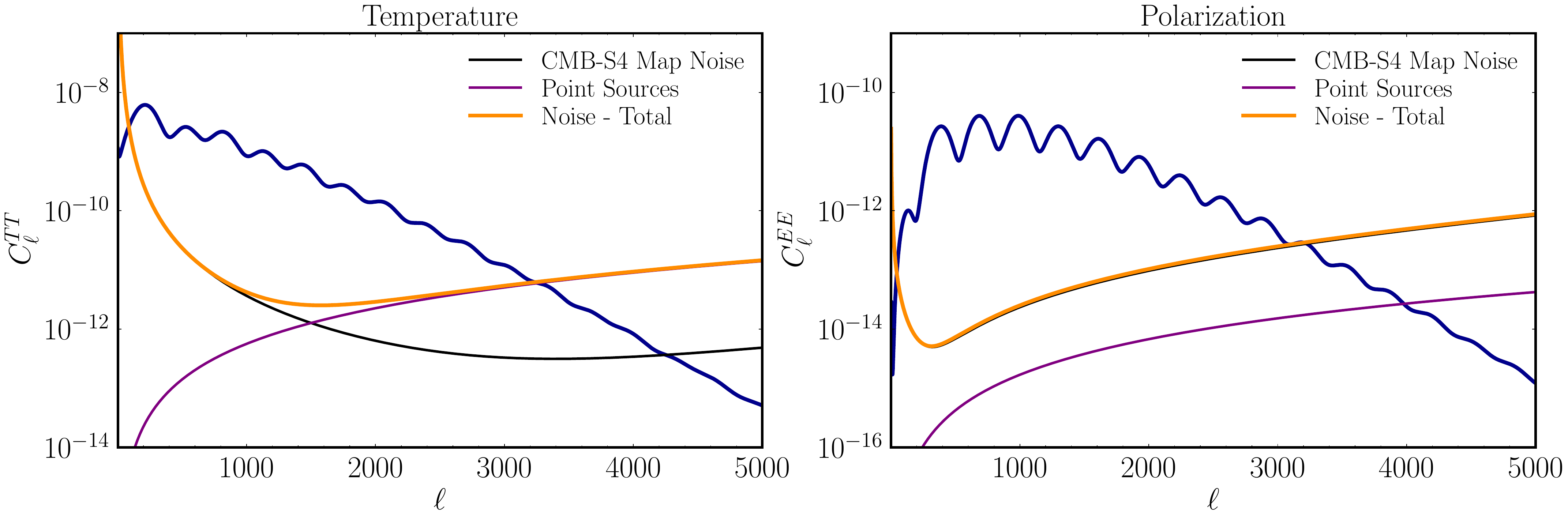}
\caption{Noise curves for temperature and polarization using similar modeling to~\cite{SimonsObservatory:2019qwx}.  The rise of the noise at low-$\ell$ is associate with noise from the atmosphere (for ground based surveys) and the point source amplitudes are can from the estimates of~\cite{Puglisi:2017lpn}.}
\label{fig:pol}
\end{figure}

From these observations, we can now understand the unique value of the polarization data. First, we see from Figure~\ref{fig:pol} that the point sources make a much small contribution to the polarization noise curves.  Concretely, in temperature, the CMB will be swamped by point sources for $\ell > 3000$.  This places a fundamental limit on our ability to recover more information from $C_\ell^{\rm TT}$ with more sensitive surveys.  Point sources are only polarized at the percent-level, which translates to a suppression of approximately $10^{-4}$ in $C_\ell^{\rm PS, EE}$ compared to $C_\ell^{\rm PS, TT}$.  As a result, the point sources are significantly more suppressed in polarization than the polarization of the CMB itself.  As a result, we see that polarization noise curves are minimally impacted by point sources with the coming generation of observations.

Because of the reduced impact of point sources, the number of $E$-modes measured in the survey is determined by the dectector noise level, at least for $\ell < 4000$. At a given noise level, the set of cosmic variance limited modes, $N_\ell < C_\ell /\sqrt{2\ell+1}$ for $\ell < \ell_{\rm max}$, define the effective number of modes in the survey, $N_{\rm modes} \approx f_{\rm sky} \ell_{\rm max}^2$. In polarization, lowering the detector noise, $C^{\rm detector}_\ell$, increases $\ell_{\rm max}$ and drives the overall sensitive of the survey.  However, because all the $C_{\ell}$ decay exponentially with $\ell$, as in Equation~(\ref{eq:dampling}), the gain in the number of modes is roughly logarithmic\footnote{Fortunately, detector sensitivity has been improving exponentially in time, in a Moore's law-type fashion, such that the number of modes has been increasing like a power-law in time.} in the decreasing noise, $\ell_{\rm max} \propto \log N^{-1}_{\ell}$.  By comparison, surveying a larger amount of the sky, or increasing $f_{\rm sky}$, gives us a linear increase in the number of modes.  As a result, most high-$\ell$ surveys are designed to optimize for $f_{\rm sky}$ (within the limitations set by the foregrounds like the galaxy or the location of your telescope on earth).

A second important result is that polarization data produces sharper acoustic peaks, as we can also see from comparing the amplitudes of the oscillations in TT and EE in Figure~\ref{fig:pol}.  This means that the uncertainties in the polarization peak locations are smaller for the polarization spectra and give cleaner measurements of cosmological parameters that affect these positions.\footnote{One can understand, from first principles, how the peak positions are effected by different physical effects~\cite{Pan:2016zla}.}  The reason for the sharper peaks is that the CMB temperature fluctuations are primarily a combination of two terms: gravitational redshift due to the overdensity at recombination, proportional to $\delta \propto \cos(c_s k \tau)$, and doppler shift, due to the line-of-sight velocity at recombination, $v_{||} \sim \sin(c_s k \tau)$. The latter is smaller because we have to average over the angle of the velocity vector relative to the line-of-sight, but is otherwise out of phase with the density fluctuations and washes out the acoustic oscillations.  Polarization is only a measurement of the temperature quadrupole at recombination and thus all contributions to the signal are in-phase.  

Using these tools, we can forecast the sensitivity of future cosmic surveys to $\Neff$.  These are shown in Table~\ref{table:Neff} along with the current best measurement of $\Neff$ from Planck~\cite{Planck:2018vyg}.  The next generation of surveys, concretely the Simons Observatory (SO)~\cite{SimonsObservatory:2019qwx} and CMB-S4~\cite{Abazajian:2019eic} will improve significantly on this measurement.  Keep in mind, the physical parameter we care about for light thermal relics is not $\Neff$ itself but the freeze-out temperate $T_F$ (or the coupling strength $\Lambda$).  In that context, the change in $\sigma(\Neff)$ from $0.07$ to $0.03$ translates into a change of $T_F > {\cal O}(1)$ GeV to $T_F > {\cal O}(100)$ GeV, as seen in Figure~\ref{fig:DeltaNeff}, for a single massless vector, complex scalar (two Goldstone bosons), or Weyl fermion.

\begin{table}[h!]

\begin{center}
\begin{tabular}{ | c || c | c |} 
 \hline  Survey & $\sigma(\Neff)$, $Y_p$-consistency & $\sigma(\Neff)$-$Y_p$ marginalized \\
 \hline\hline
 Planck~\cite{Planck:2018vyg} & 0.17 & 0.29 \\ 
\hline
 Simons Observatory~\cite{SimonsObservatory:2019qwx} & 0.07 & $\approx 0.1$ \\
 CMB-S4~\cite{Abazajian:2019eic} & 0.03 & 0.07 \\ 
 \hline
\end{tabular}
\caption{Forecasts for $\Neff$ from future CMB surveys.  The middle column using $Y_p$-consistency assumes $\Neff$ is the same at BBN and recombination (and hence we change $Y_p$ consistently).  The third column treats $Y_p$ and $\Neff$ as independent parameters so that the $\Neff$ measurement is mostly a measurement of the phase shift (as the parameters are degenerate for the damping tail). The SO forecasts with $Y_p$ marginalization is approximate, as there is no official forecast of this combination of parameters by SO. The forecasts for CMB-S4 with $Y_p$ marginalization uses delensing of the temperature and polarization~\cite{Green:2016cjr}, which sharpens the acoustic peaks and gives a better measurement of the phase shift.  Using just the lensed data, we would have $\sigma(\Neff) = 0.07\to 0.09$ instead.}\label{table:Neff}
\end{center}
\end{table}

Converting constraints on $\Neff$ to parameters of a given model will depend in detail on the specific coupling of the light particle(s) to the Standard Model.  However, in most cases, the conversion is well approximated by our formula for $T_F$ in Equation~(\ref{eq:TF_theory})~\cite{Brust:2013ova}. In a wide range of examples, future (and sometime even current) $\Neff$ measurements will provide the best constraints on couplings of a single light degrees of freedom to the Standard Model~\cite{Dvorkin:2022jyg}. The CMB has two key advantages: (1) it is potentially sensitive to $T \gg 1$ TeV and thus is a powerful probe of couplings of high dimension~\cite{Brust:2013ova} and (2) all the Standard Model particles were in equilibrium at high temperature and so it is a excellent probe of couplings to heavy fermions~\cite{DEramo:2021lgb,Green:2021hjh}. More generally, $\Neff$ is a valuable tool for searching for dark sectors, including the physics of light dark matter or light force carriers~\cite{Green:2017ybv,Knapen:2017xzo} or hidden copies of the Standard Model motivated by solutions to the hierarchy problem~\cite{Craig:2016lyx,Chacko:2016hvu,Arkani-Hamed:2016rle}. The reason $\Neff$ is so versatile is that we do not depend on a specific coupling of the particle to the Standard Model to make a detection; any coupling that thermalizes the particle at some point in the history of the Universe is enough.  Once it is produced, gravity is the force that we use to infer the presence of these particles.

\newpage

\section{Massive Neutrinos and CMB Lensing} \label{sec:lensing}

In the previous section, we discussed how the CMB encodes physics of the hot plasma around the time of recombination.  From our discussion of the dark matter, the matter power spectrum, and the BAO feature in Section~\ref{sec:coins}, we also saw how some of this information is stored in the distribution of matter at low redshifts.  In this section, we will focus on distribution of matter in the Universe after recombination and how it is sensitive to the physics of the neutrino mass and any other light but massive particles left from the early Universe~\cite{Lesgourgues:2006nd}. We will focus in particular on how the CMB is also sensitive to this physics, not from the primary CMB but from the gravitational lensing of the CMB (one of several CMB secondaries of interest). 

\subsection{Neutrino Mass and the Late Universe}

Given constraints from the primary CMB, we know the overall mass scale of the neutrinos is subject to the constraint $\sum m_\nu < 260$ meV (95\%)~\cite{Planck:2018vyg}.  On the other hand, neutrino oscillations tell use that $\sum m_\nu \geq 58$ meV, where $\sum m_\nu = 58$ meV corresponds to a single massless neutrino and a normal hierarchy of masses~\cite{Workman:2022ynf}.  This guarantees that at least 2 of the neutrino mass eigenstates will become non-relativistic at a redshift $1000 > z_{\rm NR} > 100$.   We can quantify this statement by defining $c_\nu$, the non-relativistic speed of neutrino propagation, as
\beq
c_\nu = \frac{\langle p \rangle}{m_\nu} = \frac{3 T_\nu}{m_\nu} \approx 8.7 \times 10^{-3}\, c \, \,   \frac{a_0}{a} \, \left(\frac{58 \, {\rm meV}}{m_\nu}\right) \ ,
\eeq
where $a_0 =1$ is the scale factor today.  This equation also assumes the the neutrino temperature, $T_\nu$, follows from our calculation in Equation~(\ref{eq:Tnu}).

Once the neutrinos are non-relativistic, the homogenous energy density in neutrinos is indistinguishable from any other form of matter.  As a result, the matter density at low redshifts is given by 
\beq
\bar \rho_m = \bar \rho_{\rm b} +\bar \rho_{\rm cdm} + \bar \rho_\nu \propto a^{-3} \ ,
\eeq
where 
\beq
\bar\rho_\nu =\Omega_\nu H_0^2 \ , \qquad \Omega_\nu h^2 =  6.2 \times 10^{-4} \, \left( \frac{\sum m_\nu}{58 \, {\rm meV}} \right) \ .
\eeq
We see that massive neutrinos are a small but non-zero contribution to $\omega_m$.

The baryons and cold dark matter are both cold enough that we can neglect their effective velocities, $c_{\rm b}, c_{\rm cdm} \approx 0$.  In contrast, the thermal velocities of massive neutrinos are not negligible, even though they behave like matter for the purpose of the overall expansion rate.  Neutrinos move over cosmological distances in a Hubble time and thus the evolution of their density fluctuations must include the effect of their velocity.  We can determine the evolution of the cold matter, $\delta_\mathrm{cb}= \frac{\delta\rho_\mathrm{cdm} + \delta\rho_\mathrm{b}}{\bar{\rho}_\mathrm{cdm}+\bar{\rho}_\mathrm{b}}$, and the neutrinos, $\delta_\nu = \frac{\delta \rho_\nu}{\bar{\rho}_\nu}$, as a coupled system of fluids using mass and momentum conservation 
\begin{equation}
\dot \delta_{\rm cb}(\k,t)-a^{-1} k^2 u_{\rm cb}=0 \ , \qquad
\dot \delta_{\nu}(\k,t)-a^{-1} k^2 u_{\nu}=0 \ ,
\end{equation}
and
\begin{align}
\dot u_{\rm cb}+H  u_{\rm cb} =-\frac{1}{a} \Phi \ , \qquad 
\dot u_{\nu}+H u_{\nu} =-\frac{1}{a} \Phi-\frac{c_{\nu}^{2}}{a } \delta_{\nu} \ ,
\end{align}
respectively. We have again defined $u$ as the scalar velocity potential such that $\vec v_i = \vec \nabla u_i$.  On the scales of interest, Newtonian gravity is a good approximation and 
\beq
\nabla^2 \Phi =  4 \pi G \left(\bar \rho_{\rm cb} \delta_{\rm cb} +\bar \rho_\nu \delta_\nu \right) \ .
\eeq
We can combine these equations to determine the evolution of the two over-densities, 
\begin{align}
    \ddot{\delta}_{\rm cb}+\frac{4}{3 t} \dot{\delta}_{\rm cb}&=\frac{2}{3 t^{2}}\left[f_\nu \delta_{\nu}+(1-f_\nu) \delta_{\rm cb}\right] \label{eq:nu_cmd1}\\
    \ddot{\delta}_{\nu}+\frac{4}{3 t} \dot{\delta}_{\nu}&=-\frac{2 \alpha}{3 t^{2}} \delta_{\nu}+\frac{2}{3 t^{2}}\left[f_\nu \delta_{\nu}+(1-f_\nu) \delta_{\rm cb}\right] \label{eq:nu_cmd2}
\end{align}
where
\beq
    \alpha \equiv \frac{3 k^{2} c_\nu^{2} t^{2}}{2 a^{2}}=\frac{k^2}{k_{\rm fs}^2} \ , 
    \quad f_\nu \equiv \frac{\Omega_{\nu}}{\Omega_\mathrm{m}} =  4.4\times 10^{-3} \left( \frac{\sum m_\nu}{58 \, {\rm meV}} \right) \ .
\eeq
We defined $\alpha$ in terms of a free-streaming wavenumber~\cite{Bond:1983hb}, $k_{\rm fs} = \sqrt{\frac{3}{2}}\frac{aH}{c_\nu}$ such that
\beq
    k_{\rm fs} = 0.04 \, h \, {\rm Mpc}^{-1} \times \frac{1}{1+z} \, \left(\frac{\sum m_\nu}{58 \, {\rm meV}}\right) \, . 
    \label{eq:kfs_nu}
\eeq
The phenomenology of massive neutrinos is controlled by $k_{\rm fs}$, which plays the role of an effective Jeans scale such that the neutrinos do not cluster when $k \gg k_{\rm fs}$. There is a separate $k_{\rm fs}$ for each neutrino mass eigenstate. However, in practice, near term observations will not be able to distinguish the individual mass eigenstates from cosmological observations.  For this reason, from an observational point of view, we replaced $m_\nu \to \sum m_\nu$ when discussing the suppression of small scale power. 

For free-streaming neutrinos, $\alpha$ is not a constant because $k_{\rm fs}$ depends on time.   As we vary $k$, we are interpolating between $\alpha \ll 1$ and $\alpha \gg 1$. Nevertheless, as we will explain below, the correct solutions in the limiting cases, $\alpha \ll 1$ and $\alpha \gg 1$, can be obtained by separately treating $\alpha$ as a constant in each regime. First, let us consider the case where $k$ is small and $\alpha \to 0$. The equations for cold matter and neutrinos are identical and $\delta_m = \delta_\nu + \delta_{\rm cb}$ behaves just like cold matter.  In the second case, where $k$ is large and $\alpha \gg 1$, the evolution of $\delta_\nu$ is given approximately by
\begin{align}
\ddot{\delta}_{\nu}+\frac{4}{3 t} \dot{\delta}_{\nu}\approx-\frac{2 \alpha}{3 t^{2}} \delta_{\nu} \to \delta_\nu = c t^{-\frac{1}{6}} \, \cos\left(\frac{2}{\sqrt{6}}\sqrt{\alpha} \log t \right) 
\end{align}
and therefore decays so that $\delta_\nu \to 0$ at late times.  The details of this decay are unimportant (hence justifying treating $\alpha$ as a constant), as the evolution of $\delta_{\rm cb}$, and therefore $\delta_m \approx \delta_{\rm cb}$, becomes independent of $\alpha$ when $\alpha \gg 1$.  We can therefore solve these equations analytically treating $\alpha$ as a constant throughout and still reproduce the correct limiting behavior at both large and small $k$. Following~\cite{Weinberg:2008zzc}, we make the ansatz 
\beq
\delta_{\rm c} \propto t^{\gamma}, \quad \delta_{\nu}=\xi \delta_{c}
\eeq
to find that the growing modes have
\beq
\gamma =2 / 3-\frac{2 f_\nu \alpha}{5(1+\alpha)}, \quad \xi=\frac{1}{1+\alpha}-\frac{f_\nu \alpha^{2}}{(1+\alpha)^{3}}
\eeq
to linear order in $f_\nu \ll 1$. Combining these results, we find
\beq
\delta = (1- f_\nu) \delta_{\rm c} + f_\nu \delta_\nu = (1 +(\xi-1) f_\nu )\delta_{\rm c} \propto (1 +(\xi-1) f_\nu ) t^{\gamma} \ .
\eeq
To simplify things, we can also expand the exponential
\begin{align}
\delta(\k,f_\nu) 
 &\approx \delta(\k,f_\nu=0)  (1 +(\xi-1) f_\nu )  \times \left(1 -\frac{3 f_\nu \alpha}{5(1+\alpha)} \log \frac{a}{a_{\rm NR}} \right) \label{eq:nu_suppression} \ ,
\end{align}
 where we used $a \propto t^{2/3}$ in the matter era and $a_{\rm NR} = 1/(1+z_{\rm NR}) \approx 10^{-2}$ is the scale factor when the neutrinos become non-relativistic. At large $k$, this gives rise to an overall suppression
 \beq
    P_{\sum m_\nu} (k\gg k_{\rm fs}, z ) \approx  \left(1 - 2 f_\nu  -\frac{6 }{5} f_\nu\log \frac{1+z_{\rm NR}}{1+z} \right) P_{\sum m_\nu =0}(k\gg k_{\rm fs}, z ) \label{eq:suppression}
\eeq
The factor $1-2 f_\nu$ was perhaps expected from the fact that the neutrinos don't cluster; after all, this is just correcting for the fact that only a fraction of the total matter, the cold matter, has density fluctuations on small scales.  However, the second term is the result of the change to the rate of growth of the fluctuations in cold matter itself, due to the presence of the neutrinos.  This second effect may be surprising at first sight, but we recall that we already saw that the matter density fluctuations grow much less slowly during radiation domination.  In this precise sense, the neutrino energy density is weakening the growth of structure in an analogous way to radiation.  This effect is log-enhanced because it changes the entire history of the growth of structure in the redshift range $z\in[0,100]$.  The result is that the overall suppression is approximately $(1-8 f_\nu)$.  This extra source of suppression is why we can measure sub-percent densities of neutrinos contributing to the matter density.

Equations~(\ref{eq:nu_cmd1}) and~(\ref{eq:nu_cmd2}), despite their simplicity, are actually quite accurate models of the power spectrum suppression from neutrinos at all $k$. While it is difficult to solve the coupled system of equations analytically for time-dependent $\alpha$, the equations are easily solved numerically and give excellent agreement with the results of Boltzmann codes~\cite{Green:2021gdc,Green:2021xzn}.

\subsection{Relation to Dark Sector Physics}

Given the above description, one would be naturally interested in how massive relics in dark sectors are constrained by the same measurements.  Following~\cite{Green:2021xzn}, we can introduce a dark sector of $\chi$-particles with a typical mass-scale $m_\chi$, number density $n_\chi$, and temperature $T_\chi$ at $z=0$.  We will assume this is not the primary component of the dark matter and so we require that it is a subcomponent such that   
\beq
m_\chi n_\chi < \rho_\mathrm{m} \ .
\eeq
To be distinguishable from dark matter, we will assume the mass is such that it was relativistic at BBN and therefore it contributes
\beq
\Delta \Neff^{\rm BBN} = \frac{4}{7} \, g_{\chi,\star} \left(\frac{T_\chi}{T_\nu}\right)^4 \ ,
\eeq
where $g_{\chi,\star}$ is the effective number of degrees of freedom of the $\chi$-sector. We can constrain this quantity using the primordial abundances of elements, with current limits giving $\Delta \Neff^{\rm BBN} < 0.37$ at 95\% confidence~\cite{Cyburt:2015mya}. This will change the abundance of helium according to
\beq
Y_{p} \approx 0.247+0.014 \times \Delta N_{\text {eff}}^{\mathrm{BBN}} \ ,
\eeq
which then impacts the CMB damping tail as well, even if $\chi$ is non-relativistic at recombination.  

To understand the cosmological impact of $\chi$, we can treat it like a massive neutrino and determine the redshift $z_\chi$ where $\chi$ becomes non-relativistic, namely $\langle p_\chi \rangle / m_\chi <1$.  If the particle is free-streaming, it will act like a neutrino as far as gravity is concerned for $z < z_\chi$ where
\beq
1 + z_\chi =  \frac{m_\chi}{3 T_\chi} \ .
\eeq
All observable modes of the CMB $(\ell <4000)$ enter the horizon when $z < 10^5$ and therefore if $z_\chi > 10^5$, the $\chi$ sector will behave like cold dark matter for the purposes of any CMB or LSS constraint.  

At late times, the $\chi$ field will behave like matter for the purposes of expansion and thus will also suppress power on scales smaller than the free-streaming scale of $\chi$, namely $k > k_{\rm fs}^{\chi}$.  If $k_{\rm fs}^{\chi} \lesssim k^{\nu}_{\rm fs}$, the contribution of $\chi$ to the matter density will be indistinguishable from a change to $\sum m_\nu$.  Using $\rho_\chi = m_\chi n_\chi$, the resulting effective sum of neutrino masses is given by
\beq\label{eq:mnueff}
    \sum m_{\nu,{\rm eff}} = \sum m_\nu + g_{\chi} m_\chi \left(\frac{T_\chi}{T_\nu}\right)^3 = \sum m_\nu \left(1 +  \frac{1+ z_\chi}{1+z_{\nu}} h_\chi \Delta \Neff^{\rm BBN} \right) \ ,
\eeq
where $g_\chi$ is the effective number of degrees of freedom for the number density, where we include a $3/4$ factor for fermionic particles (rather than $7/8$ for the relativistic energy or entropy density), and we have defined $h_\chi = g_\chi / g_{\chi, \star} \approx 1$ so that we can express the result in terms of $\Delta \Neff^{\rm BBN}$.

The above expression is a slight simplification, as the amplitude and shape of the effect depend non-trivially on $z_\chi$. We know the amplitude is logarithmically sensitive to $z_\chi$ using Equation~\eqref{eq:suppression}, so we are implicitly assuming $\log z_\chi / z_\nu \approx 1$. This is a reasonable assumption because we also had to assume $ k^{\chi}_{\rm fs} \lesssim k^{\nu}_{\rm fs}$ where
 \beq\label{eq:kfs_chi}
    k^{\chi}_{\rm fs}(z) = \sqrt{\frac{3}{2}}\frac{aH}{c_\chi} = k^{\nu}_{\rm fs}(z) \, \left( \frac{m_\chi}{\sum m_\nu}\right) \, \left( \frac{T_\nu}{T_\chi} \right) =  k^{\nu}_{\rm fs}(z) \, \frac{1+ z_\chi}{1+z_{\nu}} \ .
 \eeq
If we were to have $z_\chi > 1000$, the scale dependence of the signal would be visible on observable scales in current large-scale structure surveys. Therefore it is only the regime $z_\chi \ll z_\nu$ where the logarithmic factor might be important, but the amplitude of the contribution from $\chi$ is already suppressed by $(1+z_\chi) /(1+z_\nu)$ which is more important than the change to the logarithm.
 
Finally, if $z_\chi < 1100$, the energy density in $\chi$ contributes to $\Neff$ for the purposes of the CMB, namely
\beq
    \Delta \Neff^{\rm CMB} = \frac{4}{7} \, g_{\chi,\star} \left(\frac{T_\chi}{T_\nu}\right)^4 \ .
\eeq
When $z_\chi \gtrsim 1100$, we cannot simply map the parameters of this model to $\Neff$ or $\sum m_\nu$ for the purposes of the CMB and we would require a dedicated analysis.  However, constraints from $\sum m_\nu$ from the primary CMB (no lensing) constraints $\sum m_\nu < 0.26$ eV  at 95\% confidence~\cite{Planck:2018vyg} and therefore if $T_\chi \approx T_\nu$, we can anticipate a similar constraint on $m_\chi$.

\subsection{The CMB as a Backlight}

The surface of last scattering occurs at $z\approx 1100$ and has the photons arriving on earth from the furthest possible distance associated with the observable Universe.  As these photons travel towards us, they travel through all the intervening matter, radiation, and energy that fills the Universe. Many things can happen to these photons along the way that distort the appearance of the CMB.

One important effect that is always included in the $\Lambda$CDM model is the reionoization of the Universe.  As we discussed in Section~\ref{sec:coins}, the photons primarily scatter off of free electrons which are effectively absent after the Universe becomes neutral.  However, star formation introduces more high energy radiation that (re)ionizes the hydrogen and reintroduces free electrons into the Universe around $z\approx 6-7$. However, because of the expansion of the Universe, the density of electrons has diluted by a factor of $(1+z_{\rm reion.})^3/(1+z_{\rm recomb})^3  \approx 10^{-7}$, enough to make the scattering inefficient.  A small fraction of photons do scatter and the optical depth\footnote{It is common in the literature to use $\tau$ as the optical depth.  We will use $\bar \tau_{\rm optical}$ to avoid potential confusion with conformal time.} associated with this process is given by $\bar \tau_{\rm optical} = 0.054 \pm 0.007$~\cite{Planck:2018vyg}. This rescattering\footnote{There is also additional scattering of electrons by the hot gas found inside galaxy clusters.  This is known as the Sunyaev-Zeldovic effect and leads to a changed in the spectrum of photons.  This effect can be used to find and map high redshift clusters, which is an interesting tool for cosmology and astrophysics. Unfortunately, we will not have time to discuss this in detail, but see~\cite{Carlstrom:2002na,Mroczkowski:2018nrv,Basu:2019hww} for review.} of the photons reduces the amplitude of the CMB temperature fluctuations so that at high-$\ell$ the amplitude is effectively 
\beq
C_{\ell>30}^{\rm TT} \propto A_s e^{-2 \bar \tau_{\rm optical}} \ .
\eeq  
This effect is often described like frosted glass: you lose the fine details of the image of the CMB but can still see the very large scale variations. 

\begin{figure}[h!]
\centering
\includegraphics[width=4.5in]{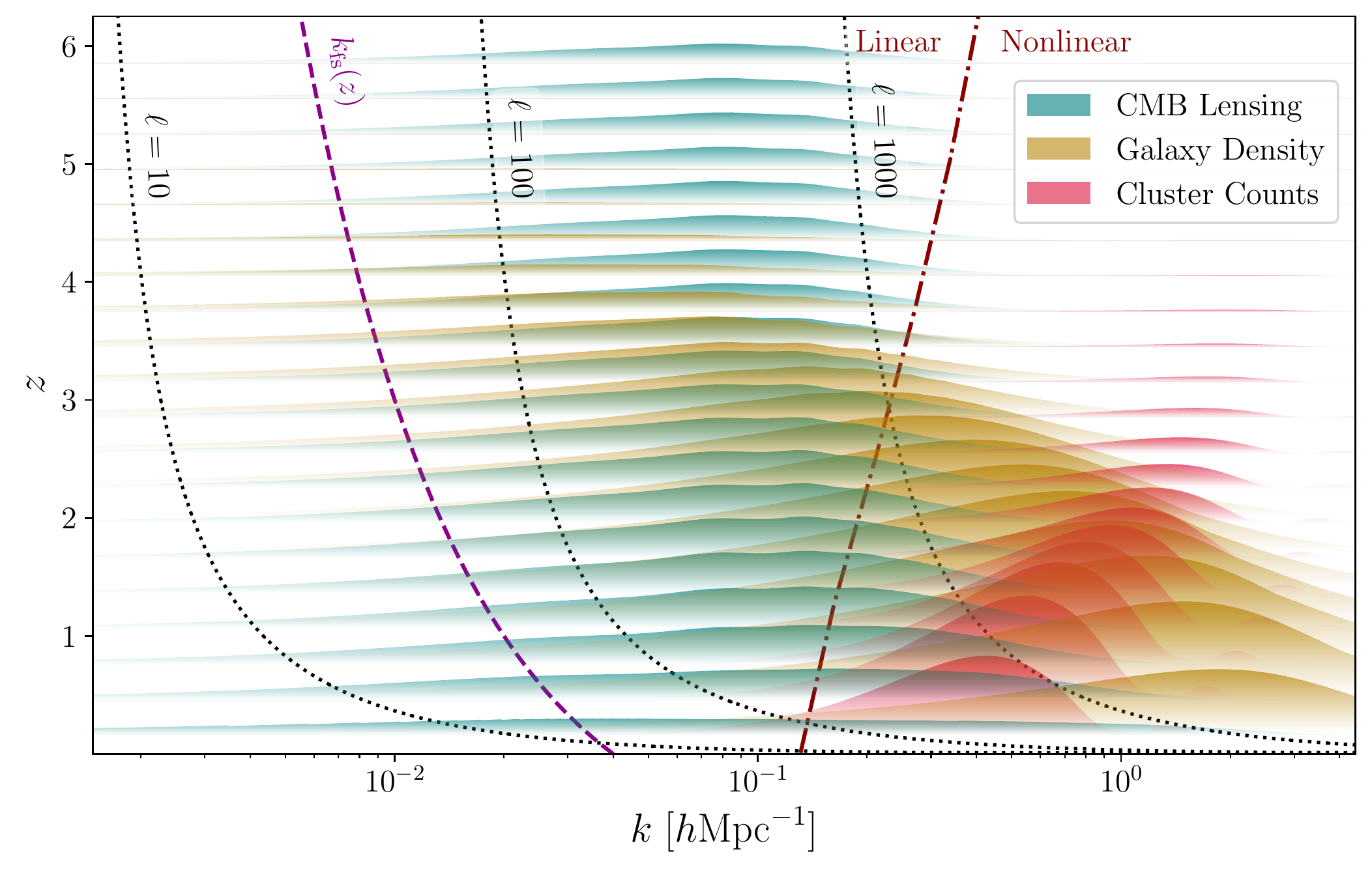}
\caption{Comparison of CMB lensing, with information for galaxy surveys reproduced from~\cite{Green:2021xzn}.  The curves represent the relative contribution from each $k$ and $z$ to the signal of each observable. Note the scale $k_{\rm fs}(z)$ is typically much smaller than the scales that contribute to these observables. Particularly due to the support at higher redshift, CMB lensing is less impacted by nonlinear physics than other LSS observables.}
\label{fig:lens_v_gal}
\end{figure}

The dark matter constitutes most of the matter in the Universe but does not interact directly with light.  However, gravitational lensing caused by all the intervening matter, including the dark matter, has a large and important effect on the CMB.  Following~\cite{Lewis:2006fu}, CMB lensing moves photons around due to the gravitational potential along the line-of-sight. Because lensing is only deflecting the photons, the lensed temperature we observe in a direction $\x$ is just the unlensed temperature that would have come from a direction $\n' = \n + \vec \alpha(\n)$, 
\beq\label{eq:T_lensed}
T_{\rm lensed}(\n) = T_{\rm unlensed}(\n + \vec \alpha(\n))
\eeq 
where $\vec \alpha = \nabla_n \phi$ is the deflection angle, $\phi$ is the lensing potential, 
\beq
\phi(\hat{\mathbf{n}}) \equiv-2 \int_{0}^{\tau_{\star}} \mathrm{d} \tau \frac{\tau_{\star}-\tau}{\tau_{\star} \tau } \Phi_+\left(\tau \hat{\mathbf{n}}, \tau_{0}-\tau\right) \ ,
\eeq
and $\Phi_+(\x, \tau)$ is the Weyl potential we defined earlier. The same kind of mapping also applies to the polarization, $Q+i U$, in each direction.  As we can see, the lensing potential is just an integral over the gravitational potential along the line-of-sight. This quantity is sometimes described in terms of the lensing convergence, $\kappa = - \nabla^2 \phi / 2$, so that $\Phi_+\to \delta_m$ and 
\beq
\kappa(\hat n) = \int dz W^{\kappa}(z) \delta_m(\tau(z) \n, z) \qquad W^{\kappa}(z)=\frac{3}{2} \Omega_{m} \frac{H_{0}^{2}}{H(z)}(1+z) \tau(z) \frac{\tau_{\star}-\tau(z)}{\tau_{\star}}
\eeq
The window function $W^{\kappa}(z)$ gives the range of redshifts that are probed by CMB lensing and the range is illustrated in Figure~\ref{fig:lens_v_gal} and compared to other low-redshift probes. The details of this expression are mostly unimportant for our conceptual understanding; we only need to recall that lensing is an integrated probe of the matter density fluctuations along the line-of-sight which gets support over a wide range of redshifts.

\begin{figure}[h!]
\centering
\includegraphics[width=6in]{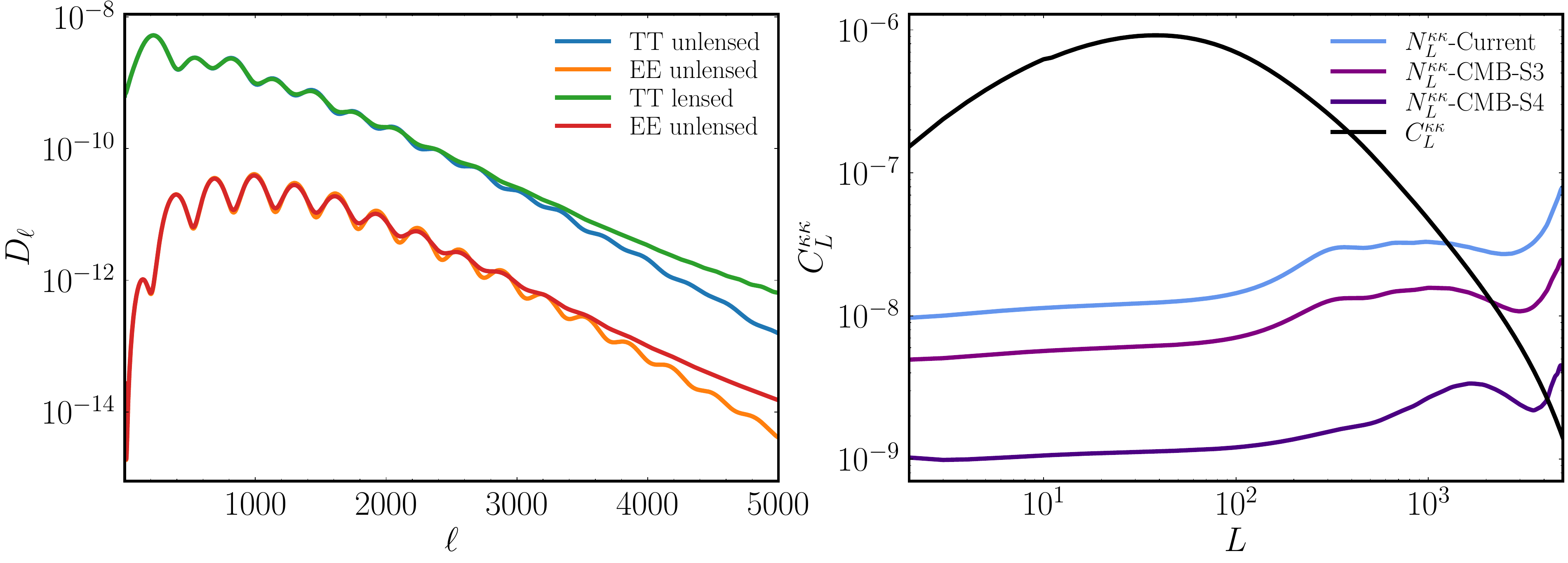}
\caption{{\it Left:} Lensed versus unlensed spectra.  We see the lensed spectra are smoothed out compared to the unlensed.  {\it Right:} The lensing convergence power spectrum, $C^{\kappa \kappa}_L$ compared to the lensing reconstruction noise $N^{\kappa \kappa}_L$ for various CMB experiments.  The noise curves were calculated assuming iterative delensing of TEB using the \texttt{CLASS\_delens}  code~\cite{Hotinli:2021umk}. }
\label{fig:lensing}
\end{figure}

There are many ways to think about what lensing does to the CMB.  The most intuitive starting point is Equation~(\ref{eq:T_lensed}), expanded in the deflection angle
\beq\label{eq:Tlensed}
T_{\rm lensed}(\n) = T_{\rm unlensed}(\n) +\nabla \phi(\n)  \cdot \vec \nabla T_{\rm unlensed}(\n)  + \frac{1}{2} \nabla_i \phi(\n) \nabla_j \phi(\n) \nabla^i \nabla^j T_{\rm unlensed}(\n) \ .
\eeq 
At high-$\ell$, $\phi(\n)$ is largely uncorrelated\footnote{At low-$\ell$ we have the integrated Sachs-Wolfe effect which is also sensitive to the low-redshift gravitational potential.} with $T_{\rm unlensed}(\n)$.  Therefore, if we calculate the leading effect on the temperature power spectrum, taking the flat sky limit $C^{TT}_\ell \to P^{TT}(\ell)$ (where $\vl$ is the Fourier transform of $\n$), we have
\beq
 P^{TT}_{\rm lensed}(\ell) \approx (1-\ell^2 C_\phi) P^{TT}_{\rm unlensed}(\ell) + \int \frac{d^2 L}{(2\pi)^2} P^{TT}_{\rm unlensed}(|\vl-\L|) ((\vl-\L)\cdot \L)^2 P^{\phi \phi}(L) \ ,
\eeq
where $P^{\phi \phi}(L)$ is the power spectrum of the Fourier transformed lensing map, $\phi(\vec L)$, and $C_\phi = \frac{1}{2} \int \frac{d^2 L}{(2\pi)^q} P^{\phi \phi}(L)$ arises from the second-order term in Equation~(\ref{eq:Tlensed}).  The latter must appear to ensure that the total power of the temperature fluctuations is unchanged (i.e.~we are only moving the fluctuations around). The consequence of this formula is that the acoustic peaks in TT are smoothed out because we convolved it with $P^{\phi \phi}(L)$, which takes sharp features in $P^{TT}(|\vl-\L|)$ and smears them over a range of $k$ weighted by $P^{\phi\phi}(L)$. The same phenomenon occurs for the polarization as well, meaning that lensing power spectra have smoother features and thus are less sensitive to imprints like the neutrino phase shift. Both effects are shown in Figure~\ref{fig:lensing}.

Fortunately, the Universe was more kind to us than to simply wash out some the CMB information~\cite{Seljak:1998aq,Hu:2001kj,Hirata:2003ka}.  Instead, the CMB actually contains the information we need to reconstruct the specific realization of $\phi(\n)$, which we can use to undo the effect of lensing (delensing) and measure $C^{\phi\phi}_L$ (here we used $L$ because it is reconstructed for temperature modes $T_\ell$ where $\ell \neq L$). To understand how this works, let us assume that there exists a single lensing Fourier mode $\phi(\L) \neq 0$.  In the presence of this mode, the Universe is no longer isotropic and homogenous and therefore we can have a non-zero off-diagonal correlator of $T(\k)$:
\beq\label{eq:off_diagonal}
\langle T_{\rm lensed}(\vl) T_{\rm lensed}(\vl') \rangle  =(\vl \cdot \L) P_{\rm unlensed}(\ell)\phi(\L) (2\pi)^2\delta(\vl+\vl'-\L) \ .
\eeq
In the absence of lensing, there are no correlations between $T(\vl)$ and $T(\vl')$ for $\vl' \neq -\vl$ and therefore by measuring this off-diagonal correlation, we determine the precise value of $\phi(\L)$ and not just some statistical average of $\phi$. Using this observation, Planck has made an all-sky map of the lensing potential and its associated power spectrum~\cite{Planck:2018lbu}.  This strategy for reconstructing the lensing mode is useful for searches for any new field that modulates the power or polarization fluctuations~\cite{Hanson:2009gu}.  For constraints on new physics, this encodes the same information as the CMB four-point function, but it gives us a more intuitive window through which to view the observational implications.

\subsection{Forecasting with CMB Lensing}

For the purpose of measuring $\sum m_\nu$, gravitational lensing of the CMB provides a measurement of $C^{\phi \phi}_L $ that we will use to determine the suppression of structure due to neutrinos.  Of course, to isolate the neutrinos we need to know all the other cosmic parameters that are important to the amplitude of the lensing power spectrum.  Fortunately, we get some of those from the primary CMB: we get a measurement of $A_s e^{-2 \bar \tau_{\rm optical}}$ from the $\ell\gg 30$ modes and a measurement of $A_s$ from $\ell \lesssim 30$ (again, think of this like frosted glass).  We usually think of the latter as a measure of $\bar \tau_{\rm optical}$ which also explains why the error bars on $\bar \tau_{\rm optical}$ are typically much larger than the other parameters of $\Lambda$CDM (after all, there aren't many modes at low-$\ell$).

We can forecast our sensitivity using the same Fisher matrix formula as before~\cite{Wu:2014hta},
\beq
F_{i j}=\sum_{\ell} \frac{2 \ell+1}{2} f_{\rm sky} \operatorname{Tr}\left(\boldsymbol{C}_{\ell}^{-1}(\theta) \frac{\partial \boldsymbol{C}_{\ell}}{\partial \theta_{i}} \boldsymbol{C}_{\ell}^{-1}(\theta) \frac{\partial \boldsymbol{C}_{\ell}}{\partial \theta_{j}}\right)
\eeq
but now where $\boldsymbol{C}_{\ell}$ is given by
\beq
\mathbf{C}_{\ell} \equiv\left(\begin{array}{ccc}
C_{\ell}^{T T}+N_{\ell}^{T T} & C_{\ell}^{T E} & C_{\ell}^{T \phi} \\
C_{\ell}^{T E} & C_{\ell}^{E E}+N_{\ell}^{E E} & 0 \\
C_{\ell}^{T \phi} & 0 & C_{\ell}^{\phi \phi}+N_{\ell}^{\phi \phi}
\end{array}\right) \ .
\eeq
The noise associated with the lensing reconstruction, $N_{\ell}^{\phi \phi}$, is somewhat more complicated to calculate and giving a complete description is beyond our scope in these lectures.  The reason is that the lensing noise is not set by fundamental limitations of the observations, but are in part determined by the method we use to measure $\phi$.  Concretely, the lensing reconstruction noise curve for CMB-S4 that is shown in Figure~\ref{fig:lensing} is the result of iteratively measuring and removing the lensing, so that it goes beyond our linearized understanding from the previous section.  In fact, in writing our formula for the Fisher matrix, we did not specify if $C^{TT,TE,EE}_\ell$ are the lensed or unlensed power-spectra.  To make the most of these maps, we should actually use the {\it delensed} TTTEEE spectra.  Although understanding these procedures and implementing them is rather involved, they can be calculated using the publicly available Delens code~\cite{Hotinli:2021umk}.

Neutrino mass specifically is a particular challenge to forecast because the impact of changing $\sum m_\nu$ on $C_L^{\phi \phi}$ is almost entirely described by a change to the amplitude.  The scale $k_{\rm fs}(z)$ is sufficiently small over the range of $z$ of the window function that there is no scale-dependent feature associated with $\sum m_\nu$.  Instead, neutrino mass is just one of three parameters that controls the amplitude of lensing,
\beq
C_L^{\phi \phi} \propto \omega_m^2 A_s (1- 8 f_\nu ) \ .
\eeq
For $\sum m_\nu = 58$ meV, which is the smallest possible value consistent with neutrino flavor oscillations, the suppression of $C_L^{\phi \phi}$ is approximately 1\% which is easily measurable given our noise curves. We see in Figure~\ref{fig:lensing} that we will measure the power spectrum to $L = {\cal O}(10^3)$ which is about $10^6$ modes with high signal to noise.  By our $1/\sqrt{N_{\rm modes}}$ counting, we would should be able to measure the overall amplitude at $10^{-3}$ precision.  

The limiting factor for determining $\sum m_\nu$ is not the precision of our measurement of $C_L^{\phi \phi}$ itself but that we also need to determine $\omega_m^2$ and $A_s$ to much better than 1\% accuracy in order to distinguish $\sum m_\nu$ from a change to these parameters in $\Lambda$CDM.  DESI or Euclid are expected to measure $\omega_m$ to the needed precision using the effect of $\omega_m$ on the expansion rate~\cite{Font-Ribera:2013rwa}, as measured by the BAO. The more fundamental limitation is our determination of $A_s$ which is degenerate with $\bar \tau_{\rm optical}$ for $\ell> 30$ in the CMB.  As a result, $\sum m_\nu$ rests on a few modes of the CMB at very large angular scales.  Planck has measured $\bar \tau_{\rm optical} = 0.054 \pm 0.007$~\cite{Planck:2018vyg}, which is sufficient for a 2-3$\sigma$ measurement of $\sum m_\nu$ at the minimal value.  In principle, all the low-$\ell$ modes contain enough information to reach $\sigma(\bar \tau_{\rm optical} ) = 0.002$ which would enable a 4-5$\sigma$ measurements.  There is also hope that $\bar \tau_{\rm optical} $ might be measurable through higher order statistics of the CMB that are not degenerate with $A_s$ and may enable a $>5\sigma$ detection of $\sum m_\nu$ with CMB-S4~\cite{Abazajian:2019eic}.

\begin{figure}[h!]
\centering
\includegraphics[width=4.25in]{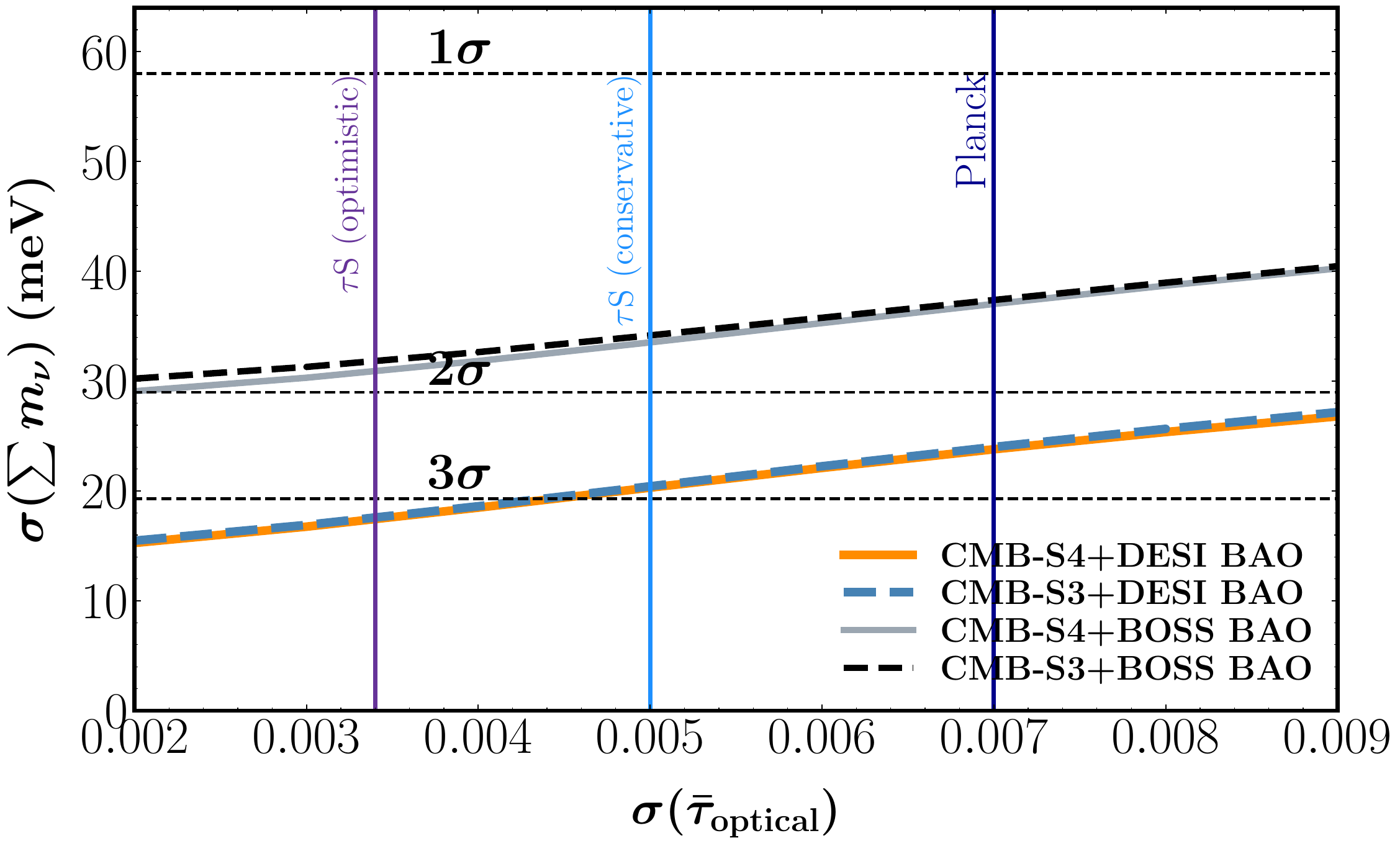}
\caption{Impact of CMB-noise levels, BAO sensitivity and $\tau$ prior on $\sum m_\nu$ forecasts. CMB-S4 forecasts follow~\cite{Abazajian:2019eic} while CMB-S3 are similar but assume a single 5$\mu$K-arcmin white noise level for the temperature maps.  BOSS and DESI BAO forecasts are based on~\cite{Font-Ribera:2013rwa}.  Finally, the Planck $\bar \tau_{\rm optical}$ prior is given the current measurement with $\sigma(\bar \tau_{\rm optical})=0.007$~\cite{Planck:2018vyg}.  Lines for optimistic and conservative balloon-borne measurements of $\bar \tau_{\rm optical}$ with $\tau$-Surveyor ($\tau$S) were derived in~\cite{Errard:2022fcm}.}
\label{fig:mnu_fore}
\end{figure}

Given these limitations, it is not surprising that forecasts are not very sensitivite to the noise level of the CMB experiment, or even the $L_{\rm max}$ that is accessible in the lensing map.  Forecasts for $\sigma(\sum m_\nu)$ are shown in Figure~\ref{fig:mnu_fore} for various combinations of CMB, BAO, and $\bar \tau_{\rm optical}$ data.  We see that the forecasts are essentially unchanged going from CMB-S3 to CMB-S4.  In contrast, the forecasts are very sensitive to the BAO measurement and the $\bar \tau_{\rm optical}$-prior, as anticipated.

More generally, CMB lensing is an important tool in the cosmologists arsenal that is potentially important for our understanding of BSM physics.  For example, constraints on dark-matter baryon interactions from CMB-S4 will be driven entirely by the measurement of lensing~\cite{Li:2018zdm}.  A natural question, however, is why should CMB lensing provide more useful information than a galaxy survey?  After all, both are measurements of the low-redshift clustering of matter and the number of modes available with lensing is smaller than both future galaxy surveys and the primary CMB.  The unique value of CMB lensing is that it is a measure of the matter distribution, $\delta_m$, in the linear regime.  The distribution of galaxies includes measurements in the linear regime but, as we will see in the next Section, the distribution of galaxies is related to the matter density by some unknown nonlinear function.  Galaxy lensing can also be used to determine $\delta_m$, but on smaller scales and lower redshifts where nonlinear effects are important.  These differences are illustrated in Figure~\ref{fig:lens_v_gal}.  Even though the statistical power of the galaxy surveys is, in principle, vastly superior to CMB lensing, the two tend to be very complimentary tools in our understanding of the Universe.

\newpage

\section{Primordial non-Gaussianity in Large Scale Structure}

We saw in Section~\ref{sec:coins} that the initial density fluctuations that gave rise to structure in the Universe could not have been produced locally during the hot Big Bang. The solution to this problem is inflation, which extends the geometry into the past (effectively inventing negative time).  This change to the geometry enables us to create the fluctuations at a much earlier time and allows them to evolve to be larger than the horizon scale.  In this way, they evolve outside the horizon for long enough to produce phase coherence and can enter the horizon around the time of recombination as needed.

\subsection{Inflation}

The inflationary epoch is defined by a period of accelerated expansion where the initial density fluctuations were created~\cite{Baumann:2009ds}.  This can be formalized by the EFT of inflation~\cite{Cheung:2007st}, which defines inflation in terms of two basic conditions (see Appendix~\ref{app:inflation} for more details):
\begin{enumerate}
\item Nearly exponential expansion, defined by a Hubble parameter $H(t) \equiv \dot a(t) / a(t)$ which satisfies
\beq
|\dot H(t) | \ll H^2(t) \ .
\eeq
\item A physical clock\footnote{Interestingly, the solid inflation~\cite{Endlich:2012pz} is not inflation by this definition, which is also why its observational predictions often differ from conventional models of inflation, e.g.~\cite{Endlich:2013jia}.} which defines the ends of inflation.  In slow-roll inflation, the clock is a scalar field $\phi$ with $\phi \approx \dot \phi t$ where $\dot \phi$ is nearly constant. More generally, we can think of this clock as a pattern of symmetry breaking where some operator ${\cal O}$ gets a time-dependent expectation value, $\langle {\cal O}(\x,t) \rangle \propto t$, that defines a preferred time coordinate. 
\end{enumerate}
The first condition gives us the required geometry to convert small-scale to large-scale fluctuations.  Since $H$ is nearly constant during inflation, modes evolve from being inside to outside the horizon,
\beq
\frac{k}{a(t_{\rm H} )} = H \approx {\rm constant} \to \frac{k}{a(t\ll t_{\rm H})} \gg H \ ,
\eeq
where $t_{\rm H}$ is the time of horizon crossing. 

The second condition not only ensures that inflation ends, but introduces the scalar metric fluctuation that we need to explain structure in the Universe.  The EFT of inflation formalizes the idea that the clock is a pattern of symmetry breaking such that the scalar density fluctuations are controlled by the Goldstone boson, $\pi$, associated with the spontaneous breaking of time translations (translations are then weakly gauged by gravity so that the Goldstone mode is eaten by the metric). Because no clock is perfect in quantum mechanics, the fluctuations of the clock (or equivalently $\pi$) must be equivalent to a change in how long inflation lasts from place to place.  As a result, we can choose a gauge where the scalar fluctuations are encoded in the metric as a field $\zeta(\x,t)$ that describes the local change to $a(t)$, namely
\beq
ds^2 =-dt^2 + a^2(t) e^{2 \zeta(\x,t)} d\vec x^2 \ .
\eeq
In single-field inflation, $\zeta(\x,t)$ is equivalent to fluctuations of the clock in terms of the Goldstone boson $\pi(\x,t)$ and it can be shown that $\zeta = - H \pi$ to leading order~\cite{Maldacena:2002vr}.  In multi-field inflation~\cite{Senatore:2010wk}, the end of inflation can be controlled by additional fields, $\sigma_i$, that are unrelated to the pattern of symmetry breaking, so that $\zeta = F[\pi; \sigma_i]$.  

The predictions of inflation are encoded in the statistical correlation functions of $\zeta(\k)$ that are inferred from the correlations of the density fluctuations at late time.  These statistical fluctuations typically arise from quantum fluctuations and can be calculated in quantum field theory using the in-in formalism~\cite{Weinberg:2005vy}.  The essential physics can be understood from a massless scalar in de Sitter (dS)~\cite{Baumann:2009ds}, with an action
\beq
S = \int d\tau d^3 x \, a^4(\tau) \left[ \frac{1}{a^2}\varphi'^2 -\frac{1}{a^2} \partial_i \varphi \partial^i \varphi \right] \ .
\eeq
In de Sitter, we can solve for the conformal time by integrating $a(t) = e^{Ht}$ and $d\tau = dt/a(t)$ to find that $a(\tau) = - 1/(H \tau)$ and $\tau \in (-\infty,0]$. The classical equations of motion for the Fourier modes, $\varphi(\k,\tau)$, are 
\beq\label{eq:classical_sol}
\varphi'' - \frac{2}{\tau} \varphi' + k^2 \varphi = 0 \ ,
\eeq
which can be solved to find
\beq
\varphi(\k, \tau) = a^{*}_\k \frac{H}{\sqrt{2} k^{3/2}} (1-i k\tau) e^{i k \tau} + a_{-\k}  \frac{H}{\sqrt{2} k^{3/2}} (1+i k\tau) e^{-i k \tau} \ ,
\eeq
where $a_\k$ are (complex-valued) constants. To quantize the field $\varphi \to \hat \varphi$, we promote the constant to operators, $a^*_\k \to \hat a^\dagger_\k$ and $a_\k \to \hat a_\k$, with the usual commutation relations
\beq
\left[\hat a_{\k}^{\dagger}, \hat a_{\k'}\right]=  (2\pi)^3 \delta\left(\k-\k'\right) \ .
\eeq
Here, I have cheated in fixing the overall normalization from the beginning.  The easiest way to have guessed the answer is to require that inside the horizon, $k \gg aH $ or $-k \tau \gg 1$, we recover the flat-space answer as a WKB approximation.\footnote{This choice, that short distances in dS are the same as flat space vacuum, is known as choosing the Bunch-Davies vacuum for the ground state. See e.g.~\cite{Green:2022ovz} for an explanation of why this vacuum is preferred.} Specifically, in flat space we define the vacuum so that we create the positive frequency modes; assuming the same will be true of the WKB solutions in the flat-space limit implies
\beq
\hat \varphi(\k_{\rm physical} \gg H) \to \hat a^\dagger \frac{1}{\sqrt{2 \omega}} e^{i \int^t dt' \omega(t')} + {\rm h.c.} \ ,
\eeq 
where $\omega(t) = k_{\rm physical}(t)= k/a(t)$ and $\int^t dt' k/a(t') =- k/(aH) = k\tau$.  Expanding our solution in (\ref{eq:classical_sol}) in the limit $\tau \to -\infty$, we find
\begin{align}
\hat \varphi(\k, \tau\to 0) &\to \hat a^{\dagger}_\k \, \frac{- i H \tau}{\sqrt{2} k^{1/2}} e^{i k \tau} +{\rm h.c.} \\
&=  -i \hat a_\k^\dagger \,  \left(-i a^{-3/2} \right) \frac{e^{-i \frac{\omega}{H}}}{\sqrt{2 \omega }}   +{\rm h.c.} \ .
\end{align}
This matches our expectations from the WKB approximation up to a phase ($-i$) and an extra factor of $a^{-3/2}$. The latter is there because we Fourier transformed with respect to comoving momenta $k = k_{\rm physical} a$ rather than $k_{\rm physical}$.  

Having determined the correct definition of the operator $\hat \varphi$, we now take the superhorizon limit, $k \tau \to 0$.  Expanding the mode functions, we find
\beq
\hat \varphi(k,\tau) \to \frac{a^{\dagger}_\k}{\sqrt{2} k^{3/2}} \left[ H(1+\tfrac{1}{2}k^2\tau^2 k^2) + \frac{i}{3} H \tau^3 k^3 \right ] + {\rm h.c.} \ .
\eeq
The leading term is a constant and gives us that famous result of a scale-invariant two-point function:
\beq\label{eq:two_point}
\lim_{\tau \to 0 }\langle \hat \varphi(\k,\tau) \hat \varphi(\k',\tau)\rangle = \frac{H^2}{2 k^3} (2\pi)^3 \delta(\k+\k')
\eeq
However, the key thing to notice is that the the imaginary part vanishes like $\tau^3 \to a^{-3}$.  This is not an accident, it is equivalent to the statement that the the equal-time commutator must be 
\beq
[ \varphi(\x,t),\dot \varphi(x',t)] = \frac{i}{\sqrt{g}} \delta(\x-\x') \to [  \varphi(\k,t), \dot \varphi(\k',t)] \to \frac{1}{a^3} (2\pi)^3 \delta(\k+\k') \ ,
\eeq
where we have dropped the $\hat \varphi \to \varphi$ for convenience.  The factor of $\sqrt{g}$ is required by diffeomorphism invariance and gives rise to the factor of $a^{-3}$ around our background.  This has two very important consequences: 
\begin{itemize}
\item The two-point function we calculated is a {\it classical} statistical correlator for all practical purposes. The commutator vanishes like $a^{-3}$ and is negligible, meaning that we have converted quantum fluctuations to classical fluctuations through the expansion of the Universe (see e.g.~\cite{Grishchuk:1990bj,Green:2020whw,Green:2022ovz} for discussion).  
\item There can only be one constant real solution outside the horizon.  This is an inevitable consequence of causality~\cite{Weinberg:2003sw} that explains the phase coherence of the sound waves in the CMB that we discussed in Section~\ref{sec:coins}.
\end{itemize}
Although we demonstrate these results for massless fields in de Sitter, these conclusions are far more general.

The above calculation, Equation~(\ref{eq:two_point}), tells us all the Gaussian correlators of the free theory.  Yet, if we want to understand is the dynamics of inflation, we will need to know the non-Gaussian correlations, e.g.~the three-point functions.  These can be calculated perturbatively using the in-in formalism~\cite{Weinberg:2005vy}, which is defined by
\beq\label{eq:inin}
\langle {\rm in}|  Q(t) |{\rm in} \rangle=\left\langle\bar{T} \exp \left[i \int_{-\infty(1+i \epsilon)}^{t}d t'  H_{\mathrm{int}}(t') \right]  \, Q_{\mathrm{int}}(t)  \, T \exp \left[-i \int_{-\infty(1-i \epsilon)}^{t} d t' H_{\mathrm{int}}(t')  \right]\right\rangle \ ,
\eeq
where $ |{\rm in} \rangle$ is the interacting vacuum at time $t$, $Q(t)$ is some operator composed of one or several fields $\varphi(\k,t)$, each with different $\k$, and $T$ ($\bar T$) denotes (anti-)time ordering.  The interactions are described by $H_{\mathrm{int}}(t)  =- \int d^3 x \sqrt{g} L_{\rm int}(\varphi_{\rm int}(\x,t))$ and $\varphi_{\rm int}$ are the interaction picture fields. We will not calculate anything in detail, but in principle everything follows from familiar techniques.

What we observe is not the fluctuations of a fundamental scalar field $\phi$, but the scalar metric fluctuation $\zeta$.  The key thing to remember is that $\zeta$ is dimensionless, so any relation to scalar field must come with a second scale.  In slow-roll inflation, this scale is set by speed of the rolling background scalar, $\dot \phi \approx$ constant, so that the relation between the fluctuations of the inflaton, $\phi$ and $\zeta$ is given by 
\beq
\langle \zeta(\k) \zeta(\k') \rangle  = \frac{H^2}{\dot \phi^2} \langle \delta \phi(\k) \delta \phi(\k') \rangle  \ ,
\eeq
where $\delta \phi$ are the fluctuations around around the background, $\phi(\x,t) \approx \dot \phi t + \delta \phi(\x,t)$.  We will often define the amplitude of the power spectrum as $\Delta_\zeta$ such that 
\beq
\langle \zeta(\k) \zeta(\k') \rangle' = P_\zeta(k) = \frac{\Delta_\zeta^2}{k^3} k^{n_s-1} \to \Delta_\zeta^2 = A_s 2 \pi^2 \ ,
\eeq 
where $A_s = (2.1\pm 0.03) \times10^{-9}$ (68\% confidence interval) is the $\Lambda$CDM parameter measured by Planck and $\langle ... \rangle =\langle ... \rangle' (2\pi)^3 \delta(\k +\k')$ as before.  The scalar spectral index, $n_s$, is also measured by Planck, $n_s =0.965\pm0.004$ (68\%), which at our level of approximation means $n_s-1 \approx 0$. In the late Universe, the Newtonian potential is given by $\Phi = -3 \zeta /5$ so that $\Delta_\Phi = 3 \Delta_\zeta / 5$.

The three point function in Fourier space, also known as the bispectrum, is restricted by symmetries to take the form
\beq
\langle \zeta(\k_1) \zeta(\k_2) \zeta(\k_3) \rangle = \fnl B(k_1, k_2,k_3) (2\pi)^3 \delta(\k_1 +\k_2+\k_3) \ ,
\eeq
where $\fnl$ is the amplitude normalized so that $k_\star^6 B(k_\star, k_\star,k_\star)=\frac{18}{5} \Delta_\zeta^4$ for some reference scale $k_\star$ (usually called the pivot scale). This is a slightly funny normalization because an order-one non-Gaussian correlator would have $B={\cal O}( \Delta_\zeta^3)$ or $\fnl \Delta_\zeta ={\cal O}(1)$, which in our Universe is $\fnl = {\cal O}(10^4)$. Said differentially, $\fnl = {\cal O}(1)$ would be a $10^{-4}$ departure from Gaussianity and would require roughly $10^8$ modes to measure. It is also important that $B(k_1,k_2,k_3)$ is a function only of the length $k_i =|\vec k_i|$.  Since the three vectors $\k_i$ sum to zero, they form a closed triangle where the bispectrum depends only on the length of each side.

The bispectra are often characterized as ``shapes"~\cite{Babich:2004gb} that reflect how each triangle contributes to the signal-to-noise,
\beq
\left( \frac{S}{N} \right)^2 = V \int \frac{d^3 k_1 d^3 k_2 d^3 k_3}{(2\pi)^9} \frac{\fnl^2 B(k_1,k_2, k_3)^2}{6 P(k_1) P(k_2) P(k_3) } (2\pi)^3 \delta(\k_1+\k_2 +\k_3) \ ,
\eeq
where $V \approx k_{\rm min}^{-3}$ is the survey volume, and we have assumed that we can measure the bispectrum directly\footnote{When the Universe is linear, this isn't a bad approximation: $\delta_m(\k) = -\frac{3}{5} \zeta_\k  \T(k)$ for some transfer function $\T(k)$ that is essentially fixed by the measurement of the power spectrum. This is a precise example of the benefit of factorization, as we noted in Equation~(\ref{eq:factorization}), that enables us to separate inflationary and recombination-era physics. } in terms of $\zeta$ (which is the optimal estimate of $\fnl$). As the spectra are usually scale invariant, it is often useful to factor out the overall momentum scale so that (see Appendix~\ref{app:SN}) 
\beq
\Big(\frac{S}{N} \Big)^2 = V \fnl^2  \int  \frac{d^3 k_1}{(2\pi)^3}  \frac{1}{(4\pi^2)}  \int_{1/2}^1 d x_2 \int_{1-x_2}^{x_2} dx_3 x_2^4 x_3^4 B(1,x_2,x_3)^2/ \Delta_\zeta^6 \ ,
\eeq
where $x_i = k_i / k_1$, and we have used the permutation invariance of $B(k_1, k_2, k_3)$ to enforce $k_1 \geq k_2 \geq k_3$ without loss of generality.  The shape of non-Gaussianity is characterized by the function
\beq
S(x_2,x_3) \equiv (x_2 x_3)^2  B(1,x_2,x_3) \ .
\eeq
This is preferable to using $B(1,x_2,x_3)$ directly, as the shape $S(x_1,x_2)$ accounts for the suppression of the signal to noise in limits like $x_3 \to 0$, which arise because of the small number of modes with $k_3 \to k_{\rm min}$.  Finally, notice that the overall scaling of the signal to noise, after integrating over $k_1$ will be $(S/N)^2 \propto V k_{\rm max}^3 \approx N_{\rm modes}$, so our error bars will scale like $1/\sqrt{N_{\rm modes}}$ as expected.

\begin{figure}[h!]
\centering
\includegraphics[width=3.in]{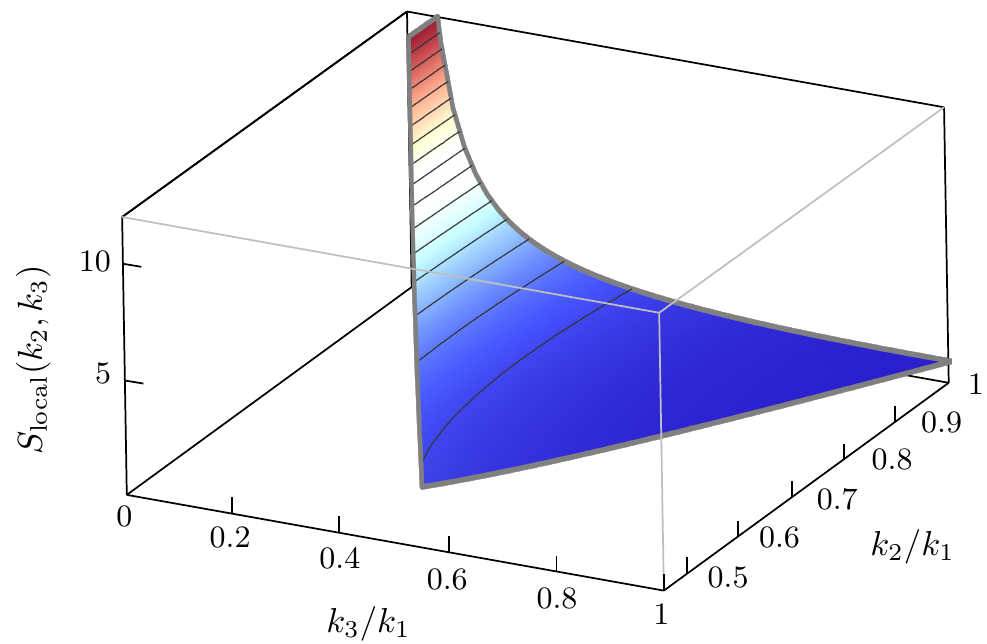}
\includegraphics[width=3.in]{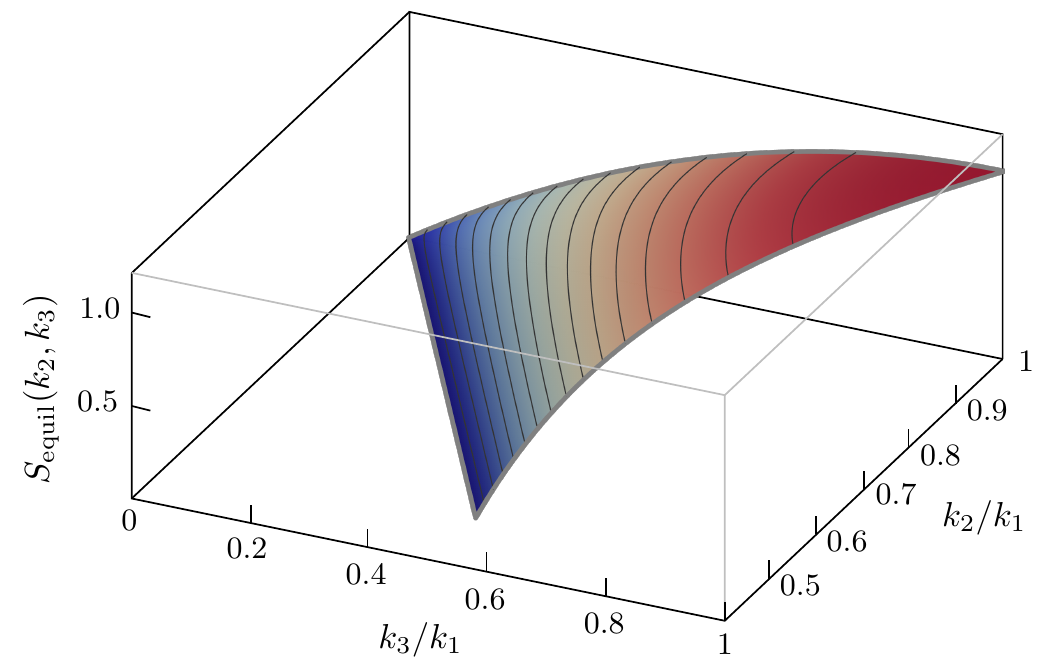}
\caption{The shape functions, $S(x_2,x_3)$ for local ({\it left}) and equilateral ({\it right}) non-Gaussianity.  Reproduced from \cite{Baumann:2022mni}. }
\label{fig:shapes}
\end{figure}

There are a wide variety of possible bispectra and higher-point correlation functions that encode the physics of inflation in different ways. It would be an entire set of lectures to explain how this works in detail~\cite{Chen:2010xka,Baumann:2018muz}.  Instead, I will focus on two particular shapes, local and equilateral (shown in Figure~\ref{fig:shapes}), and how they are measured.  Even without being able to do all the calculations, these two examples are a good window into the relationship between the physics of inflation and the strategies to search for non-Gaussianity.

\subsection{Local Non-Gaussianity}\label{sec:local}

\subsubsection*{Theoretical Background}

Local non-Gaussianity was the first model used in observational searches for non-Gaussianity~\cite{Komatsu:2001rj}.  It is often expressed in terms of the Newtonian potential $\Phi$ as 
\beq\label{eq:local_ans}
\Phi(\x)= \phi_{g}(\x) - \fnlloc (\phi_g(\x)^2 - \langle\phi_g(\x)^2\rangle ) \ ,
\eeq
where $\phi_g(\x)$ is a Gaussian random field.  Using $\Phi = -3 \zeta /5$ and $\varphi_g = - 5\phi_g/3$, we can also write this as
\beq
\zeta = \varphi_{g}(\x) + \frac{3}{5}\fnlloc \varphi_g(\x)^2  \ .
\eeq
From Wick contractions of the Gaussian field, we can see that $P_\zeta(k) \approx P_\varphi(k)$ and
\beq
B(k_1,k_2,k_3) \approx \frac{6}{5} \fnlloc\left( P_\zeta(k_1) P_\zeta(k_2) + P_\zeta(k_1) P_\zeta(k_3) + P_\zeta(k_2) P_\zeta(k_3)\right) \ .
\eeq
The most common way to generate local non-Gaussianity is through an additional light scalar field (i.e.~a field with $m\ll H$ that is not the inflaton) that changes the duration of inflation or time to reheating.  Popular models of this kind include the curvaton~\cite{Lyth:2001nq} or modulated reheating~\cite{Dvali:2003em,Zaldarriaga:2003my}.  We can understand the general idea as follows: suppose we have a light scalar $\sigma(\x)$ that controls some physical parameters of the theory, decay rates, masses, etc.  Suppose these parameters control how long inflation lasts, or the amount of time it takes to reheat the Universe.  This means that the number of e-folds of expansion between the start of inflation and the end of reheating (for example), depends on the local value of $\sigma(\x)$.  Now notice that if we define the amount of expansion up to a time $t$ with $\sigma(\x)= 0$ by $a(t)$, then expansion at time $i$ with $\sigma(\x) \neq 0$ can be written as $a(t) e^{\zeta(\x)}$~\cite{Starobinsky:1985ibc,Salopek:1990jq,Lyth:2004gb,Sugiyama:2012tj}. Hence,
\beq
\zeta(\x) = N_e(\sigma(\x)) - N_e(0) = \frac{dN_e}{d\sigma} \sigma(\x) +\frac{1}{2} \frac{d^2N_e}{d^2\sigma} \sigma^2(\x) + \ldots \ ,
\eeq
where $N_e = \log [a(t_{\rm end}) / a(t_{\rm begin})]$ is the local number of e-folds of inflation.  Therefore, any nonlinear relationship between $\sigma(\x)$ and $N_e$ will give rise to local non-Gaussianity where 
\beq
\varphi_g(\x) \equiv  \frac{dN_e}{d\sigma} \sigma(\x) \ .
\eeq
In more complicated multifield models, $\varphi_g(\x)$ may corresponds to a model-dependent linear combination of the additional fields.

One of the most important characteristics of local non-Gaussianity is that it does not\footnote{This statement may not hold away from the Bunch-Davies vacuum~\cite{Flauger:2013hra} or around a non-attractor solution~\cite{Namjoo:2012aa,Chen:2013aj}. } arise in single-field inflation~\cite{Maldacena:2002vr,Creminelli:2004yq}. In single-field inflation, $\zeta = {\rm constant}$ is equivalent to a homogenous FRW background up to a change of coordinates: if we shift $\zeta$ by a constant, $\zeta \to \zeta +c$, and rescale the coordinates, $\x \to \x e^{-c}$, the metric is unchanged. This implies that the bispectrum must obey a Ward identity~\cite{Assassi:2012zq,Hinterbichler:2013dpa}, also known as a single-field consistency condition~\cite{Maldacena:2002vr,Creminelli:2004yq},
\beq\label{eq:sfcc}
\lim_{k_1 \to 0} B(k_1,k_2,k_3) = \left(-\left(\frac{d}{d \log k_1}+3\right) +{\cal O}(k^2_1) \right) P_\zeta(k_1) P_\zeta(k_3)  \ .
\eeq
This can be stated as $\fnlloc = -5 (n_s-1)/3$ in single-field inflation, although the full predictions of single-field inflation cannot be reproduced by a local ansatz~\cite{dePutter:2016moa} like~(\ref{eq:local_ans}).  In fact, even for the bispectrum alone this statement is somewhat misleading because for this special case where $\fnlloc = -5 (n_s-1)/3$, the bispectrum actually introduces no physical effects on small scales~\cite{Pajer:2013ana}, as it is equivalent to a change of coordinates (up to gradients).  For these reasons,   it is very useful to understand the limit of one vanishing momentum (here $k_1\to 0$) in the bispectrum (or any correlator), which is also given the name {\it the squeezed limit}.

At face value, the above description is quite confusing.  The symmetry we defined above is always a symmetry of the action, regardless of the field content of the theory; it is just a large diffeomorphism of the FRW background~\cite{Weinberg:2003sw}.  However, we made an implicit assumption when relating the symmetry to the observable correlator, namely that an observable fluctuation $\zeta(\vec k \to 0)$ is the same as $\zeta = {\rm constant}$.  This is not a trivial assumption in general but especially in a model with multiple fields where $\zeta(\k \to 0) = F[{\pi ; \sigma_i}, t]$.  For example, a second field can cause $\zeta$ to change over time, $\dot \zeta(\k \ll aH) \propto \sigma(\k \ll aH,t)$ in which case the long wavelength field is not equivalent to a constant rescaling of the coordinates because $\zeta$ depends on time~\cite{Assassi:2012et}.  Remarkably, this occurs even if $\sigma$ is massive such that it has redshifted away by the end of inflation~\cite{Chen:2009zp,Baumann:2011nk}.  For example, a massive scalar with $m^2> 0$ that interacts with the inflaton can introduce a non-trivial correction in the squeezed limit
\beq\label{eq:massive_B}
\lim_{k_1 \to 0} B(k_1,k_2,k_3) = \left(\frac{d}{d \log k_1} + c \left(\frac{k_1}{k_3}\right)^{\frac{3}{2} - \nu} \right) P_\zeta(k_1) P_\zeta(k_3)\ ,  \qquad \nu = \sqrt{\frac{9}{4} - \frac{m^2}{H^2}} \ .
\eeq
Notice that ${\rm Re}\left( \frac{3}{2} - \nu\right) < 2$ for all $m^2$. This means that taking $m \gg H$ does not simply reduce the results to single-field inflation. Instead, in this regime the amplitude, $c$, becomes exponentially suppressed~\cite{Noumi:2012vr,Arkani-Hamed:2015bza}, so that the dominant signal is controlled by single-field inflation.

More generally, non-Gaussianity offers an opportunity to measure the spectrum of particles present during inflation in terms of their masses (in $H$ units) and their spins~\cite{Lee:2016vti}.  The fact that these additional fields break the single-field consistency conditions is particularly useful as it gives us a tool to observationally separate classes of models.  Yet, all inflationary models tend to produce slightly different correlation functions at some level, such that our ability to distinguish them will be hard to appreciate without understanding the limitations of realistic cosmic surveys. Any characteristic differences between models will only be valuable observationally if they can be exhumed from actual maps of the Universe.

\subsubsection*{Observational Signatures}

The best current constraints on $\fnlloc$ are from the CMB (Planck), namely $\fnlloc=-0.9\pm5.1$ at $1\sigma$~\cite{Planck:2019kim}.  There is not  much to explain in their process: they measure the three-point functions of their maps and infer $\fnlloc$ the same way you would measure the variance of the primordial fluctuations ($A_s$) from the two-point statistics ($C_\ell$).  However, our interest is how these constraints will improve with the next generation of surveys.  Although maps of the CMB will continue to improve, we saw the number of modes grows slowly with the noise level.  In detail, the CMB constraints on primordial non-Gaussianity improve more slowly than just $1/\sqrt{N_{\rm modes}}$ which fundamentally limits to improvements possible in the CMB~\cite{Kalaja:2020mkq}. Instead, it is galaxy surveys that promise to improve measurements on $\fnlloc$ by an order of magnitude.

Local non-Gaussianity impacts the formation of galaxies is a surprisingly non-trivial way~\cite{Dalal:2007cu}, which is a major reason for the optimistic forecasts.  The easiest way to understand this is to remember that galaxy formation is an essentially local process (by cosmologists' standards) that happens on some short distance scale. In contrast, we want to measure correlations in the number of galaxies that are separated by cosmological distances.  For the Gaussian fields, the small and long wavelength fluctuations are uncorrelated, so we can write $\phi_g = \phi_S + \phi_L$ to distinguish modes that live on galactic ($k_S$) and cosmological scales ($k_L$) respectively.  However, non-Gaussianity mixes these modes so that the long wavelength field will influence the amplitude of the small scale fluctuations, 
\beq
\langle \Phi(\k_S)\Phi(\k'_S) \rangle  = \langle \phi(\k_S) \phi(\k'_S) \rangle (1+ 2 \fnlloc \phi_L(\x)) \ .
\eeq
We see that the amplitude of small scale fluctuations of $\Phi(\k_S) $ are modulated by the value of the long wavelength Gaussian field, $\phi_L(\x)$. This is still true after we apply the linear transfer function to $\Phi(\k_S)$ to  determine the local matter density, 
\beq
\delta_m(\vec{k}, z)=\frac{2 k^{2} T(k) D(z)}{3 \Omega_{m} H_{0}^{2}} \Phi(\vec{k}) \equiv \T(k, z) \Phi(\vec{k}) \ ,
\eeq
where $T(k)$ is the transfer function and $D(z) = D(a(z))$ is the growth function. Again, the small-scale matter fluctuations are modulated according to
\beq\label{eq:local_mod}
\langle \delta_m(\k_S)\delta_m(\k'_S) \rangle  = \langle \delta_{m,{\rm G}}(\k_S) \delta_{m,{\rm G}}(\k'_S) \rangle (1+ 2 \fnlloc \Phi_L(\x)) \ ,
\eeq
where $\delta_{m,{\rm G}}(\k) = \T(k) \phi_g(\k)$ is the Gaussian matter fluctuation.  This implies that the amplitude of fluctuations of the matter density (a physical quantity) on small scales is modulated by the local value of the long wavelength Newtonian potential, $\Phi_L(\x)$.  This is usually forbidden by the equivalence principle, but this has occurred here because of the presence of a light scalar during inflation.

We can understand the implications for cosmology while being largely agnostic about how galaxy formation works in detail.  Instead, we will only need to assume that galaxy formation is local in space and that it is a non-linear process.  Since the metric is not a local observable, this tells us the density contrast of galaxies should be a nonlinear function of the matter density contrast,
\beq
\delta_{\rm gal}(\x) = b_1 \delta_m(\x) +b_2 \delta_m^2(\x) + \ldots \ ,
\eeq
where $b_i$ are constants in space but may depend on time ($z$). This is known as the {\it bias expansion} or {\it biasing}~\cite{Desjacques:2016bnm} because it tells use that $\delta_{\rm gal}$ is a biased (in the statistical sense) measurement of $\delta_m$.  So far, we haven't accounted for the Newtonian potential that appears in~(\ref{eq:local_mod}).  Since it alters the amplitude of small scale fluctuations, $\langle \delta^2(\x) \rangle \propto  (1+ 2 \fnlloc \Phi_L(\x))$, we should really add the Newtonian potential since it can affect the number of galaxies
\beq
\delta_{\rm gal}(\x) = b_1 \delta(\x) +b_2 \delta^2(\x) + 2 b_\Phi \fnlloc \Phi_L(\x) +\ldots  \ .
\eeq
Importantly $b_\Phi \neq 0$ has been seen in many simulations and can be derived from popular models of halo formation~\cite{Dalal:2007cu}.  Taking the Fourier transform and keeping only the linear terms, we find
\beq
\delta_{\rm gal}(\k) = \left(b_1 + b_\Phi \fnlloc \frac{3\Omega_m H_0^2}{k^2 T(k) D(z)} \right) \delta_m(\k) \ .
\eeq
The transfer function is defined such that $T(k\to 0) \to 1$, so that at small $k$ (large distances), $\delta_{\rm galaxy}(\k) \propto \delta(\k) / k^2$.  This tells of that this signal\footnote{We still have to determine if the noise will allow us to observe this part of the signal.} of local non-Gaussianity peaks at the largest possible scales ($k \to 0$), far away from the nonlinear regime.  

This phenomena, sometimes called {\it scale-dependent bias}, is a property of the squeezed limit of the bispectrum, $\lim_{k_1\to 0} B(k_1, k_2,k_3)$.  If we interpret his limit as telling us how long and short modes are correlated then we can determine the effect of scale-dependent bias for any bispectrum. For example, with massive fields the bispectrum is given in~(\ref{eq:massive_B}) and scale-dependent bias arises in the form
\beq
\delta_{\rm gal}(\k) = \left(b_1 + b_\Phi \fnl^{\nu} \frac{3\Omega_m H_0^2 k^{3/2 -\nu}}{k^2 T(k) D(z)} \right) \delta_m(\k) \ .
\eeq
Notice that I have not included the $(n_s-1)$ contribution here because it is equivalent to a change of coordinates and has no effect on the number of galaxies~\cite{dePutter:2015vga}.  In contrast, the scale dependent effect from these massive fields changes the scaling behavior of the galaxy power spectrum in a way that allows us to extract the mass of the particle~\cite{Gleyzes:2016tdh}.

From our discussion of signal-to-noise in the CMB, one should worry that a signal that lives at small $k$ is actually limited by cosmic variance.  After all, we have always argued that our errors are fundamentally limited by the number of modes available, and there are very few modes with small $k$. However, the nature of this signal is somewhat counter-intuitive and the forecasts are better for two important reasons: 
\begin{itemize}
\item Cosmic variance is really set by the fluctuations of the Gaussian theory $P_{\rm gal} = b_1^2 P_m(k)$ and therefore the signal, $P_{\rm gal}(k) \propto P_m(k) k^{3-2\nu}/k^4$ will be enhanced for small enough $\nu$.
\item It is possible to measure the individual Fourier modes of $\delta_m(\k)$ directly, using CMB lensing (as we discussed around Equation~(\ref{eq:off_diagonal})) or the gravitational lensing of high-redshift galaxies.  If we know both $\delta_m(\k)$ and $\delta_{\rm gal}(\k)$ for the same $\k$, then the relationship is completely deterministic and has no cosmic variance~\cite{Seljak:2008xr}. This is possible because the scale-dependent bias was really the result of nonlinear physics at the scale $k_S$ which converted a three-point function involving $k_s$ into a two point function at $k_L$. As a result, one should count the number of modes at $k_S$ to determine the fundamental limit set by cosmic variance. In detail, one finds that this ``no cosmic variance" result is utilizing most but not all of the information one would expect from this kind of mode counting~\cite{dePutter:2018jqk}.  In this sense, the non-linear physics of galaxy formation is actually measuring a lot of modes for us but processing it in a way where we can recover most of its statistical value without actually resolving these modes ourselves.
\end{itemize}
For these reasons, the prospects of measuring local non-Gaussianity, or any non-Gaussianity with a non-trivial squeezed limit, is quite promising in the next generation.  Concretely, the SPHEREx satellite~\cite{Dore:2014cca} is specifically designed to go after these low-$k$ mode for the purpose of measuring the scale-dependent bias.  Their forecasts , or future ground based surveys~\cite{DESI:2022lza}, for the local shape (i.e.~$m=0$ or $\nu =3/2)$) suggest that $\sigma(\fnlloc) = 0.2-0.5$ could be achievable in the relatively near future, which translates into a factor of 10-25 improvement over the current Planck constraints.  By combining maps of galaxies with CMB lensing from CMB-S4, it is possible that surveys like LSST/Vera Rubin could achieve similar results~\cite{Schmittfull:2017ffw}.  For additional massive particles (intermediate shapes) will be more sensitive than Planck to additional particles with $m \lesssim H$~\cite{Gleyzes:2016tdh}.

\begin{figure}[h!]
\centering
\includegraphics[width=4.5in]{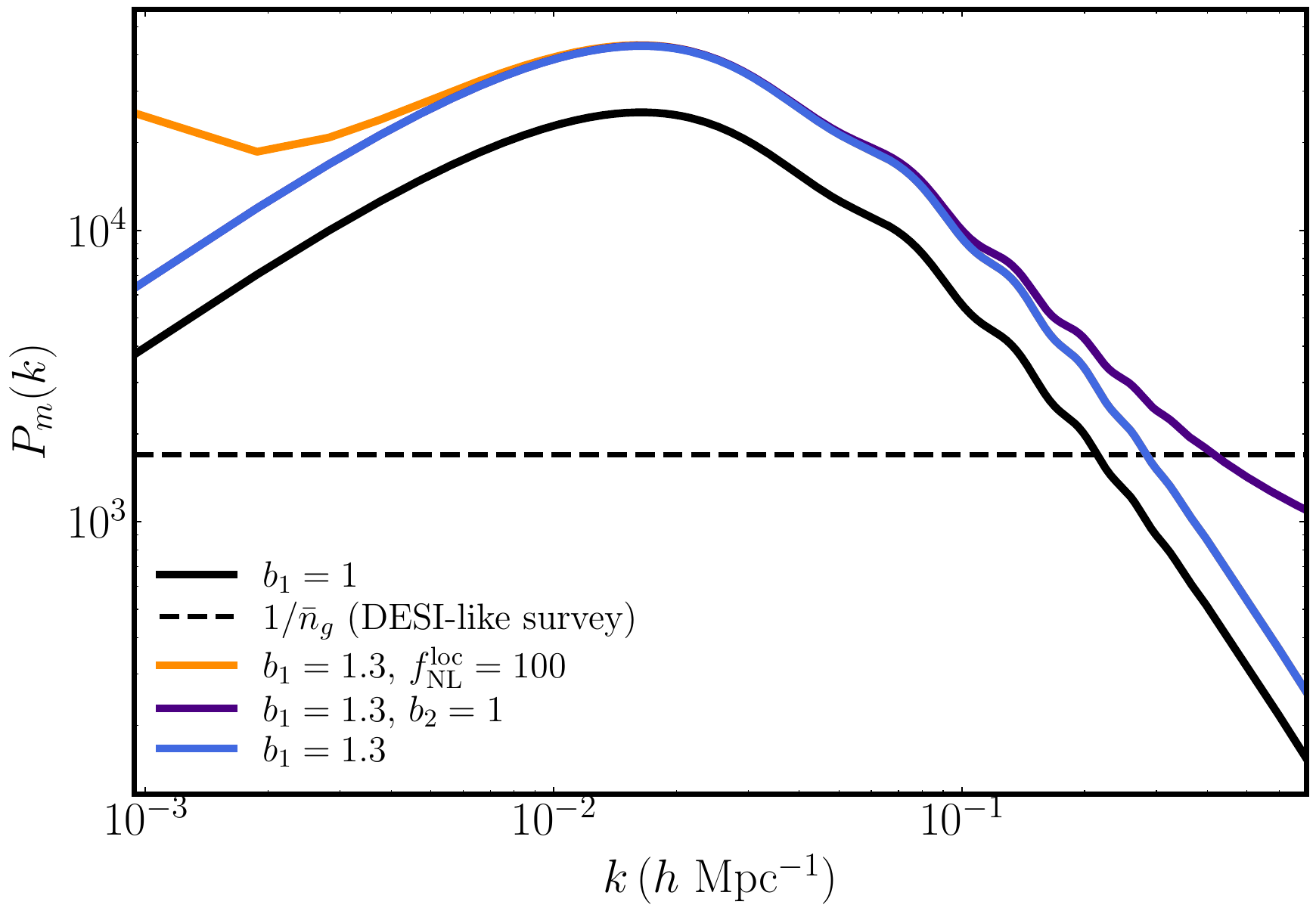}
\caption{Galaxy power spectrum for difference choices of bias parameters and including local non-Gaussianity.  We see that adding $\fnlloc$ changes the low-$k$ behavior, while $b_2$ (and other bias parameters) changes the high-$k$ behavior. The shot-noise of a DESI-like survey is shown for comparison.  Notice that the signal to noise is strongly affected by $b_1$ for a fixed density of objects. }
\label{fig:fnlloc}
\end{figure}

The fundamental limitation of these surveys turns out to be ``shot noise" rather than the detector noise that we saw limited that available modes in the CMB. We measure the three dimensional position of some fixed number of galaxies, $N_{\rm gal}$, and reconstruct a galaxy density field $\delta_{\rm gal}(\x) = \rho_{\rm gal}(\x) / \bar \rho_{\rm gal}$.  However, notice that the matter density contrast $\delta_m$ is effectively continuous, while the density contrast of galaxies is discrete.  This disconnect between the continuous underlying density and the discrete locations of galaxies gives rise to shot noise.  The galaxies follow the background density field, $\delta_m(\x)$, but their exact location is still random within any small region and uncorrelated with the other galaxies we might measure (this is sometimes implement explicitly in the form of ``stochastic bias").  In position space, this would imply the correlation between any two galaxies at positions $\x$ and $\x'$ contains a term proportional to $\delta(\x-\x')$, reflecting the non-deterministic component of galaxy formation.  In the galaxy power spectrum, the shot noise manifests itself as an additional constant term
\beq\label{eq:gal_cov}
P_{\rm gal} (k) \approx b_1^2 P_m(k) + \frac{1}{\bar n_{\rm gal}} \ ,
\eeq
where $\bar n_{\rm gal} = N_{\rm gal} / V$. This tells us that if we want to measure the structure of $P_{m}(k)$ at a given scale $k$, then we need the $\bar n_{\rm gal} \times (b_1^2 P_m(k) ) \gg 1$.  Given this source of noise, the Fisher matrix for a the power structure in a LSS survey will take a form familiar form from the CMB, Equation~(\ref{eq:fisher_CMB1}), 
\beq
F_{ij} = V \int \frac{d^3 k}{(2\pi)^3} \frac{\partial P_{\rm gal}(k)}{\partial \Theta_i}  \frac{\partial P_{\rm gal} (k)}{\partial \Theta_j} \frac{1}{\left(b_1^2 P_{\rm m}(k) + \frac{1}{\bar n_{\rm gal}} \right)^2} \ .
\eeq
This formula is approximate as we have neglected the fact the measurements are in redshift and angles (not three dimensional positions), and we dropped higher order terms in the noise covariance (i.e.~we use Equation~(\ref{eq:gal_cov}) when calculating the noise). Importantly, the parameters $\Theta_i$ now include the bias parameters $b_1, b_2, ...$ as we typically do not know them from first principles and therefore must also measure them in the data.  It is often assumed that $b_\Phi \propto (b_1-1)$ is something we know from first principles, and therefore we do not have to fit $b_\Phi$ independently; this assumption has come under some recent scrutiny~\cite{Barreira:2022sey} and is surely overly optimistic.

\subsection{Equilateral Non-Gaussianity}

\subsubsection*{Theoretical Background}

The story of inflation is often told in terms of a fundamental scalar field $\phi$ with a canonical kinetic term and a potential $V(\phi)$.  Inflation is a period during the early Universe when the energy density associated with the scalar field is dominated by its potential energy, $T_{\mu\nu} \approx g_{\mu \nu} V(\phi)$, where $V(\phi) \approx {\rm constant}$ for at least 50-60 e-folds.  However, there is no reason to assume that inflation is described by a weakly coupled (fundamental) scalar.  As we saw above, we can define inflation as a pattern of symmetry breaking where the long wavelength adiabatic scalar density fluctuations are the Goldstone boson of the spontaneously breaking time translations\footnote{The fact that this is also a metric fluctuation is the familiar fact that the metric eats the Goldstone when time translations are weakly gauged by gravity.}.  Naturally, we can ask if this pattern of symmetry breaking is the result of a weakly coupled scalar (like a fundamental Higgs field) or something more dynamical (like Technicolor). Like electroweak symmetry breaking, writing a specific model of inflation in terms of a weakly coupled scalar and a $V(\phi)$ is subject to a number of questions about fine tuning~\cite{Baumann:2014nda}, which has motivated an exploration of more dynamical models of inflation (e.g.~\cite{Alishahiha:2004eh,Dimopoulos:2005ac}).  Interesting, as we will see, these models typically produce novel observational signatures as a consequence.

Just like particle physics, one way to test for new physics in the inflationary sector is to study the interactions of the mode we have observed\footnote{This strategy is similar to a Higgs factory approach to searching for BSM physics at colliders.}, $\zeta$ (or $\pi$). At low energies, these fields can have Lorentz-violating derivative interactions like $\dot \zeta^3$ that give rise to a bispectrum, in terms of $\Phi$, 
\beq
 f_{\mathrm{NL}}^{\mathrm{eq}}  B_{\Phi}^{\mathrm{eq}}\left(k_{1}, k_{2}, k_{3}\right)=162 f_{\mathrm{NL}}^{\mathrm{eq}} \frac{\Delta_{\Phi}^{5}}{k_{1} k_{2} k_{3}\left(k_{1}+k_{2}+k_{3}\right)^{3}} \ ,
\eeq
where we have taken the limit $n_s \to 1$.  Notice that expanding in $k_1 \ll k_2, k_3$ this obeys the single-field consistency condition in Equation~(\ref{eq:sfcc}).  By dimensional analysis, the amplitude of non-Gaussianity is set by $\fnleq \Delta_\zeta \propto H^2/\Lambda^2$ where $\Lambda$ is the UV cutoff of the EFT of Inflation or, said differently, the energy scale where perturbative unitarity fails~\cite{Baumann:2011su,Baumann:2015nta,Grall:2020tqc}.  

\begin{figure}[h!]
\centering
\includegraphics[width=5.5in]{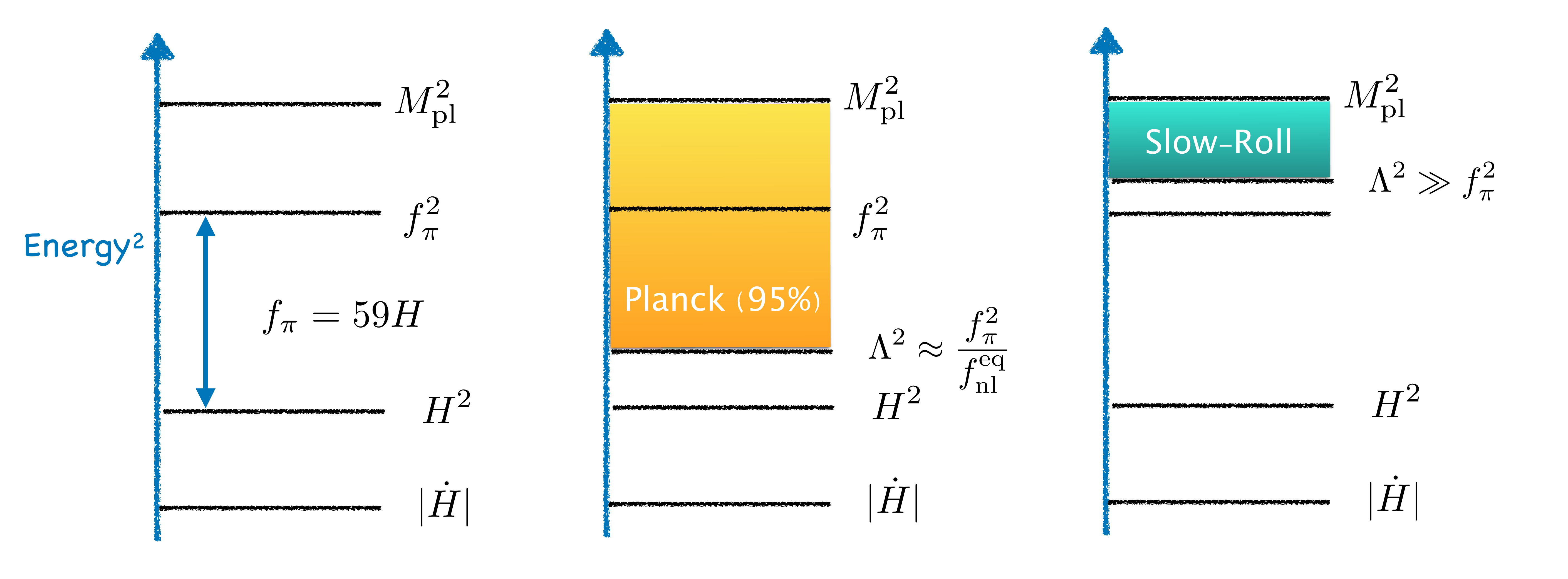}
\caption{The energy scales that control single-field inflation: $H$ is the scale relevant for the freeze-out of the scalar fluctuations ($\pi$), $f_\pi$ defines the scale of the clock, $\Lambda$ defines the strength of self-interactions of the Goldstone boson, and $\dot H$ is the deviation from an exact de Sitter background. We have assumed all the scales are small that then reduced Planck mass, $\Mpl$, so that they are in the regime of control of perturbative quantum gravity.}
\label{fig:E_scales}
\end{figure}

In order to make sense of the constraints on $\fnleq$ we need to understand the other energy scales in the problem~\cite{Baumann:2011su}, as illustrated in Figure~\ref{fig:E_scales}.  The scale where the symmetry is broken can be defined by the Goldstone boson decay constant $f_\pi$ which in canonical slow roll models is given by $f_\pi^4 = \dot \phi^2$.  Interestingly, in the EFT of inflation, this scale controls the amplitude of scalar fluctuations, $4 \pi^2 A_s = H^4 /f_\pi^4$, so that we know $f_\pi = 59 H$ from the value of $A_s$ measured by Planck, $A_s \approx 2.1\times 10^{-9}$.  In order for the model of inflation to be described by a weakly coupled scalar, the UV scale $\Lambda$ must be larger than the scale of the background $\Lambda \gg f_\pi$ or $\fnleq \ll 1$~\cite{Creminelli:2003iq}.  In contrast, any measurement of $\fnleq > 1$ would exclude canonical slow roll inflation.  In addition, there is a conjecture that $\fnleq \ll 1$ is always described by a weakly coupled fundamental scalar~\cite{Baumann:2015nta}.  In this precise sense, $\fnleq$ allows us to distinguish mechanisms of inflation.

There are a number of ways to generate $\fnleq>1$ in controlled examples.  The $P(X)$ models offer a simple class~\cite{Chen:2006nt}, where we still have a single scalar field $\phi$ but now the action is
\beq
{\cal L} = P(X) - V(\phi) \qquad X\equiv \frac{1}{2} \partial_\mu \phi \partial^\mu \phi \ .
\eeq
DBI inflation, where $P(X) = \Lambda^4 \sqrt{1- X/\Lambda^4}$, is one such example where the full non-linear structure of $P(X)$ can be derived from a UV completion~\cite{Alishahiha:2004eh}. However, the equilateral shape arises far more generally as the result of interactions of the inflaton with heavy fields~\cite{Chen:2009zp,Tolley:2009fg,Achucarro:2010da,Baumann:2011su,Assassi:2013gxa}.  Although these additional fields will introduce a deviation from the single field consistency conditions in the soft-limit for physically producing the extra mode, the signal to noise is dominated by the self-interactions of the inflaton that  are generated by integrating out the extra field so that $\Lambda^2$ is controlled by the gap\footnote{In non-relativistic theories, the gap is not solely determined by the mass so this distinction can be important.} in the spectrum of fluctuations.

One of the general characteristics of equilateral-type non-Gaussianity that makes it a particularly interesting theoretical target is that it usually appears with a pole in the total energy~\cite{Maldacena:2011nz,Raju:2012zr,Baumann:2022jpr}, $(k_1+k_2+k_3)^{-n}$ .  The remarkable thing about this pole is that the residue is the flat space scattering amplitude for the associated field.  In this sense, the bispectrum contains the same dynamical information as we would associated with the $S$-matrix.  This connection can be understood if we think of the pole as the cosmological analogue of the energy conserving $\delta$-function we would get from scattering $0 \to 3$ particles.  The fact that there are no poles at physical momenta, (e.g. $(k_1+k_2-k_3)^{-n}$) is a direct consequence of the fact these are quantum vacuum fluctuations~\cite{Green:2020whw,Green:2022fwg} such that all the fluctuations (modes) have positive energy (frequencies).  This observation also opens the possibility of testing the quantum origin of structure from inflation via equilateral non-Gaussianity~\cite{Green:2020whw}.

\subsubsection*{Observational Signatures}

Equilateral non-Gaussianity is the hallmark of a probe driven by the number of observed modes, as the signal is essentially scale-invariant and statistical in nature.  The best current constraints come from the CMB (Planck), namely $\fnleq = -26 \pm 47$~\cite{Planck:2019kim}.  However, if we are going to reach $\sigma(\fnleq) ={\cal O}(1)$ we will need roughly 100-1000 times more modes than Planck, something that is simply not possible in the CMB.  We must therefore turn to the three-dimensional volume of the Universe for more modes, via galaxy surveys or some other LSS probe~\cite{Chang:2022lrw}.

The difference between equilateral and local non-Gaussianity is that the signal-to-noise of the equilateral bispectrum is maximized by including the largest $k_{\rm max}$ possible.  Very schematically, the number of modes in the survey is $V k_{\rm max}^3$, so at a fixed survey volume, $V$,  you will gain/lose a huge amount of information from a relatively small change in $k_{\rm max}$.  The problem is that $k_{\rm max}$ is often not limited by the noise of the survey (the shot noise discussed previously) but by the limitations in modeling nonlinear evolution~\cite{Baldauf:2016sjb}.  To understand the challenge, let us again return to our biasing model, ignoring the scale-dependent bias~\cite{Desjacques:2016bnm}
\beq
\delta_{\rm gal}(\x) = b_1 \delta(\x) +b_2 \delta^2(\x) + b_{{\cal G}_2} (\nabla_i \nabla_j \Phi)^2 + b_3 \delta^3(\x) +\ldots 
\eeq
We have introduced the tidal tensor bias terms, starting with $b_{{\cal G}_2}$, as theses term are consistent with the equivalence principle~\cite{McDonald:2009dh} but greatly expand the number of free parameters of our model.

The signal we are trying to measure is 
\beq
\langle \delta_{\rm gal}(\k_1)\delta_{\rm gal}(\k_2)\delta_{\rm gal}(\k_3) \rangle_{\rm  f_{\mathrm{NL}}^{\mathrm{eq}} }  =b_1^3 \T(k_1) \T(k_2) \T(k_3)  f_{\mathrm{NL}}^{\mathrm{eq}}  B_{\Phi}^{\mathrm{eq}}\left(k_{1}, k_{2}, k_{3}\right) (2\pi)^3 \delta(\k_1+\k_2+\k_3) \ .
\eeq
However, we notice that there are lots of new contributions to the bispectrum of galaxies, e.g.
\beq
\langle \delta_{\rm gal}(\k_1)\delta_{\rm gal}(\k_2)\delta_{\rm gal}(\k_3) \rangle \supset b_1^2 b_2 \left(P_{\rm gal}(k_1)  P_{\rm gal}(k_2) + {\rm permutations} \right)(2\pi)^3 \delta(\k_1+\k_2+\k_3) \ .
\eeq
We can define how similar two bispectra are by defining an inner product,
\beq
B_i \cdot B_j \equiv  \int \frac{d^3 k_1 d^3 k_2 d^3 k_3}{(2\pi)^9} \frac{B_i(k_1,k_2, k_3) B_j(k_1,k_2, k_3)}{6 P_{\rm gal}(k_1) P_{\rm gal}(k_2) P_{\rm gal}(k_3) } (2\pi)^3 \delta(\k_1+\k_2 +\k_3) \ ,
\eeq
and their cosine,
\beq
\cos( B_i, B_j ) = \frac{B_i \cdot B_j}{\sqrt{B_i \cdot B_i} \sqrt{B_j \cdot B_j}} \ .
\eeq
This cosine is defined so that when $|\cos(B_i, B_j )| \approx 1$, the two bispectra are very degenerate making the error bars on either parameter increase significantly when marginalizing over the other.  Alternatively, when $|\cos(B_i, B_j )| \ll 1$, we can effectively measure both parameters independently at the same time.  In proper forecasts, $P_{\rm gal}(k)$ includes the shot-noise term (and potentially higher order stochastic effects) discussed around Equation~(\ref{eq:gal_cov}).  However, we will assume that the shot noise is low enough (which is under our experimental control with the design of the survey) so that we are primarily limited by our understanding of non-linear effects.

If we were only to measure the bispectrum and power spectrum, the cosine between the bispectra controlled by $\fnleq$ and $b_{n}$ respectively is close to one for most $n$. This means that a bispectrum analysis is very sensitive to higher-order nonlinear effects ($\delta^n$). If we marginalize over $b_n$ then it will increase our error bars on $\fnleq$ significantly.  Alternatively, if we hold $b_n$ fixed, a small error in the value of the $b_n$ will lead to a bias, in the statistical sense, in our measurement of $\fnleq$ (because the measurements of $\fnleq$ and $b_n$ are highly correlated).  It was argued in~\cite{Baldauf:2016sjb} that one can reduced the potential for bias by introducing theoretical uncertainty as a component of the error of the survey, at the cost of reducing the overall sensitivity.

\begin{figure}[h!]
\centering
\includegraphics[width=5.in]{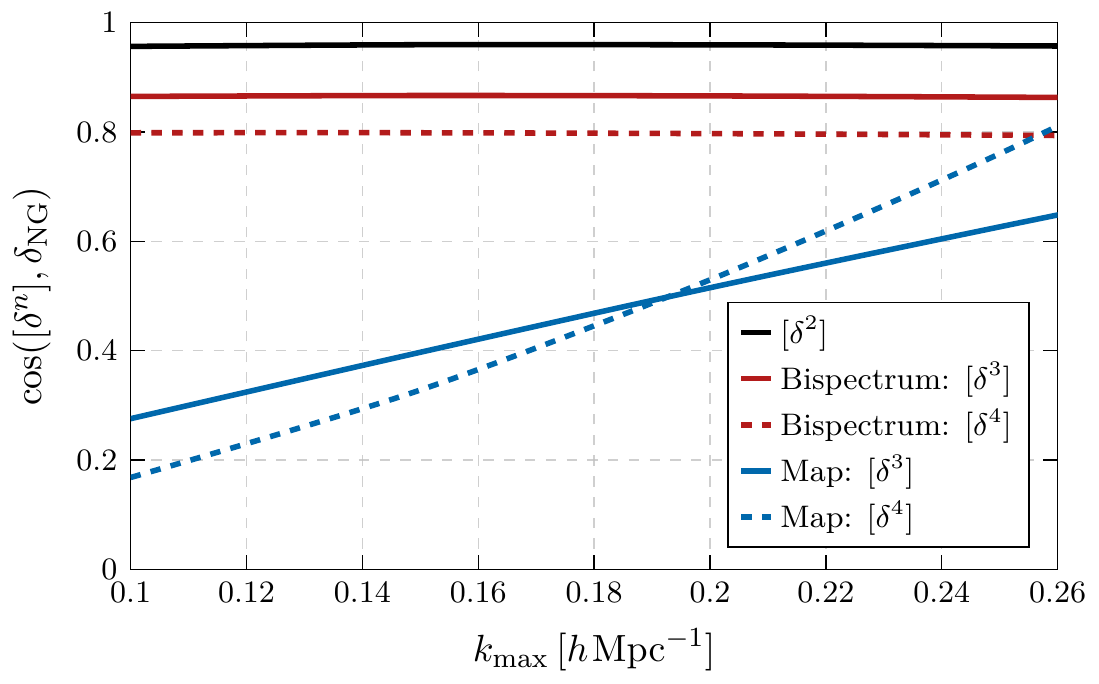} 
\caption{{\it Left:} Cosine between $\fnleq$ and the bias parameter $b_{n >1}$, as defined by the $(S/N)^2$, for the bispectrum-only and map level constraint respectively. Here the $\fnleq$, $b_n$ changes to the map is labeled by $\delta_{\rm NG}$ and $\delta_n$ respectively. Note that $b_2 \delta_2$ primarily affects the bispectrum of the map and therefore the bispectrum and map-level cosines are the same.  Reproduced from~\cite{Baumann:2021ykm}.}
\label{fig:fnleq}
\end{figure}

Fortunately, the bispectrum and power spectrum do not contain all the information in the maps produced by LSS surveys~\cite{Baumann:2021ykm}.  To see this, we note that adding a parameter $b_{n>2} \neq 0$ not only alters the bispectrum, it also introduces a non-zero $(n+1)$-point function in the map,
\beq
\langle \delta_{\rm gal}(\k_1)..\delta_{\rm gal}(\k_{n+1}) \rangle \supset b_1^n b_n \left(P(k_1).. P(k_n)+ {\rm permutations} \right) (2\pi)^3 \delta\left(\sum_{i=1}^{n+1} \k_i \right) \ .
\eeq
In the same way that we defined a cosine from the signal-to-noise for the bipsectrum, we can generalize a cosine to include these higher point correlators.  Figure~\ref{fig:fnleq} shows how the degeneracy between $\fnleq$ and $b_{n>2}$ is dramatically reduced when all the information in the maps is included. This suggests that making a competitive measurement of $\fnleq$ may be less sensitive to nonlinearity than previously thought.

The reality is that we are far from having demonstrated that a realistic survey can achieve the kinds of sensitivities we would need to improve upon the measurement of $\fnleq$ from the CMB.  The first proof of principle for a measurement of $\fnleq$ using galaxy survey data was only performed in 2022 and had to make a number of very conservative choices~\cite{Cabass:2022wjy,DAmico:2022gki}.  Modeling the nonlinear evolution in sufficient detail to measure the bispectrum is technically challenging, and it is complicated even further by the fact that we observe galaxies in angle and redshift rather than three-dimensional positions.  Redshift space distortions become non-perturbative on surprisingly large scales and strategies to mitigate them is still a work in progress. A selection of current measurements and published forecasts for $\sigma(\fnleq)$ is shown in Table~\ref{table:fnleq} to give a sense of the current state of the art.

\begin{table}[h!]

\begin{center}
\begin{tabular}{ | c || c | c |} 
 \hline
 Survey & $\sigma(\fnleq)$ & $k_{\rm max}$ \\
 \hline 
 Planck~\cite{Planck:2019kim} & $-26 \pm 47$&  -- \\ 
  BOSS~\cite{Cabass:2022wjy} & $260 \pm 300$ & $0.08 \, h \, {\rm Mpc}^{-1}$ \\
 BOSS~\cite{DAmico:2022gki} & $2 \pm 212$ & $0.22 \, h \, {\rm Mpc}^{-1}$ \\
 \hline\hline
 CMB-S4~\cite{Abazajian:2019eic} & 22 & -- \\ 
Euclid~\cite{Amendola:2016saw} & 35 & $0.16 – 0.24 \, h \, {\rm Mpc}^{-1}$ ($0.7\leq z \leq 2$) \\ 
Mega-Mapper~\cite{Cabass:2022epm}& 17 & $0.2 \, h \, {\rm Mpc}^{-1}$ at $z=3$\\ 
PUMA~\cite{PUMA:2019jwd}& 4.5 & $0.1/D(a) \, h \, {\rm Mpc}^{-1}$ ($0.3\leq z \leq 6$)\\
 \hline
\end{tabular}
\caption{Current measurements and forecasts for $\fnleq$ for a selection of CMB and LSS surveys. The assumptions on $k_{\rm max}$ and other aspects of the LSS surveys vary wildly, but the final column gives a rough indication of how $k_{\rm max}$ was chosen in the respective bispectrum analyses or forecasts. Here $D(a)$ is the growth function discussed in Section~\ref{sec:intro_lss}. }\label{table:fnleq}
\end{center}
\end{table}

Despite these challenges, there is also reason to be optimistic that we can speed up simulations of the maps of the Universe to perform this analysis without the same reliance on a perturbative understanding of the modes~\cite{Jung:2022gfa}. There is evidence from multiple perspectives that the $\fnleq$ signal is encoded in these maps in a way that is distinguishable from late time non-linear evolution~\cite{Baumann:2021ykm,Biagetti:2020skr,Biagetti:2022qjl}.  There are more than enough modes in LSS surveys to make a competitive measurement. Preliminary work using these novel techniques suggests that measurements that exploit these features may be possible in the near future.

\newpage
\section{Conclusions}

In these lectures, we only scratched the surface of how cosmological data can be used to understand physics that manifests itself in map of the Universe. We are now entering the third decade of high precision cosmology and a wide range of ideas have been explored in actual analyses of cosmology data that take advantage of novel analytic, numerical, and statistical techniques. Naturally, for many models (or even vague ideas) you come up with, analyses constraining the relevant parameters have been performed, at least in a model that looks qualitatively similar to the specific example of interest.  In this sense, you can learn a lot from cosmological data without performing the analyses yourself.  Yet, at the same time, the data is public and new and interesting analyses are being performed by theorists all the time~\cite{Green:2022hhj}.

Our goal here was to see how the microphysics of a specific model manifests itself in the way we analyze our data to look for signals of beyond the Standard Model physics.  For particles with sub-MeV masses, the entropy of these particles cannot be hidden and will show up in the CMB and/or large scale structure in ways that depend on the characteristics of the particles and their interactions.  The most universal impact of these particles arise from gravity, via $\Neff$, $\sum m_\nu$, etc.  Direct interactions of new particles with themselves and/or the Standard Model are more model dependent but can be equally important.  Given that gravitational effects are detectable, a wide range of seemingly weak interactions can have measurable cosmological effects as well.  The same is true of the physics of inflation, where quantum fluctuations of the vacuum produce observable effects that can distinguish between models of inflation or probe the masses and spins of particles coupled to the inflaton.

One of the most exciting aspects of theoretical cosmology is the sense that a deeper understanding of the space of models, or their role in the history of the Universe, can directly impact our ability to test these models with current and future surveys.  Local non-Gaussianity offers a beautiful illustration of this interplay, where our deepening understanding of the space of inflationary non-Gaussian signals led directly to the discovery of scale-dependent bias and numerous advances in the statistical tools we use to extract this signal from data.  This pattern continues to this day, as new directions in cosmology are being driven by theorists expanding how we look at data. I hope these lectures help inspire you to want to learn this full picture and maybe even get in on the fun yourself.

\paragraph{Acknowledgements}

I am grateful to Daniel Baumann, Tim Cohen, Nathaniel Craig, Kyle Dawson, Olivier Dor\' e, JiJi Fan, Raphael Flauger, Vera Gluscevic, Yi Guo, Jiashu Han, Yiwen Huang, Misha Ivanov, Tongyan Lin, Aneesh Manohar, Joel Meyers, Chia-Hsien Shen, Oliver Philcox, Surjeet Rajendran, An\v ze Slosar, and Benjamin Wallisch for helpful discussions. I would also like to thank Daniel Baumann, Tim Cohen, Joel Meyers, and Ben Wallisch for detailed comments on these lecture notes. DG is supported by the US~Department of Energy under Grant~\mbox{DE-SC0009919}.


\newpage
\appendix

\section{Thermodynamics}\label{app:thermo}
\subsection{Some Important Formulas}

Reminder of some basic definitions and results:
\beq
n = \int \frac{d^3 p}{(2\pi)^3}  f_\pm(\p) \qquad \rho =\int \frac{d^3 p}{(2\pi)^3} E(p) f_\pm(\p) 
\eeq
where $E(p)$ is the energy of particle with three-momentum $\p$ and the distribution functions are Fermi-Dirac (+) or Bose-Einstein (-)
\beq
f_{\pm} = \frac{g_s}{e^{E(p)/T} \pm 1} \ .
\eeq 
For massless particles we have
\bea
n &=& a_{\star} \frac{\zeta(3)}{\pi^2} T^3 \\ 
\rho &=& g_\star \frac{\pi^2}{30} T^4
\eea
where $a_\star = g_\star = g_s$ for bosons and $g_\star = \frac{7}{8} g_s$, $a_\star = \frac{3}{4} g_s$ for fermions, and $g_s$ is the number of independent spin degrees of freedom.  Finally, during radiation domination, the Hubble parameter is given by
\beq
3 \Mp^2 H^2 = \rho = g_\star \frac{\pi^2}{30} T^4 \ .
\eeq
These definitions are useful shorthands for cosmology, as the simply represent the results of integrals we would otherwise have to do all the time.

\subsection{Proof of Entropy Conservation}

In deriving the approximate value of $\Neff =3$ in the Standard Model, used the fact that the total entropy is conversed in the expanding universe. We will derive this result here.  First, recall from thermodynamics that 
\beq
T d(s V) = d(\rho V) + p dV \ .
\eeq
In order to match the coefficient of $dV$ on both sides of the equations, we have $Ts = \rho+p$.  After making this identification, we then have $TVds = Vd\rho$.  Applying our definition of $s$ we get
\beq
ds = T^{-1} (d\rho +   dp) - s \frac{dT}{T}  = ds + dp - s\frac{dT}{T} \to dp = (\rho +p) dT \ .
\eeq  
Now, as a last step, we recall that conservation of the energy momentum tensor for a homogenous fluid requires that $\dot \rho +3 H(\rho+p) = 0$. 

Combining all the observations, we can derive the rate of change in the total entropy
\bea
a^{-3} \frac{\partial}{\partial t} (a^3 s) &=&3 H s + T^{-1} [ \dot \rho + \dot p - \frac{\dot T}{T}(\rho+p) ] \nonumber\\
&=& 3 H s- T^{-1}3 H(\rho+p) +  \dot p - \frac{\dot T}{T}(\rho+p) \nonumber \\
&=&   \dot p - \frac{\dot T}{T}(\rho+p) \nonumber \\
&=& 0 \ .
\eea
This shows that in thermal equilibrium, the total entropy is conserved.

\section{Inflation} \label{app:inflation}

In this appendix, we will review some basic facts about inflation to complement the discussion in the main text.
\subsection{Slow roll inflation}

The canonical example of inflation is {\it slow-roll inflation}.  The idea is very simple, we have a scalar field $\phi$ with a Lagrangian
\beq
{\cal L} = -\tfrac{1}{2} \partial_\mu \phi \partial^\mu \phi - V(\phi)
\eeq
For some range of field values, we would like the potential to be flat enough that it is potential energy dominated.  Here, we can use Einstein's equation to find
\beq
3 M_{\rm pl}^2 H^2 = \tfrac{1}{2}\dot \phi^2 + V(\phi) \ .
\eeq
For the RHS to be potential energy dominated, we require that $\dot \phi^2 \ll V(\phi)$ over a sufficiently long period of time.  In practice, this means that we need $\dot \phi$ to be nearly constant as well.  The equations of motion for $\phi$ are given by
\beq
\ddot\phi + 3 H \dot \phi + V'(\phi) = 0 \ .
\eeq
Demanding the slow-roll solution $\dot \phi \sim - V' / 3H$.  Using this solution with potential energy domination tells us that $M_{\rm pl}^2 (V'/V)^2 \ll 1$.  This inequality is equivalent to $|\dot H| \ll H^2$.  This ensures that we have $a(t) \sim e^{H t}$ and $H$ is nearly constant.  These meet our requirements for creating superhorizon fluctuations. 

Maintaining this solution requires that $\eta = M_{\rm pl}^2 V'' /V \ll 1$.  If we expand in $\phi(t,x) = \phi(t) + \varphi(x,t)$, then we see that $m_{\varphi}^2 \ll H^2$ is the same as demanding small $\eta$.  Therefore, we expect to see thermal fluctuations of $\varphi$.
\vskip 6pt
{\it Ending Inflation :} \hskip 4pt  The motivation having slow roll inflation is that there is a easy mechanism to end inflation.  This can be done in several ways by engineering the potential and/or interactions with other fields.  The main feature is that there exists a value of $\phi = \phi_*$ above or below which the slow roll conditions no longer hold.  This could happen because a field $\chi$ has a mass that depends on $\phi$ and becomes negative at this critical value (hybrid mechanism).  Alternatively, the potential may simply cease to be sufficiently flat.

Although ending inflation is the prime motivation for inflation rather than meta-stable de Sitter, the details do not much matter for single-field inflation.  Because $\zeta$ in conserved outside the horizon in single-field inflation, the predictions to do depend on how inflation ends.  However, in multifield inflation, the situation we be very different as we saw in Section~\ref{sec:local}.

\subsection{Breaking Time Translations}

The important feature of slow roll is that the field $\phi(t)$ plays the role of a physical clock~\cite{Cheung:2007st}.  This is the important feature of the background evolution of $\phi$ that allows inflation to end everywhere in the Universe at once.  However, what we observe is not $\phi$ but the fluctuations around this background.  Therefore, if one is only interested in the predictions for observations, it makes sense to construct the theory for the fluctuations directly without having to write an explicit model for the background.

The general idea is that a physical clock defines a preferred notion of time.  If we were in a non-dynamical (rigid) flat background, this would spontaneously break the time translation symmetry, $t \to t+c$.  In the rigid limit, there must be a goldstone boson $\pi$ associated with this broken symmetry.  Because this symmetry is spontaneously broken, this goldstone must non-linearly realize the time translation so have an invariant action. It is easiest to write the action with this property is in terms of an operator that linearly realizes the symmetry $U=t+\pi$.  In terms of $U$, any scalar function fill produce a symmetric action and therefore $S = \int d^4 x f(U=t +\pi)$ will be time translation invariant. However, we must impose an addition non-trivial constraint by demanding that the background does not contain a tadpole for $\pi$, or $\langle \pi \rangle = 0$.

We will not be working in flat space, and the time translations will be (weakly) gauged by gravity.  Because the time translation is now a gauge symmetry, we are free to choose many different gauges.  The two that will be most important to use are: 
\vskip 4pt
{\bf $\pi$-gauge :} In this gauge, the metric is takes the form
\beq
ds^2 = - N^2 dt^2 + a(t)^2\delta_{ij} (d x^i + N^i dt) (d x^j N^j dt) \ .
\eeq
We will be ignoring the tensor fluctuations for now, but they are easy to put back in.  In this gauge, $\pi$ is the propagating scalar degree of freedom, and we must solve the ADM constraints to determine $N$, $N^i$ as a function of $\pi$.  During inflation, $\pi$ gauge is made convenient because of an analogue of the goldstone boson equivalence theorem.
\vskip 4pt
{\bf $\zeta$-gauge :} 
In this gauge (also called {\bf unitary gauge}), the metric is takes the form
\beq
ds^2 = - N^2 dt^2 + a(t)^2 e^{2 \zeta} \delta_{ij} (d x^i + N^i dt) (d x^j N^j dt) \ .
\eeq
In this gauge, we have set $\pi = 0$ to that the scalar fluctuation appears only in the metric. 

To related these gauges, recall that under a diffeomorphism, $\pi \to \pi - \xi(t,x)$.  Therefore, the diffeomorphism that relates $\zeta = 0$ ($\pi$ gauge) and $\pi = 0$  ($\zeta$-gauge) is $\zeta = -H \pi + \ldots$~\cite{Maldacena:2002vr}.  This expression is very useful because we can do calculations in $\pi$ gauge, where the action is easier to write down, and convert back to $\zeta$ gauge at horizon crossing.  Converting back is important because $\zeta$ is the perturbation that is exactly constant outside the horizon.
\vskip 8pt
For a number of reasons, I would like to work in $\pi$-gauge throughout these lectures.  The original papers derive many of their results in unitary gauge~\cite{Cheung:2007st}.  While it is sightly easier to derive a few results in unitary gauge, $\zeta$ is like the longitudinal mode of a gauge boson and, just like a massive gauge field, for understanding physics to short distances (high energies), it is easier to work with $\pi$~\cite{Baumann:2011su}.

\subsection{Actions}

Let us construct the action for the fluctuations around an FRW solution.  In $\pi$ gauge, recall that under the diffeomorphism $t \to t + \xi(x,t)$ we transform $\pi \to \pi - \xi(x,t)$ to that the combination $t + \pi$ transforms as a scalar.  To write a fully diffeomorphism invariant action, we may write
\begin{align}\label{eq:EFT_action}
S = \int d^4 x \sqrt{-g} \bigg[ &\, \tfrac{1}{2} M_{\rm pl}^2 R - F(t+\pi) - c(t+\pi) \partial_\mu (t+\pi) \partial^\mu(t+\pi) \\
& + {\cal O} \Big( (\partial_\mu (t+\pi) \partial^\mu(t+\pi) +1)^2 \Big) \, \bigg] \nonumber \ ,
\end{align}
where we have contracted the indices with the metric $g^{\mu \nu}$.  Here the last term means ``quadratic in fluctuations".  As written, it may not be clear why we have organized terms in this way.  Any action that is a function of $t+\pi$ would have been equally good, however, we still need to impose our no-tadpole condition.  To accomplish this goal, we will demand that the solutions to the equations of motion permit an FRW solution of the form
\beq
ds^2 = - dt^2 + a(t)^2 (\frac{dr^2}{1-K r^2} + r^2 d \Omega^2) \ ,
\eeq
with $\langle \pi \rangle =  0$. We will start by solving Einsteins equation around this background with a general $a(t)$.  All the terms after the Einstein-Hilbert term ($R$) in Equation~(\ref{eq:EFT_action}) represent contributions to $T^{\mu \nu}$.  Differentiating with respect to $g^{\mu \nu}$ one finds that
\bea
3 \Mp^2 (H^2 + \frac{K }{a^2})&=& c(t) + F(t) \\
3 \Mp^2 (\dot H + H^2) &=& F(t) - 2 c(t)
\eea
Putting these equations together, we see that $c(t)= - \Mp^2 (\dot H - \frac{K}{a^2}) $ and $F(t) = \Mp^2(3 H^2 + \Mp^2 \dot H + 2 \frac{K}{a^2})$.  Putting these terms back in the action, we find
\begin{align}
S = \int d^4 x \sqrt{-g} [ \, \tfrac{1}{2} M_{\rm pl}^2 R &- \Mp^2(3 H^2 + \Mp^2 \dot H + 2 \frac{K}{a^2})\\ &+\Mp^2 (\dot H + \frac{K}{a^2}) \partial_\mu (t+\pi) \partial^\mu(t+\pi) + \ldots \, ] \ .
\end{align}
For a generic FRW solution, this construction will hold (so my title is a little misleading -- this is the EFT of any FRW, inflation or otherwise).  However, for a generic FRW, there is no advantage to working in $\pi$ gauge as perturbations will become non-linear and are more easily understood in Newtonian gauge (as we did in the first lecture). At this point, $H(t)$ is some unknown function although we usually impose $\dot H \leq 0$ by the null-energy condition.

Our primary motivation for this construction was to study the perturbations during inflation.  We will take as our definition of inflation a period during which $H^2(t) \gg |\dot H|$. This is the same as our first slow roll condition for the explicit model of using scalar fields.  Since $H = \dot a / a $, we will define the number of e-folds as $\log(a_e/a_i) \equiv N_e = \int^{t_e}_{t_i} H(t) dt$.  We require 60 e-folds of inflation to solve the horizon and flatness problems (assuming a high reheat temperature).

In the context of inflation, we can neglect the curvature, $K /a^2 \to 0$, as it vanishes exponentially quickly.  Since $\dot H$ is small, it may not be obvious that curvature cannot be important, but we will see that the hierarchy of scales required in the model ensures that $K/a^2$ is negligible.  If we ignore higher order terms, we have simply rewritten slow roll inflation.  To see this, note that our action is given by
\bea
S = \int d^4 x \sqrt{-g} [ \, \tfrac{1}{2} M_{\rm pl}^2 R- \Mp^2(3 H^2 + \Mp^2 \dot H)+\Mp^2 \dot H \partial_\mu (t+\pi) \partial^\mu(t+\pi) ]\ .
\eea
If we work in unitary ($\zeta$) gauge, this is simply
\bea
S_{\rm unitary} &=& \int d^4 x \sqrt{-g} [ \, \tfrac{1}{2} M_{\rm pl}^2 R- \Mp^2(3 H^2 + \Mp^2 \dot H)+\Mp^2 \dot Hg^{00} ]\\
&=&  \int d^4 x \sqrt{-g} [ \, \tfrac{1}{2} M_{\rm pl}^2 R- V(\phi)+\tfrac{1}{2}\dot \phi^2 g^{00} ] \to  \int d^4 x \sqrt{-g} [ \, \tfrac{1}{2} M_{\rm pl}^2 R-\tfrac{1}{2}\partial_\mu \phi \partial^\mu \phi - V(\phi) ] \nonumber
\eea
In this precise sense, we see that slow-roll inflation is equivalent to keep just these universal terms in the EFT of inflation.
\vskip 6pt
Now it should be clear how to extend the EFT of inflation to describe models beyond the regime of slow roll.  Slow roll is simply the universal part of the action for the goldstone boson, but we are free to add terms that are quadratic, or higher, in the fluctuations without altering our background solution.  For example, we can write 
\begin{align}
S = \int d^4 x \sqrt{-g} \bigg[& \, \tfrac{1}{2} M_{\rm pl}^2 R- \Mp^2(3 H^2 + \Mp^2 \dot H)+\Mp^2 \dot H \partial_\mu (t+\pi) \partial^\mu(t+\pi) \\ \nonumber
& + \sum_{n=2}^{\infty} \frac{1}{n!} M_n^4(t+\pi) (\partial_\mu (t+\pi) \partial^\mu(t+\pi)+1)^n  \bigg] \label{equ:PX} \ .
\end{align}
Construction, $M_n^4$ will contribute to the action at order $\pi^n$. We may also include terms with additional derivatives acting on $\pi$. Again, we can use the operator $U = t+\pi$ that $U$ transforms linearly under diffeomorphisms and simply add higher-derivatives of $U$ to the action.  However, in curved space, we must use covariant derivatives in curved spacetime which may cause addition complications in counting powers of $\pi$ for a given interaction.  It is therefore useful to introduce quantities such as 
 \beq
\label{equ:def}
   \Big[\nabla_\mu \nabla_\nu U  \Big]  \equiv  \Big(\nabla_\mu \nabla_\nu U - H(g_{\mu \nu} + \nabla_{\mu} U \nabla_{\nu} U) \Big) \ ,
 \eeq
 so that $\Big[\nabla_\mu \nabla_\nu U  \Big]  = {\cal O}(\pi)$.

If we keep the $M_2^4$ term, we can see that $M_2^4 = -2\Mp^2 \dot H (1-c_s^2)c_s^{-2}$ and the quadratic Langrangian takes the form
\beq
{\cal L}_{\rm quad} = -\frac{\Mp^2 \dot H}{c_s^2} (\dot \pi^2 - \frac{c_s^2}{a^2} (\partial_i \pi)^2 ) \ .
\eeq
We see that $c_s$ controls the speed of propagation so we require $c_s\leq 1$, The additional terms $M_{n>2}^4$ only contribute to non-Gaussian correlation functions and are included perturbatively.  

So far, nothing I have said ever required that I picked any gauge.  The action with $t+\pi$ is actually gauge invariant!  This was clear because I went to $\zeta$ gauge by setting $\pi = 0$.  All the complications of each gauge come from picking a gauge fixing condition {\it and} solving the ADM constraint equations.  Neither of these steps have been performed so far.  We will not go through the algebra but will simply quote the result.  For this case, the ADM constraints in $\pi$ gauge give us~\cite{Cheung:2007sv}
\beq
\delta N = \epsilon H \pi \qquad {\rm and} \qquad \partial^i N_i = - \epsilon \frac{H \dot \pi}{c_s^2} \ ,
\eeq
where $\epsilon = - \dot H / H^2 \ll 1$. Substituting back into the action gives us the quadratic terms.
\beq
{\cal L}_{\rm quad} = -\frac{\Mp^2 \dot H}{c_s^2} (\dot \pi^2 - \frac{c_s^2}{a^2} (\partial_i \pi)^2 + 3 \epsilon H^2 \pi^2 ) \ .
\eeq
As we established previously, what we actually observe are the fluctuations in $\zeta$ so one should change variables.  Plugging $\pi \approx -\zeta/H$ we find that we get
\beq
{\cal L}_{\rm quad} = -\frac{\Mp^2 \dot H}{H^2 c_s^2} (\dot \zeta^2 - \frac{c_s^2}{a^2} (\partial_i \zeta)^2 ) \ .
\eeq
The cancelation of the ``mass term" is not surprising, as it is a requirement that $\zeta(\k\to 0, t$ is a constant (conservation of the adiabatic mode), which is only possible if $\zeta$ is exactly massless.

It is noteworthy that the coupling to gravity was suppressed by $\epsilon \ll 1$.  This is a related of a goldstone boson equivalence theorem, which requires that the Goldstone decouples from dynamical gravity in the limit $\Mpl \to \infty$, $\dot H \to 0$, holding $\Mpl^2  \dot H$ fixed.  The decoupling limit is useful because it tells us, up to terms suppressed by $\epsilon$ or $H^2/\Mpl^2$, the action of $\pi$ {\it without} dynamical gravity captures the dynamics of the scalar mode of the metric.  See~\cite{Baumann:2011su} for further explanation.

\section{Bispectrum Signal-to-Noise} \label{app:SN}

In the main text, we calculated the $(S/N)^2$ for the bispectrum using a change of variable $k_2 , k_3 \to x_2, x_3$.  Here we sill show how that change of variables was performed.  

In general, when we integrate over all the modes in the bispectrum, we have the following measure of integration
\beq
\int  \frac{d^3 \k_1}{(2\pi)^3}  \frac{d^3 \k_2}{(2\pi)^3}  \frac{d^3 \k_3}{(2\pi)^3} (2\pi)^3 \delta^3(\k_1 +\k_2 + \k_3) \ .
\eeq
To change variables, we first integral over the $\delta$-function integral to
\beq
\int  \frac{d^3 \k_1}{(2\pi)^3}  \frac{d^3 \k_2}{(2\pi)^3}  \ ,
\eeq
with $\k_3 = -\k_1-\k_2$. Now, we are left with an integral over $\k_1$ and $\k_2$.  By rotational invariance, the $(S/N)^2$ will depend on one of the angular integrals.  However, there will remain one $d \cos(\theta)$ integral is non-trivial as it represents the angles between the vectors $\k_1$.  To simplify this integral, we work in a polar coordinate basis with respect to $\k_1$ so that $\theta$ is the angle between $\k_1$ and $\k_2$.  In this basis, we have
\beq\label{eq:cos_def}
\cos(\theta) = (k_3^2 - k_2^2 -k_1^2) /(2 k_1 k_2)  \ ,
\eeq
which follows from $k_3^2 = (\k_1+\k_2)^2 = k_1^2 + k_2^2 +2k_1 k_2 \cos(\theta)$. Now let us rewrite our integral as
\beq
\int  \frac{d^3 \k_1}{(2\pi)^3}  \frac{1}{(4\pi^2)} \frac{1}{k_1^3} \int d\cos(\theta)\int d x_2 x_2^2
\eeq
where $x_2 = k_2/k_1$ and $x_3 = k_3 / k_1$.  We don't want to integrate over angles but over the length of the sides of the triangle made by $\k_i$. We can do this with the change variables in Equation~(\ref{eq:cos_def}) and $d \cos(\theta) \to x_3 d x_3 / x_2$. In this new basis, our integration measure becomes
\beq
\int  \frac{d^3 \k_1}{(2\pi)^3}  \frac{1}{(4\pi^2)} \frac{1}{k_1^3} \int x_3 x_3 \int d x_2 x_2
\eeq
Now we are left with the question, what is the domain of integration of $x_3$ and $x_2$?  Using permutations of the $\k_i$s, we can multiply our measure by a factor of 6 and then enforce $1 > x_2 > x_3$.  Solving for the end points of the cosine we first find that $x_3 > x_2 - 1$ (which corresponds to $\cos = -1$). Finally, the triangle inequality tell us $x_2 + x_3 > 1$ but $x_2 > x_3$ and therefore $x_2 > 1/2$.  Altogether, our final result is
\beq
\int  \frac{d^3 \k_1}{(2\pi)^3}  \frac{3!}{(4\pi^2)} \frac{1}{k_1^3}  \int_{1/2}^1 d x_2 x_2\int_{1-x_2}^{x_2} d x_3 x_3 \ .
\eeq
From here we get that the single to noise is
\beq
\left( \frac{S}{N} \right)^2 = V \int  \frac{d^3 \k_1}{(2\pi)^3}  \frac{3!}{(4\pi^2)}  \int_{1/2}^1 d x_2 \int_{1-x_2}^{x_2} dx_3 x_2^4 x_3^4\,\frac{ B(1,x_2,x_3)^2}{ \Delta_\zeta^6} \ ,
\eeq
where we have assumed exact scale invariance, $P(k) = k^{-3} \Delta_\zeta^2$.

\clearpage
\phantomsection
\addcontentsline{toc}{section}{References}
\bibliographystyle{utphys}
\bibliography{TASI_refs}

\end{document}